\documentclass[modern]{aastex61}

\renewcommand{\vec}{\mathbf}

\newcommand{\jgra}{{\jgr}}

\received{in press} \revised{} \accepted{}

\submitjournal{ApJS}

\shorttitle{Small-scale Magnetic Flux Ropes} \shortauthors{Hu et
al.} 

\begin{document}

\title{Automated detection of small-scale magnetic flux ropes in the solar wind: First results from the Wind spacecraft measurements}

\correspondingauthor{Qiang Hu} \email{qiang.hu@uah.edu}

\author[0000-0002-7570-2301]{Qiang Hu}
\affiliation{Department of Space Science and CSPAR \\
The University of Alabama in Huntsville \\
Huntsville, AL 35805, USA}

\author{Jinlei Zheng}
\affil{Department of Space Science \\
The University of Alabama in Huntsville \\
Huntsville, AL 35805, USA}

\author{Yu Chen}
\affil{Department of Space Science \\
The University of Alabama in Huntsville \\
Huntsville, AL 35805, USA}

\author{Jakobus le Roux}
\affil{Department of Space Science and CSPAR \\
The University of Alabama in Huntsville \\
Huntsville, AL 35805, USA}

\author[0000-0003-3936-5288]{Lulu Zhao}
\affil{Department of Physics and Space Sciences\\
 Florida Institute
of Technology\\
 Melbourne, FL 32901, USA}



\begin{abstract}
We have developed a new automated small-scale magnetic flux rope
(SSMFR) detection algorithm based on the Grad-Shafranov (GS)
reconstruction technique. We have applied this detection algorithm
to the Wind spacecraft in-situ measurements during 1996 - 2016,
covering two solar cycles, and successfully detected a total
number of 74,241 small-scale magnetic flux rope events with
duration from 9 to 361 minutes. This large number of small-scale
magnetic flux ropes has not been discovered by any other previous
studies through this unique approach.  We perform statistical
analysis of the small-scale magnetic flux rope events based on our
newly developed database, and summarize the main findings as
follows. (1) The occurrence of small-scale flux ropes has strong
solar cycle dependency with a rate of a few hundreds per month on
average. (2) The small-scale magnetic flux ropes in the ecliptic
plane tend to align along the Parker spiral. (3) In low  speed
($<$ 400 km/s) solar wind, the flux ropes tend to have lower
proton temperature and higher proton number density, while in high
speed ($\ge$ 400 km/s) solar wind, they tend to have higher proton
temperature and lower proton number density. (4) Both the duration
and scale size distributions of the small-scale magnetic flux
ropes obey a power law. (5) The waiting time distribution of
small-scale magnetic flux ropes can be fitted by an exponential
function (for shorter waiting times) and a power law function (for
longer waiting times). 
(6) The wall-to-wall
time distribution obeys double power laws with the break point at
60 minutes (corresponding to the correlation length). (7) The
small-scale magnetic flux ropes tend to accumulate near the
heliospheric current sheet (HCS). The entire database is available at
\url{http://fluxrope.info} and in machine readable format
in this article.

\end{abstract}




\section{Introduction}\label{introduction}

Magnetic flux ropes, a type of space plasma structures,
characterized by their spiral magnetic field line configurations,
have long been studied in heliophysics. In particular, the
relatively large-scale flux rope structures, known as magnetic
clouds (MCs), have been well identified and modeled from in-situ
spacecraft measurements. They are generally believed to correspond
to coronal mass ejections (CMEs), originating from the Sun. On the
other hand, the relatively small-scale magnetic flux ropes
(SSMFRs) of duration ranging from tens of minutes to a few hours
in the solar wind at 1 astronomical unit (AU), have also been
identified and shown to have different occurrence rates and
possibly different origination mechanism from their large-scale
counterparts, i.e., MCs. As we will review below, focusing on
well-recognized efforts in building event databases, a number of
studies concerning the identification and characterization of
these structures in the solar wind, most were based on manual
operations and simplified models. We present our latest effort in
employing a totally different approach, based on the
state-of-the-art flux rope model, in an automated manner, to
detect and analyze the flux rope structures in the solar wind of a
wide range of duration/sizes. The purpose is to formally publish
the extensive event database,
 summarize the main findings, and to stimulate further studies.




\citet{Cartwright2010} carried out an earlier study of identifying
and characterizing the small-scale flux ropes in the solar wind
with duration $\geq$ 10 minutes, similar to the range of duration
we have examined. They surveyed small-scale flux ropes between
heliocentric distances 0.3 and 5.5 AU, from 1974 to 2007. They
found that the occurrence rate of small-scale magnetic flux ropes
has a negative solar cycle dependency. More events tend to occur
during solar minimum rather than solar maximum. They found that
the averaged monthly occurrence counts of small-scale flux ropes
had a negative correlation with the average annual sunspot number.
However, significant uncertainty or variability existed in that
correlation study. Although their database included small-scale
flux rope events from multiple spacecraft missions with
heliocentric distances ranging from 0.3 to 5 AU, the event
occurrence rate was still very low, about 1 per month. The total
number of events in Cartwright and Moldwin's database is far too
few.

\citet{Feng2007, Feng2008} investigated the small to intermediate
size magnetic flux ropes that had duration  beyond the range we
have examined. They suggested that the small-scale flux ropes are
interplanetary manifestations of small CMEs, originating from weak
solar eruptions and forming in the solar corona, just like
magnetic clouds. \citet{Feng2008} found a positive correlation
between the occurrence rate of small- and intermediate-scale
magnetic flux ropes and the occurrence rate of magnetic clouds
from 1995 to 2005. Therefore, they called these small- and
intermediate-scale magnetic flux ropes as small magnetic clouds
(SMCs). However, the occurrence trend of SMCs shown by
\citet{Feng2008} (Figure 4 in their paper) is not consistent with
the trend of sunspot numbers. Additionally, as pointed out by
\citet{Cartwright2010}, \citet{Feng2008} did not exclude the
Alfv\'enic waves or structures in their database. Another caveat
weakening their conclusion is that the total number of events in
their database is also very small.

\citet{Yu2014,2016JGRA..121.5005Y} performed careful analysis of
small transients (STs) at 1 AU, based on certain criteria, using
both Wind and STEREO spacecraft measurements. They identified
events with duration between 0.5 and 12 hours, which include flux
rope type based on minimum variance analysis on the magnetic field
\citep[MVAB; see, e.g.,][]{Sonnerup1998}, a traditional and simple
way of identifying flux ropes. The event occurrence rate was found
to be no more than 100 a year for solar cycles 23 and 24. They
also found that such occurrence rate of STs anti-correlates with
the sunspot number and these structures occurred more often (over
80\%) in the slow wind (with speed $<$ 450 km/s).  Traditionally
it has long been shown \citep[e.g.,][]{Gopalswamy2015} that there
is a positive correlation between the annual number of magnetic
clouds events and the the sunspot numbers. It is because that the
magnetic clouds originate from solar eruptions, as is widely
accepted. One important conclusion from the studies on STs is that
the plasma $\beta$ value is not always negligible for these
events, especially during solar minimum years. In addition, no
significant enhancement of iron charge states was found. Lately,
\citet{2018ApJ...859....6H} attempted the global numerical
simulation study of the generation of flux rope type and
pseudo-flux rope (torsional Alfv\'en wave) structures directly
from the Sun. The former was primarily released from streamer
belts in the heliospheric current sheet (HCS), while the latter
from polar regions in that study. Since that simulation was still
confined to low corona near the Sun, the direct relation between
these structures and in-situ spacecraft measurements, e.g., out at
1 AU, has yet to be established. As discussed in length in that
study, these two types of structures can be distinguished based on
in-situ plasma and magnetic field measurements, as we will
describe below, which constitutes a critical criterion for our
flux rope event detection.


The solar cycle dependency of small-scale flux rope occurrence
rate provides weak support for the hypothesis that both
small-scale flux ropes and magnetic cloud are created by solar
eruptions. In the following analysis, we are going to investigate
more comprehensive statistical characteristics of the small-scale
flux ropes, and seek for more clues on their origination and
formation mechanism. We would like to stress that what
distinguishes our analysis from all previous studies is that we
have far more number of flux rope events, about a few hundreds per
month as opposed to about $\sim$1 per month, on average.
Therefore, our analysis results will be based on much better
statistics with a significantly large sample of events. {We will
use the in-situ measurements of interplanetary magnetic field and
plasma parameters from the Wind spacecraft, the flagship mission
of the International Solar Terrestrial Physics (ISTP) program.
Specifically for years 1996-2016, we use the 1-minute cadence
datasets from the Magnetic Field Investigation (MFI)
\citep{1995SSRv...71..207L} and the Solar Wind Experiment (SWE)
\citep{1995SSRv...71...55O} instruments. All data are accessed via
the NASA Coordinated Data Analysis Web (CDAWeb)
 including the electron temperature measurements from SWE
when available. }


Additionally, a highly relevant study by \citet{Borovsky2008}
analyzed the properties of 65,860 ``flux tubes" bounded by
boundaries/discontinuities identified from 1998 to 2004 observed
by the ACE spacecraft. The wall-to-wall distances of their flux
tubes were ranging from about $10^5$ km to about $10^7$ km, which
is similar to the scale size range of the small-scale flux ropes
in our database. They found that most flux tubes tend to align
their axial directions with the Parker spiral. In general, the
small-scale flux ropes may be considered equivalent to or to be  a
subset of the flux tubes. Given that the number of events
identified by \citet{Borovsky2008} is of the same order of
magnitude as ours but via a totally different approach without
explicit identification of flux ropes, we discuss the relevance of
their results in the last section.

This report is organized as follows. A brief description of the GS
reconstruction technique is given in Section~\ref{sec:GSeqn}, in
which the emphasis is put on the basic features of the GS equation
and its relation to a cylindrical flux rope configuration.
Detailed descriptions about the applications of the technique are
provided elsewhere (especially for a recent comprehensive review,
see \citet{Hu2017GSreview}). Section~\ref{sec:algorithm} provides
the technical details of the automated detection algorithm,
including a flowchart illustrating the essential elements and a
detailed description of the core procedures. This serves, to
certain extent, as a detailed recipe, for interested readers to be
able to implement their own code, following these technical
descriptions. Sections~\ref{sec:database} and \ref{sec:WTD}
present the database and associated statistical results derived
from the database built from the Wind spacecraft measurements.
While the former presents a generic set of histograms for various
bulk parameters, the latter singles out the waiting time
distribution that can be related to other relevant studies.
Various elements of the database are readily available and mostly
self-explanatory at \url{http://fluxrope.info}.
Section~\ref{sec:HCS} discusses a simple timing analysis of the
occurrence of SSMFR events with respect to the heliospheric
current sheet, which may have implications for their generation
mechanism. The last section summarizes our findings and attempts
at an initial discussion about the origination of SSMFRs based on
our results.

\section{The Grad-Shafranov Reconstruction Technique}\label{sec:GSeqn}

The Grad-Shafranov (GS) reconstruction technique, based on the
plane GS equation, represents a unique method in characterizing
space plasma structures from in-situ spacecraft measurements
\citep{Sonnerup1996,Hau1999,Hu2000,Hu2001,Hu2002}. The GS equation
describes two-dimensional (2D) configuration in magnetohydrostatic
equilibrium, reduced from the equation $\nabla
p=\mathbf{J}\times\mathbf{B}$, where the plasma pressure ($p$)
gradient is balanced by the usual Lorentz force. The magnetic
induction $\mathbf{B}$ satisfies the solenoidal condition and the
current density is given by the Amp\`{e}re's law,
$\mu_0\mathbf{J}=\nabla\times\mathbf{B}$. In a Cartesian
coordinate system, $(x,y,z)$, the GS equation is written, with $z$
being the ignorable coordinate, i.e., $\partial/\partial z=0$
(e.g., being the axis of a cylindrical flux rope),
\begin{equation}
\frac{\partial^2 A}{\partial x^2}+\frac{\partial^2 A}{\partial
y^2}=-\mu_0\frac{d}{dA}(p+\frac{B_z^2}{2\mu_0})=-\mu_0
J_z(A).\label{eq:GS}
\end{equation}
The cross section of a cylindrical flux rope is therefore fully
characterized by the scalar flux function $A(x,y)$, varying on the
$x$-$y$ plane. The equi-value contours of $A$ represent the
transverse magnetic field lines on the $x$-$y$ plane, because it
can be shown $\nabla A\cdot\mathbf{B}_t=0$ with the transverse
field  $\mathbf{B}_t=(\partial A/\partial y, -\partial A/\partial
x)$. Although all physical quantities are functions of $(x,y)$,
there are a number of field-line invariants which are
single-variable functions of $A$ only. Namely they include the
axial field $B_z$, the plasma pressure $p$, the axial current
density $J_z$, and subsequently the transverse pressure,
$P_t=p+B_z^2/2\mu_0$. Thus a typical magnetic flux rope
configuration is characterized by nested iso-surfaces of $A$ with
the field lines spiraling on the distinct surfaces. The set of
field-line invariants also varies among these surfaces while
remaining constant on each distinct surface with a distinct $A$
value.

This important feature of the 2D configuration composed of nested
cylindrical flux surfaces enables the development of the GS
reconstruction technique, using the single-spacecraft
measurements. A single path across such a configuration enables
the sampling of such a set of nested flux surfaces or a set of
closed loops as viewed on the 2D $x$-$y$ plane, and the evaluation
of associated field-line invariants, based on quantitative
spacecraft measurements, including both magnetic field and plasma
parameters. The critical steps are to obtain the values of $A$
along the spacecraft path, and the trial-and-error procedure to
determine the optimal $z$ axis orientation. Both steps are to be
described in Section~\ref{sec:detection}. A comprehensive review
on the development and application of the GS reconstruction
method, including a set of metrics in assessing the satisfaction
of model assumptions, was given by \citet{Hu2017GSreview}. In the
current analysis, we follow the same well-established procedures
in determining the flux rope interval and the optimal $z$ axis
orientation, but without carrying out the final step of solving
the GS equation to obtain the  solution of the cross section for
each flux rope interval identified.

\section{Automated Detection Algorithm Based on the GS Method}\label{sec:algorithm}
\citet{Hu2002} developed the approach to determine the flux rope
axial orientation base on the GS equation, i.e., the requirement
that the function $P_t$ \textit{versus} $A$ be single-valued and
double-folded through a flux rope interval. In such a
configuration of nested flux surfaces as described before, the
flux function $A$ usually reaches an extremum near the center
along the single spacecraft path, while each flux surface is
crossed twice, along the path toward and away from the center,
resulting in two halves (or so-called ``branches") of the $P_t(A)$
plot. Since the quantity $P_t$ is a single-variable function of
$A$, each pair of the same $A$ values from the crossings of the
same flux surface ought to correspond to a pair of the same $P_t$
values. This feature requires a double folding behavior of $P_t$
\textit{versus} $A$ plot, i.e., one branch of $P_t(A)$ folds back
and overlaps with the other branch. In this approach, a residue is
defined to  assess the goodness of the aforementioned
double-folding behavior of $P_t(A)$, in order to determine an
optimal $z$-axis orientation through a trial-and-error process. If
a trial $z$-axis deviates from the optimal axis, the two branches
of $P_t$ \textit{versus} $A$ will not be overlapping, and the
corresponding residue will be larger than a threshold value. In
short, a trial-and-error process was devised to search for the
optimal $z$ axis orientation in the whole parameter space. An
appropriately spaced search grid is constructed on the upper-half
hemisphere of a unit sphere in a spherical coordinate system, on
which each grid point represents one trial $z$-axis orientation
(represented by the polar angle $\theta$ and the azimuthal angle
$\phi$). All grid points are traversed  to find the $z$-axis
orientation with the minimum
 residue of $P_t(A)$, which is usually taken as the optimal flux
rope $z$-axis orientation. This forms the basis for our automated
detection algorithm of SSMFRs.

In our flux rope detecting algorithm, we extend the usage of Hu
and Sonnerup's approach. This approach is not only used for
determining the optimal $z$ axis of a flux rope, but also used for
checking flux rope candidacy. Since in the case of a locally
cylindrical flux rope, an optimal $z$-axis with a reasonably small
residue will surely be found. Conversely, small residue of $P_t$
\textit{versus} $A$ is taken as a main criterion for identifying
flux rope candidates, together with additional criteria to be
presented in the following subsections.

\subsection{Automated Flux Rope Detection
Algorithm}\label{sec:detection}

In the present study, we detect flux ropes with duration from
about 10 minutes to 360 minutes. We split this task into multiple
iterations separated by different-width search windows applied to
the time-series data with 1-minute cadence. These iterations are:
10-15 minutes, 15-20 minutes, 20-25 minutes, ..., and 355-360
minutes. Each iteration identifies flux rope candidates with the
duration falling in the time range specified. For example, when we
run the 10-15 minutes iteration, we set a sliding window width to
15 minutes, and the lower limit of flux rope duration to 10
minutes. With this setting, the program checks the data segment in
a sliding 15-minute window, and if the data segment corresponding
to the double folded part of $P_t(A)$ is shorter than 10 minutes,
the program will discard it. When this iteration is done, the flux
ropes with duration longer than 10 minutes and shorter than 15
minutes will be discovered and recorded. After all the given data
is scanned by the 15-minute window, we rewind and start the next
iteration to find flux ropes with other sizes. In practice, we
extend the boundaries of each iteration by 1 minute to make them
overlap with
 two adjacent iterations. Then the iterations become: 9-16
minutes, 14-21 minutes, 19-26 minutes, ..., and 354-361 minutes.

There are several reasons why we use the multiple iteration
strategy. The first reason is that we need different levels of
smoothing  to detect  flux ropes of different sizes. For example,
a small fluctuation which is significant for  a 10 minutes flux
rope interval may be negligible for a 60 minutes interval flux
rope. Moreover, because the long data segment tends to have more
small fluctuations than short data segment, it is more likely to
be rejected by the detecting algorithm due to multiple turn points
in $P_t$ \textit{versus} $A$. To guarantee both quality and
quantity in the detecting process, we need to apply strong
smoothing process to make the large size flux ropes survive, while
we need to apply a weak smoothing process to make short flux ropes
of bad quality be rejected. As to be explained, we use the third
order Savitzky-Golay filter \citep{2007nrca.book.....P} to smooth
the $A$ array, and set the smoothing window width to be one half
of the lower limit associated with each detection window width.
The second reason to use multiple window strategy is that we want
to keep the coding logic and program architecture as simple as
possible. If we try to use a large fixed-width sliding window to
detect flux ropes of all sizes, we have to deal with many
problems, especially when multiple flux rope structures exist in
the window, such as trimming data segment, splitting multiple
structures, moving window forward,  so on and so forth.
Furthermore, in the multiple flux rope structure, each individual
flux rope may have different axial orientation, which makes
fitting all of them in one window impossible. The code will be
much more simpler if we just seek for the flux ropes with
pre-defined range of duration in one iteration. The third reason
is that we want to make the database easily expandable. If we
decide to add flux ropes with additional sizes to our database,
what we need to do is to just run one more iteration with a new
search window width, instead of repeating the entire process.

Therefore, the ground rules we play by are:  (1) as many as
possible flux rope candidates are to be identified by  an
exhaustive sifting process through the time-series data, and (2) a
dominant single flux rope is to be identified for each
interval/event.

Now we are ready to introduce the flux rope detection algorithm.
Our automated detection algorithm is based on the fact that the
magnetic flux function $A(x, y)$ is a field line invariant, and
the transverse pressure $P_t$ is a single-valued function of $A$,
based on the  GS equation described in Section~\ref{sec:GSeqn}. To
reiterate, as a spacecraft passes through the cross section of a
magnetic flux rope with closed transverse field lines, it firstly
crosses some transverse magnetic field lines in its first-half
path toward the center, then it crosses exactly the same set of
transverse field lines, but in reverse order, in its second-half
path. Therefore, along the spacecraft path, the measured magnetic
flux function $A$ associated with the field lines traversed twice
by the spacecraft shows a double-folded pattern, or contains a
turn point where an extremum is reached. Since the transverse
pressure $P_t$ is a single-valued function of $A$, the two
branches of the data points along the first and second halves of
the spacecraft path for $P_t$ \textit{versus} $A$ should coincide
as well. Conversely, given a specific time interval of interest,
the transverse pressure $P_t$ and the flux function $A$ are
calculated from the in-situ spacecraft data.  Then we check
whether the $P_t$ \textit{versus} $A$ curve has the double-folded
feature and how good the overlapping is. Later we will define the
selection criteria. If the criteria are satisfied, the structure
under checking is considered as a flux rope candidate.

\begin{figure}[p]
\vspace*{-1.7cm} \centering
\includegraphics[width=1.0\textwidth]{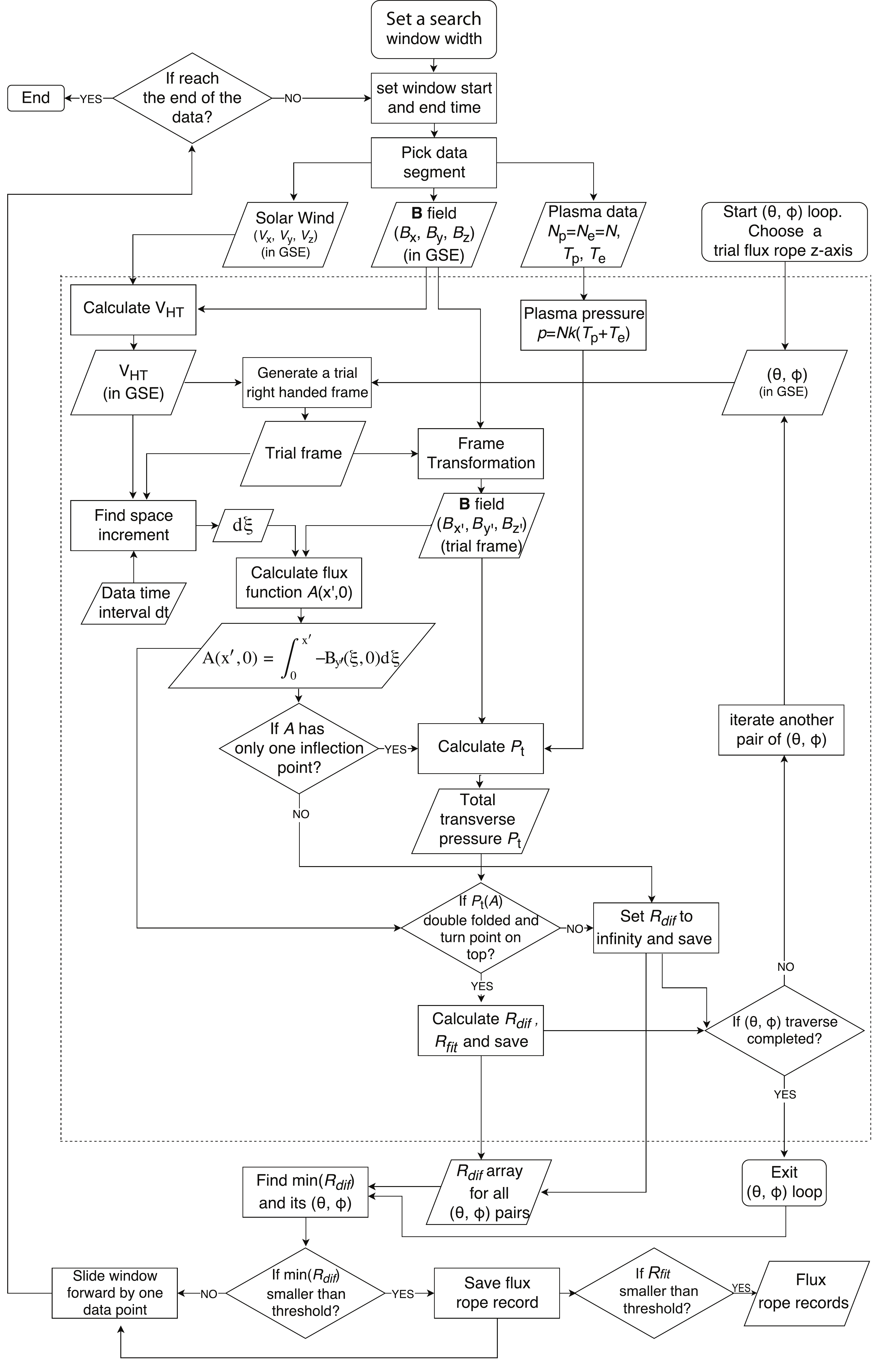}
\caption{\label{flowchart}Flux rope detecting algorithm flowchart.
A rounded rectangle represents the beginning or end of a loop; a
rectangle represents a process; a diamond represents a conditional
judgement; and a parallelogram represents data input or output.
The core loop is enclosed by the dashed lines.}
\end{figure}

A flowchart with control flow and data flow
(Figure~\ref{flowchart}) is prepared to show the detailed
technical procedures. The flowchart illustrates the flux rope
detecting process with a fixed window width for one of the
iterations described earlier. The inner loop over directional
angles $(\theta,\phi)$ is enclosed by the dashed lines, which
shows the flux rope axial orientation determination process. The
outer loop starting with the top rounded rectangle (``Set a search
window width") shows the sliding window process, which moves the
window forward by one data point each time (by 1 minute in the
present study) to scan the entire time series data.

A  sliding window with the preset width is used to select data
segment for analysis. The window width defines the maximum
duration of the flux rope to be detected during this iteration. We
also define a lower limit of the  flux rope duration. The flux
ropes only with their sizes between the lower and upper limits
will be processed during this iteration. Subsequently we are going
to run multiple iterations with different window widths to detect
flux ropes with different sizes. Specifying the lower limit will
avoid the duplication among the windows with different widths. For
the time series magnetic field data within a given window, to make
the $P_t$ \textit{versus} $A$ curve double folded, the $A$ array
must have one and only one inflection point (or turn point),
defined as the place where the magnetic field component $B_{y'}$
changes sign (see Equation~\ref{eq:A_chap3}) or equivalently the
$A$ value reaches an extremum. As discussed in
Section~\ref{sec:GSeqn}, since the transverse pressure $P_t$ is a
single-valued function of $A$, the only way to make $P_t$
\textit{versus} $A$ curve double folded is that the $A$ array has
to fold onto itself so that the $P_t$ \textit{versus} $A$ curve is
split into two branches at the inflection point. If the $A$ array
has more than one inflection point, the corresponding $P_t$
\textit{versus} $A$ curve will be multiple folded, which may not
meet the ground rule (2) for a single flux rope configuration.
When such a situation occurs, the window may contain more than one
flux rope structures. We just need to narrow down the window size
to make it contain only one single flux rope structure. On the
other hand, a narrow window cannot detect the flux ropes with the
duration longer than the window width. Therefore, to detect flux
ropes with different sizes, we have adopted the strategy of using
multiple windows with different widths.

In the detection program, the number of inflection points is
examined by checking the number of extrema of $A$ values in the
data interval. The segments with more than one extreme $A$ values,
excluding the boundaries, will be discarded.  In practice, the
local extrema in the $A$ array may be caused by measurement
uncertainty  or small fluctuations. To remove the effect of small
local extrema, the smoothed $A$ array is used to check the number
of inflection points. The third order Savitzky-Golay filter is
applied to smooth the $A$ array. The width of the smoothing window
needs to be specified for the Savitzky-Golay filter to apply the
smoothing. If the smoothing window size is too small, smoothing
process can not remove most of the small fluctuations. In this
case, some real flux ropes with small fluctuations will be
discarded due to their multiple inflection points. If the
smoothing window size is too large, the smoothing process will
force to remove the large local extrema, which will cause the low
quality structures to be labeled as flux rope candidates. After
extensive experiments, we have found that the one half of the
lower limit associated with each search window width is an
appropriate choice of width for its smoothing window. Because the
smoothing window width in Savitzky-Golay filter needs to be an odd
number, if the one half of the lower limit is not an odd number,
we will round it up to the nearest odd number.

The core procedure, corresponding to the inner loop denoted in
Figure~\ref{flowchart}, consists of the following two major steps:

\begin{itemize}
\item{\textbf{Step 1.}}
As the sliding search window moves forward, we calculate the $A$
array along the projected spacecraft path (at $y'=0$) by (the
prime symbol denotes the trial coordinates for a trial flux rope
$z$-axis)
\begin{equation} A(x',0)=\int_0^{x'} -B_{y'}(\xi,0) d\xi.\label{eq:A_chap3}
\end{equation}
The spatial increment is converted from the temporal sampling
interval via $d\xi=-\mathbf{V}_{HT}\cdot\hat{\mathbf{x}}' dt$,
where a constant frame velocity $\mathbf{V}_{HT}$ is taken as the
usual deHoffmann-Teller frame velocity \citep{Khrabrov1998}. The
remaining flow $\mathbf{v}'=\mathbf{V}_{sw}-\mathbf{V}_{HT}$ in
such a co-moving frame of reference would generally become
negligible, where the solar wind velocity $\vec{V}_{sw}$ is
measured in the spacecraft frame, e.g., the Geocentric Solar
Ecliptic (GSE) coordinates.  In practice, to speed up the
searching process and noting that there is little difference
between $\mathbf{V}_{HT}$ and average solar wind velocity over the
data segment, the latter is usually used in lieu of the former.
Apparently the inflection point corresponds to the point along the
spacecraft path where the field component $B_{y'}$ changes sign
and a point at which the extreme value in $A$ is reached.
 If we find the calculated $A$ array within the window is monotonic
 or has more than one distinct inflection points, we will do nothing but simply move the window forward by one data point.
  A monotonic $A$ array contains no extrema and  can never lead to a double-folded $P_t$ \textit{versus} $A$ curve.
  An $A$  array with more than one inflection points indicates that the current window  may contain multiple flux rope structures.
  For the former situation, there is no further action needed to be done, and for the latter,
  a smaller size search window in another run will take care of it.
  Once an $A$ array with only one inflection point is discovered, the $A$ array will be split into two branches
  at the inflection point. Then the two branches will be trimmed to have the same $A$ values at the boundaries.
  Note that the trimming is not according to the number of data points, but is according to the $A$ value, because physically the boundary of
  a flux rope is identified by a flux surface with the same $A$ value. After trimming, the two branches may not have the same
  length, but must have the same or similar boundary $A$ values.
  After getting two branches of $A$ values with  the same boundaries, we can calculate $P_t$ values corresponding to each $A$ value.
   With both $P_t$ and $A$ values obtained along the two halves of the spacecraft path, we can evaluate the agreement between the corresponding
    two sets of  $P_t$ \textit{versus} $A$ values.

\item{\textbf{Step 2.}}
The next step is to examine how well  the two halves (branches) of
the $P_t$ \textit{versus} $A$ curve overlap. Before doing this,
another check process can be performed to reduce the further
workload. As a physical nature of the flux rope under
consideration, the total transverse pressure $P_t$ must reach its
maximum in the flux rope center. Reflected in the two branches of
the $P_t$ \textit{versus} $A$ curve, the turning point
(corresponding to the inflection point in the $A$ array) must be
on the top. We choose to remove the cases with turning points not
on top. Taking into account the measurement uncertainty  and small
fluctuations, we introduce the tolerance. With tolerance, we
require that the $P_t$ value at the turning point be in top 15\%
of all $P_t$ values. If the data segment would survive the check
after undergoing all the aforementioned procedures, we are ready
to  obtain two more metrics as defined below to check the
double-folding quality:

\begin{equation}\label{equ:R_dif}
R_{dif} = [\frac{1}{2N} \sum_{i=1}^{N}({(P_t)}_{i}^{1st} -
{(P_t)}_{i}^{2nd})^2]^{\frac{1}{2}}/|\max(P_t) -\min(P_t)|,
 \end{equation}
 and
\begin{equation}\label{equ:R_fit}
R_{fit} = [\frac{1}{L} \sum_{i=1}^{L}(P_{t}(x_i,0) -
P_{t}(A(x_i,0)))^2]^{\frac{1}{2}}/|\max(P_t) - \min(P_t)|.
 \end{equation}
 We determined that the conditions $R_{dif}\leq0.12$ and $R_{fit}\leq0.14$ could guarantee good flux rope quality while keeping
 a significant number of candidates.
 \end{itemize}

Equation~(\ref{equ:R_dif}) is modified from Equation (5) in
\citet{Hu2002}, and Equation~(\ref{equ:R_fit}) is taken from
\citet{Hu2004}. The residue $R_{dif}$ represents the point-wise
difference between two branches, in which both ${(P_t)}_{i}^{1st}$
and ${(P_t)}_{i}^{2nd}$ are calculated from observational data. We
find the $A$ array and the corresponding $P_t$ array in the first
branch,  then use these $A$ values to look up the corresponding
$P_t$ values in the second branch. If there is no correspondence
in the second branch for some points in the first branch, we use
linear interpolation to create a match. Then we repeat the same
process for the second branch. Finally, each $P_t$ in one branch
has a counterpart in the other branch. We insert the two
interpolated $P_t$ arrays into Equation~(\ref{equ:R_dif}) to
calculate $R_{dif}$.  Only $R_{dif}$ alone is not sufficient to
decide if a data segment is a good flux rope candidate or not,
because a small $R_{dif}$ can only guarantee the good
double-folding of the two branches of $P_t$ \textit{versus} $A$
curves, no matter what the shape of the folded curve is. A
reliable threshold for $R_{dif}$ is hard to set for acceptable
flux rope candidates. To help with this, we obtain an additional
fitting residue by using a 3rd order polynomial to fit the data
points of $P_t$ \textit{versus} $A$. This fitting ignores the time
sequence of the data points and merges two branches into one. Its
fitting residue is defined in Equation~(\ref{equ:R_fit}) and
denoted as $R_{fit}$, where $P_{t}(x_i,0)$ is calculated from
measured data and $P_{t}(A(x_i,0))$ is calculated from the fitting
function. Note that, there is a fraction factor $1/2N$ in
$R_{dif}$'s definition, but in $R_{fit}$'s definition, this factor
is $1/L$. This is because in $R_{dif}$, the number of terms under
the summation operator is only about half of the number in
$R_{fit}$, i.e., $L\approx 2N$, if the two branches of $P_t$
\textit{versus} $A$ curve have the same number of data points. We
use two different factors to make the two metrics, $R_{dif}$ and
$R_{fit}$, comparable in magnitude. However, the two branches of
$P_t$ \textit{versus} $A$ curve must have the same $A$ value
range, but not necessarily have the same number of data points. In
practice, even with the inclusion of these  scaling factors, we
still set different threshold values for $R_{dif}$ and $R_{fit}$.
Besides the number of data points, the range of $P_t$ value may
also affect $R_{dif}$ and $R_{fit}$. So we normalize $R_{dif}$ and
$R_{fit}$ by the range  $|\max(P_t) - \min(P_t)|$. In the flux
rope searching process, we use only $R_{dif}$ to look for optimal
$z$-axis orientation as illustrated in Figure~\ref{flowchart}. For
a given data segment, the value $R_{dif}$ for each trial $z$-axis
orientation is calculated, then the $z$-axis orientation with the
minimum $R_{dif}$ is taken as the optimal axial orientation. With
the determined optimal axial orientation, if both $R_{dif}$ and
$R_{fit}$ satisfy our criteria, this data segment will be labeled
as a flux rope candidate. The threshold values for $R_{dif}$ and
$R_{fit}$ are selected empirically by examining thousands of data
segments with double-folding features in $P_t(A)$, and are given
in Table~\ref{Criteria}.

\subsection{Cleanup and Post-Processing} \label{sec:cleanup}
When one sliding window process with one preset window width is
finished, we will obtain one record list of identified flux rope
intervals. However, this record list has many overlapped records
or intervals. We use an example to illustrate how overlapping
happens. We imagine a true flux rope starting from 8:00 am and
ending at 10:00 am. When a sliding window with the width of 2
hours just covers the entire flux rope time range, this flux rope
will be recognized and recorded. If the window moves forward and
covers the time range from 8:05 am to 10:05 am, the detecting
algorithm will likely recognize a flux rope from 8:05 am to 9:55
am (after trimming the two branches of $P_t$ \textit{versus} $A$
curve). As the window moves on by one data point each time, as
long as it still covers the turn point located in the middle of
the flux rope interval from 8:00 am to 10:00 am (the largest flux
rope), and the detected flux rope is longer than the lower limit,
the program will pick a part of the largest flux rope as a new
flux rope. These flux ropes share the same turn point, and we call
them a flux rope cluster. To clean up such a cluster from  the
record list, we usually pick the segment  with the smallest
$R_{dif}$ value and discard the others. We have considered
possible improvements to avoid the overlapping. One possible
method is to move the entire sliding window out of the time range
of a flux rope once it is detected. However, we cannot guarantee
that the one we  picked is the best one in a flux rope cluster.
Eventually we decided to slide the search window continuously to
guarantee the detection of the maximum number of flux ropes.

The other situation of overlapping occurs when additional
iterations, i.e., additional lists of records, are completed. When
all iterations with different window widths are finished, we need
to combine all flux rope records. With the program setting
introduced in the beginning of Section~\ref{sec:algorithm}, we
will end up with 70 event lists from 70 iterations. We cannot
simply merge these lists into one, since a long flux rope may
cover one or more short flux ropes from different lists.
In fact, this is an interval scheduling problem in computer
science. The optimal solution is to accommodate as many flux rope
records as possible. We use the greedy algorithm to find the
optimal solution. The main idea of the greedy algorithm in
interval scheduling problem is to firstly accommodate the event
with the earliest end time.

  The cleanup procedures proceed with the
following steps: 1) Clean the overlapped records in the event list
with the longest flux rope duration. We sort the records by the
flux rope end time. Firstly, find all the flux ropes that overlap
with the first record, then keep the first flux rope and discard
others. Then process the next record until all overlapped records
are removed. This step results in a temporary list with ``empty
slots" in-between flux rope intervals. 2)  Then we pick the
suitable records from the event list with the second longest flux
rope duration, and insert them into the corresponding slots. Each
slot may contain some overlapping intervals. We still use the
greedy algorithm illustrated  in step 1) to accommodate them. 3)
Continue with the next longest, and so on, until all flux ropes
are accommodated.

The last check step is the standard Wal\'en test to rule out
possible Alfv\'enic structures \citep{Paschmann2008}. An Alfv\'en
wave or structure may show the similar magnetic field profile to a
magnetic flux rope \citep[e.g.,][]{2010AIPC.1216..240M}. The scale
size of small-scale flux ropes is comparable with that of
Alfv\'enic structures. We have to do further test to remove
Alfv\'en waves from our flux rope database in order to comply with
the GS equation in the present study. The Wal\'en slope is defined
as the slope of the linear regression line between the remaining
flow velocity $\mathbf{v}'$ in the co-moving frame of reference
and the local Alfv\'en velocity, component-wise. We remove the
records whose absolute value of Wal\'en test slope is greater than
or equal to 0.3 (indicating significant remaining flows in the
reference frame co-moving with the structure). Up to this point,
we have finished all major steps toward building the flux rope
database. At last, we have the option to apply one more criterion.
Considering that the average magnitude of magnetic field in the
ambient solar wind is about 5 nT, we remove the flux rope records
whose average magnitude of magnetic field $\bar{B}$ is less than 5
nT.

In summary, Table~\ref{Criteria} lists the set of metrics and
criteria we use in the identification of small-scale magnetic flux
rope events from the in-situ Wind spacecraft measurements. The
duration $9\sim361$ minutes covers the range of most small-scale
flux ropes, and this range is easily to be expanded to
intermediate or large size flux ropes, e.g., with duration
$6\sim12$ hours or more, to bridge the gap between small-scale
flux ropes and MCs. The condition $\bar{B}\geq5$ nT excludes small
fluctuations in the solar wind. The thresholds on the  metrics
$R_{dif}$ and $R_{fit}$ guarantee the good quality of small-scale
flux ropes, complying with the GS equation. And the Wal\'en test
slope threshold condition ($\leq0.3$) acts to remove  the
Alfv\'enic structures or waves in the present study.

\begin{table}[h]
    \caption{Small-scale Magnetic Flux Rope Detection Metrics and Criteria}\label{Criteria}
    \centering
    \begin{tabular}{ccccc}
        \hline
        Duration & $\bar{B}$ & $R_{dif}$ & $R_{fit}$ & Wal\'en test slope\\
\hline
        $9\sim361$ (minutes) & $\geq5$ (nT) & $\leq0.12$ & $\leq0.14$ & $\leq0.3$\\
\hline
    \end{tabular}
\end{table}

\begin{figure}[htb] 
\centering
\includegraphics[width=1.0\textwidth]{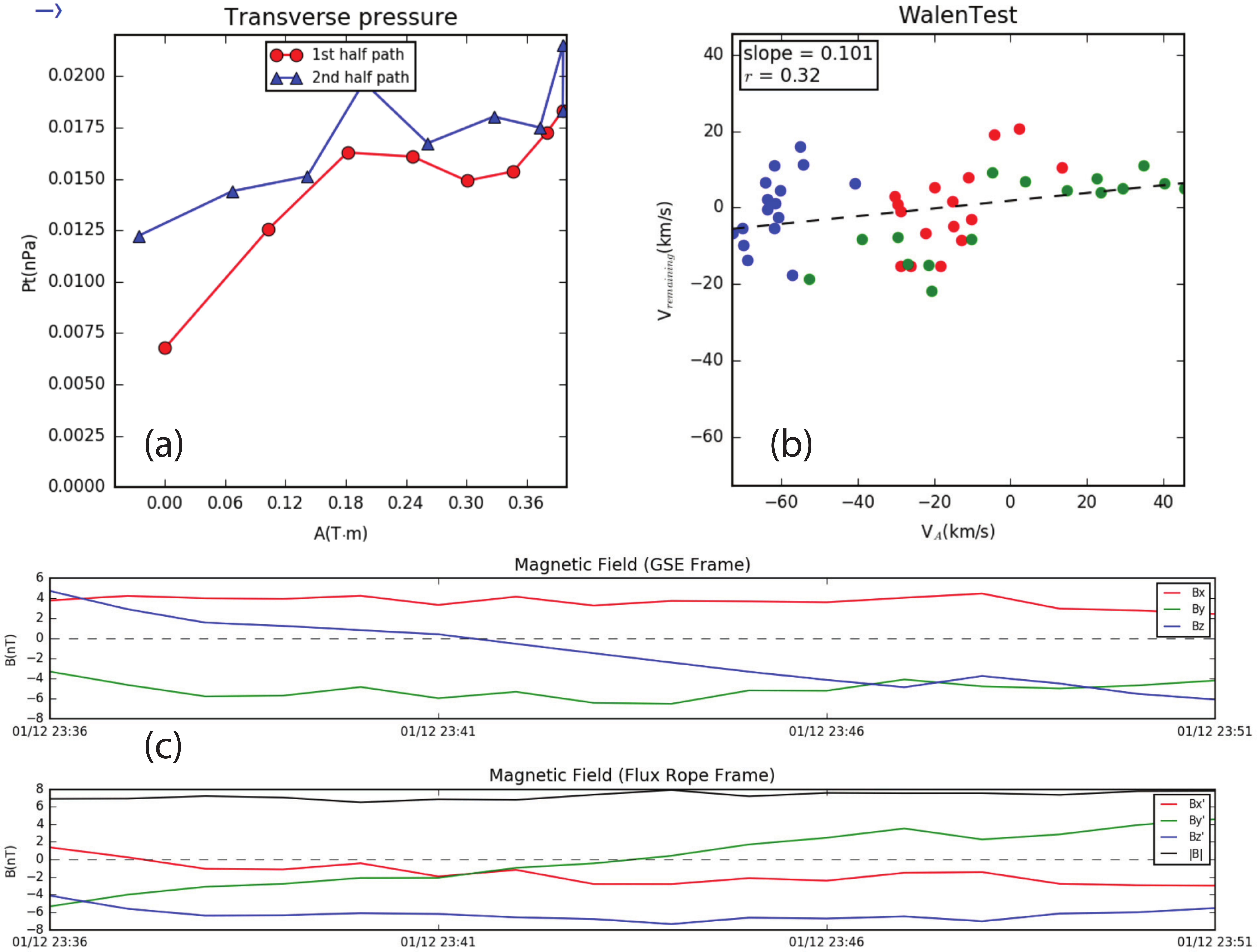}
\caption{A collage of selected plots available on the SSMFR
database website for one particular event. (a) The $P_t$
\textit{versus} $A$ data points along the 1st half (red dots) and
the 2nd half (blue triangles) of the spacecraft path. (b) The
Wal\'en test plot: $\mathbf{v}'$ \textit{versus} the Alfv\'en
velocity $\mathbf{V}_A$ in GSE components (colored dots). Dashed
line marks the linear regression line with the slope and the
correlation coefficient given in the legend. (c) The magnetic
field components (in red, green and blue) projected onto the usual
GSE (upper panel) and the local flux rope (lower) coordinates,
respectively. }\label{event181}
\end{figure}

To conclude this section, we present an illustrative example for a
particular event No.~181 in Year 2016 from the database website.
Figure~\ref{event181} shows the corresponding  $P_t(A)$, Wal\'en
test and the magnetic field components plots directly extracted
from the event web page. The flux rope interval spans the time
period 2016/01/12 23:36 - 2016/01/12 23:51 UT with a duration 16
minutes. We also obtain the following parameters for the flux rope
interval: $R_{fit}\approx 0.14$, $\bar{B}=7.25$ nT,   $B_{max}=
7.88$ nT, $\beta=\beta_p= 0.94$, the average solar wind speed 537
km/s, the average proton temperature $T_p= 0.28\times10^6$K, and
the optimal $z$-axis orientation $(\theta,\phi)=(80, 140)$
degrees. The $z$-axis direction translates into GSE cartesian
coordinates as  a unit vector $(-0.7544, 0.6330, 0.1736)$.
Figure~\ref{event181} (a) shows the double-folding feature of
$P_t(A)$ in which the 2nd half (branch) folds back onto the 1st
one. The deviation between the two branches was evaluated by the
two metrics and they fell below the thresholds given in
Table~\ref{Criteria}. Figure~\ref{event181} (b) shows the Wal\'en
test result with a slope 0.101 from the linear regression line.
Figure~\ref{event181} (c) shows the magnetic field components in
the flux rope interval, especially in the lower panel, after
projected onto the flux rope frame $(x',y',z')$ for the optimal
$z$-axis orientation determined. It shows a clear rotation in the
$B_{y'}$ component, and a unipolar axial field $B_{z'}$,
consistent with a flux rope configuration.
 For additional information, we encourage interested readers to explore the database
website to learn about the variability and extensiveness of such
an  event database.

\section{Small-scale Magnetic Flux Rope Database from Wind Spacecraft
Measurements}\label{sec:database} We apply the flux rope detection
algorithm based on the GS reconstruction technique to the Wind
spacecraft measurements during 1996-2016, covering nearly two
solar cycles. We successfully detected a large number of
small-scale magnetic flux ropes with more general configurations,
including non-force-free and non-axisymmetric configurations under
the assumption of 2D quasi-static equilibrium.
Table~\ref{FR_count_table} lists the number of flux ropes detected
by our algorithm in each year. There are a total number of 74,241
small-scale magnetic flux ropes detected, with an average number
more than 3,500 per year. This database provides sufficient number
of samples for interested researchers to carry out statistical
analysis, correlate with other structures, and examine some
special cases in detail.

\begin{figure} 
    \begin{center} (a) \\
            \includegraphics[width=1\textwidth]{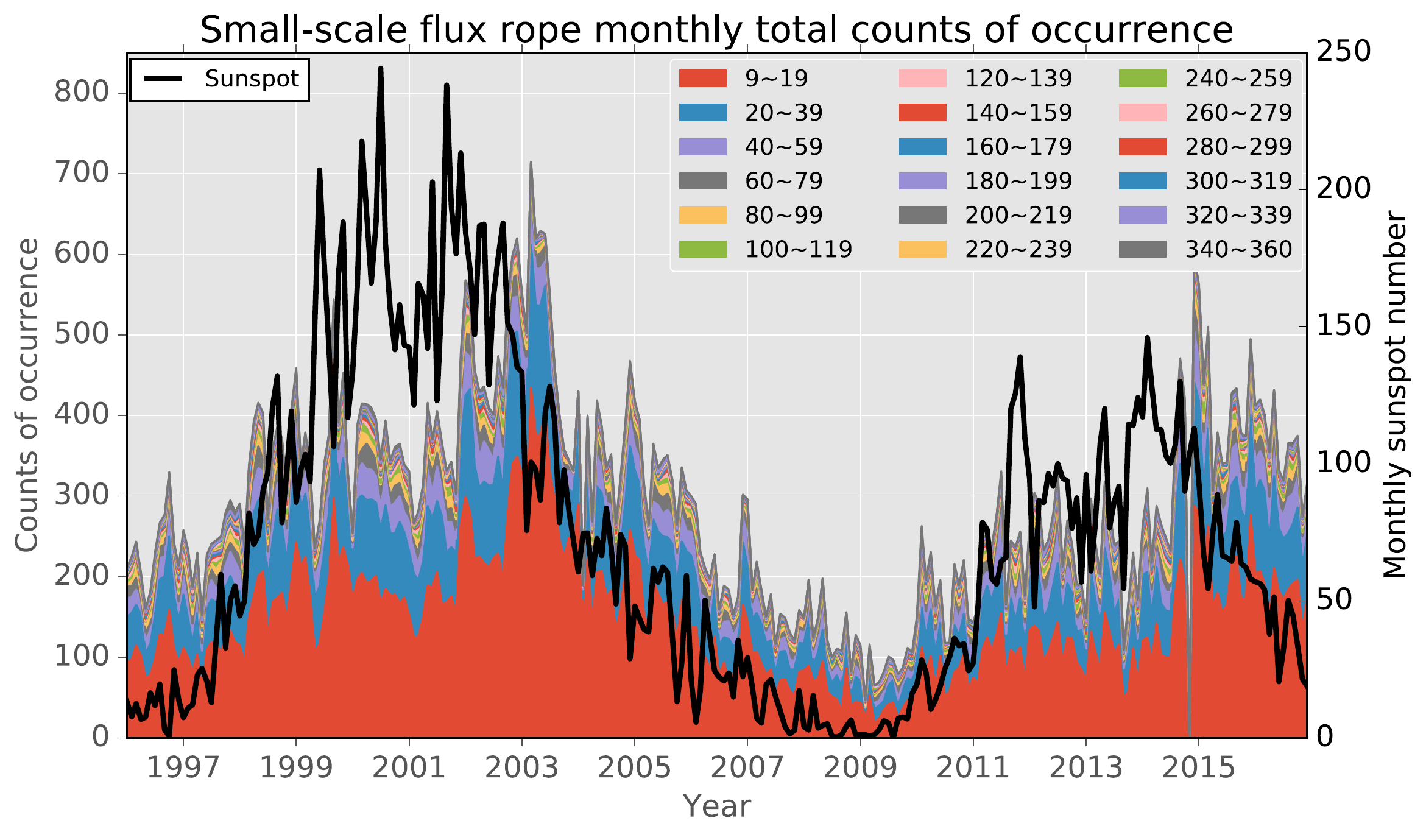}
    (b)\\
                \includegraphics[width=1\textwidth]{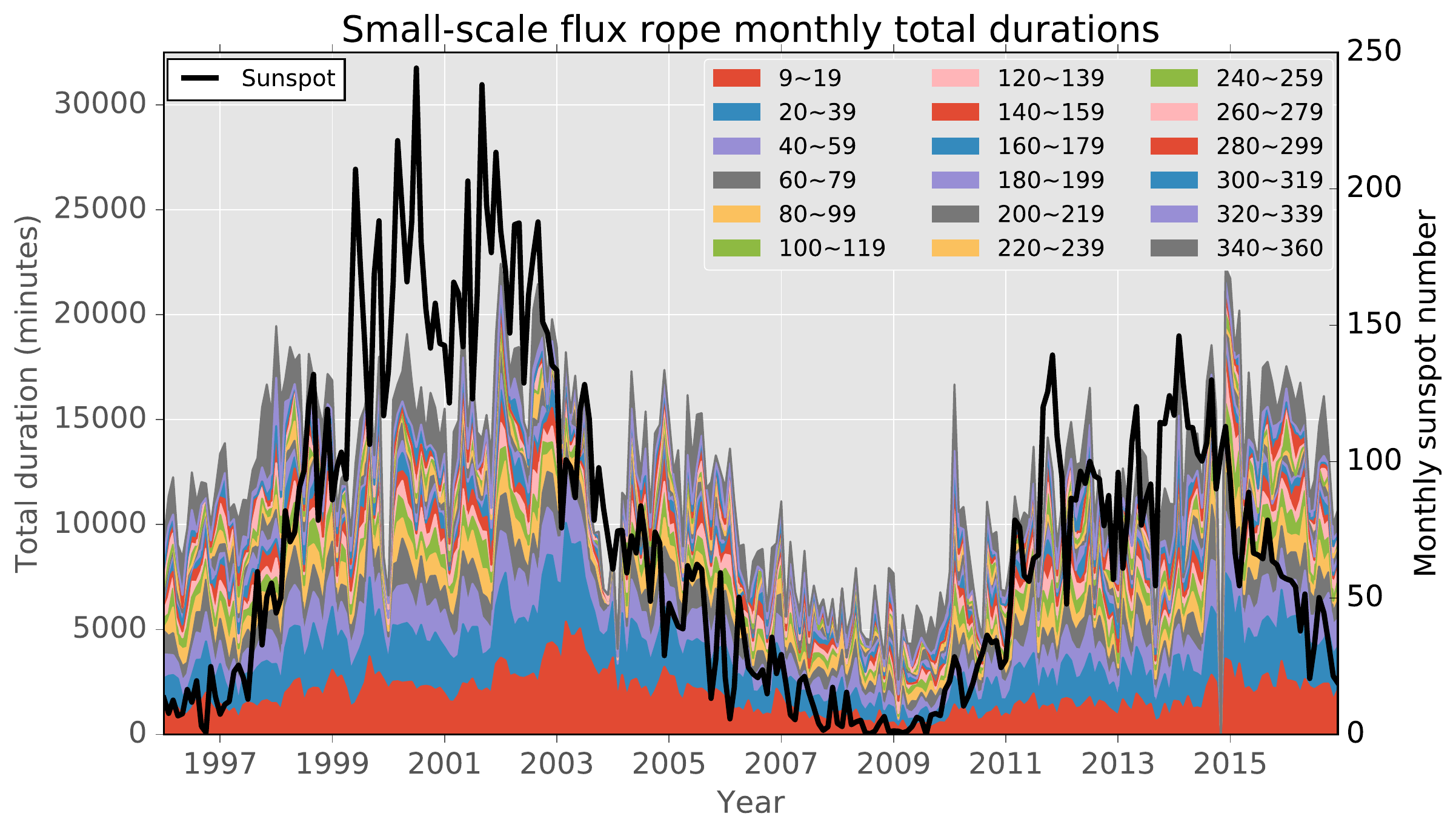} 
    \end{center}
    \caption{(a)  The monthly counts of flux rope events occurrence (left axis) and the monthly sunspot numbers
    (black curve; right axis) during 1996-2016. Colors represent counts of events with different duration as indicated.
    (b) The monthly flux rope total duration (left axis) and the monthly sunspot numbers (right axis) during 1996-2016.
    The format is the same as in (a). {The sunspot numbers are courtesy of WDC-SILSO, Royal Observatory of Belgium, Brussels.}}\label{STA_fluxrope_monthly_counts_duration}
\end{figure}

 Figure~\ref{STA_fluxrope_monthly_counts_duration} (a) shows the monthly occurrence counts of
 small-scale flux ropes from 1996 to 2016, covering solar cycles 23 (from May 1996 to December 2008),
 and 24 (which began in December 2008, reached its maximum in April 2014, and may have ended in early 2017).
  The different colors represent different duration of small-scale flux ropes (from 9 minutes to 360 minutes),
  and the thick black line is the corresponding monthly sunspot number. The events of smaller duration generally have greater
  rates of occurrence. Clearly the total counts including all events of variable duration follow the monthly sunspot numbers,
  hinting at solar-cycle dependency of these events. Note that the occurrence counts also vary cycle by cycle.
  From the sunspot numbers we can see that the solar activity in cycle 23 is more intense than that of cycle 24,
  and accordingly, the overall small-scale flux rope occurrence counts in cycle 23 are greater than those of cycle 24, approximately proportional to sunspot numbers.  The peaks of occurrence counts tend to appear in the declining phase of each solar cycle.
Figure~\ref{STA_fluxrope_monthly_counts_duration} (a) shows
variations in a way consistent with the change of magnetic cloud
occurrence rate as shown by, e.g., \citet{Gopalswamy2015}. During
solar maximum, there are more events occurring, and there are
fewer during solar minimum. One may argue that both magnetic
clouds and small-scale flux ropes originate from solar eruptions,
since they seem to share the similar dependency pattern with solar
activity. However, this fact is still not a sufficient condition
such that the small-scale flux ropes originate from the Sun. On
one hand, there may be two populations of small-scale flux ropes
that have different origins and different solar cycle
dependencies, i.e., one population has solar cycle dependency but
the other does not. When they are mixed, the solar cycle
dependency may still appear. On the other hand,  even if all
small-scale flux rope events have solar cycle dependency as
magnetic clouds do, we still cannot conclude that they originate
from the Sun, since magnetic clouds have clear solar eruption
correspondences but small-scale flux ropes do not. There may be
other plasma dynamic processes far away from the Sun that could
create these small-scale flux ropes and are also modulated by the
solar activity cycle.

 Figure~\ref{STA_fluxrope_monthly_counts_duration} (b) is the monthly total duration for each group in (a) from 1996 to 2016.
 The format is the same as  Figure~\ref{STA_fluxrope_monthly_counts_duration} (a).
 In Figure~\ref{STA_fluxrope_monthly_counts_duration} (b), the vertical axis represents the total small-scale flux rope duration
 in minutes for each similarly color-coded group. Although in the monthly duration plot, the area taken up by smaller flux ropes is suppressed, the solar cycle dependency shown in  Figure~\ref{STA_fluxrope_monthly_counts_duration} (a) still appears in  Figure~\ref{STA_fluxrope_monthly_counts_duration} (b). This plot shows that the solar cycle dependency is not only attributed to smaller flux ropes, but also to larger flux ropes.
{The sudden dips in event count and duration occur for the month
of  November 2014 due to a prolonged data gap, during which no
event was identified.}

\begin{table}[h]
    \caption{The Number of Detected Small-scale Magnetic Flux Ropes in Each Year}\label{FR_count_table}
    \centering
    \begin{tabular}{cccccccccccc}
\hline

        \textbf{Year} & 1996 & 1997 & 1998 & 1999 & 2000 & 2001 & 2002 & 2003 & 2004 & 2005 & 2006\\
        \textbf{Counts} & 2787 & 2878 & 4182 & 4454 & 4425 & 4203 & 5930 & 6086 & 4229 & 4017 & 2620\\
\hline
        \textbf{Year} & 2007 & 2008 & 2009 & 2010 & 2011 & 2012 & 2013 & 2014 & 2015 & 2016 & \textbf{Total}\\

        \textbf{Counts} & 2040 & 1620 & 1076 & 2209 & 2731 & 3051 & 2658 & 3690 & 4987 & 4368 & \textbf{74241}\\
 \hline
    \end{tabular}
\end{table}

\begin{figure} 
  \begin{center}
    \includegraphics[width=1.0\textwidth]{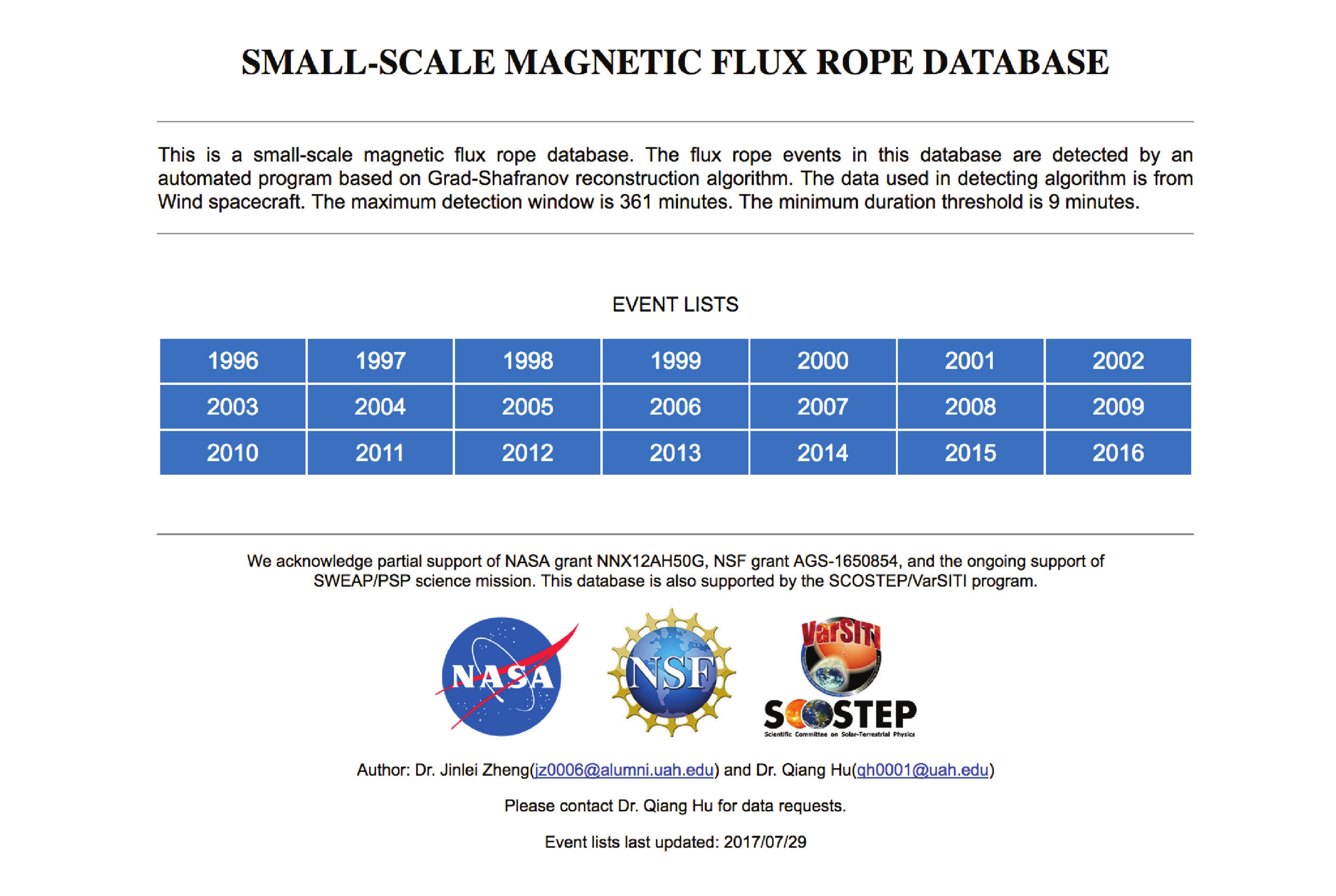}
  \end{center}
  \caption{The home page of the small-scale magnetic flux rope database website.}\label{homePage}
\end{figure}

\subsection{On-line Database}
We have established a website (\url{https://fluxrope.info}) to
host this database online. We make the database open to the public
and keep it up to date. More flux rope events with longer duration
and at higher latitude locations will be added in near future.
Figure~\ref{homePage} is a snapshot of the home page of the
small-scale magnetic flux rope database website. When clicking on
any year on the ``EVENT LISTS'' table, the annual event list page
will show up. The full table is available in machine readable format.
Table 3 gives an example of its form and content.
The event list page lists
every detected flux rope event in one year in chronological order.
For each flux rope record in each row, some basic characteristics
are listed including the time range in UT, duration in minutes,
fitting residue ($R_{fit}$), average magnetic field strength,
maximum magnetic field strength, average plasma $\beta$, average
proton plasma $\beta_p$, average solar wind speed
$\overline{V}_{sw}$, average proton temperature, and flux rope
axial orientation. More information on magnetic field and plasma
profiles is stored offline and can be put online in new versions
of the website. These quantities make it very convenient for us
and other researchers to further apply more selection criteria to
pick desirable subset of events for different purposes.

\begin{table}
\caption{Flux rope events}
\begin{tabular}{lll}
\hline
Column & Label & Explanation \\
\hline
1 & Num & Index identifier \\
2 & Start & Observation start; mm/dd/yyyy hh:mm \\
3 & End & Observation end; mm/dd/yyyy hh:mm \\
4 & Time & Observation duration \\
5 & $R_{dif}$ & Residue, see text \\
6 & $<B>$ & Average magnetic field strength (nT) \\
7 & Bmax & Maximum magnetic field strength (nT) \\
8 & $<{\beta}>$ & Average plasma {beta} value \\
9 & $<{\beta}_p>$ & Average proton {beta} value \\
10 & V$_{sw}$ & Solar wind velocity (km/s) \\
11 & $<T_p>$ & Average proton temperature (MK) \\
12 & $\theta$ & Polar angle (deg) \\
13 & $\phi$ & Azimuthal angle (deg) \\
14 & zAxis$_0$ & Z-axis orientation \\
15 & zAxis$_1$ & Z-axis orientation \\
16 & zAxis$_2$ & Z-axis orientation \\
17 & Size & Scale size (AU)  \\
18 & $N_p$ & Mean proton number density (cm$^{-3}$) \\
\hline
\multicolumn{3}{c}{Table 3 is published in its entirety in the electronic edition} \\
\multicolumn{3}{c}{of the {\it Astrophysical Journal}.  A portion is shown here } \\
\multicolumn{3}{c}{for guidance regarding its form and content.}\\
\end{tabular}
\end{table}

%
%
%
The time range for each record is a clickable hyperlink which will
navigate to the detailed flux rope information page which is
demonstrated in part in Figure~\ref{event181} for one particular
event. They help judge  the ``double-folding'' quality of
$P_t(A)$, and the result from the Wal\'en test. Generally
speaking, a spread of the data points horizontally in the latter
plot indicates satisfaction of the detection criterion based on
the magnetohydrostatic theory. Additionally, the usual hodograms
of the flux rope magnetic field components from the MVAB are also
displayed, which visualizes the movement of the end points of
magnetic field vectors. Generally a smooth rotation in one or two
magnetic field components, typical of a flux rope configuration,
may be inspected in these plots. Such signatures consistent with a
flux rope configuration are further displayed in panels of various
time series data of magnetic field and solar wind parameters,
complemented with the pitch angle distribution of suprathermal
electrons, the temperature of protons and electrons, when
available, and plasma $\beta$, etc.


This website provides a large number of small-scale magnetic flux
rope events and a good deal of associated information, which can
benefit the relevant studies on these and other structures in the
solar wind. Before we delve into detailed statistical analysis, we
show the basis for an usual classification of events, based on the
solar wind speed.

\begin{figure}[ht] 
  \begin{center}
    \includegraphics[width=0.9\textwidth]{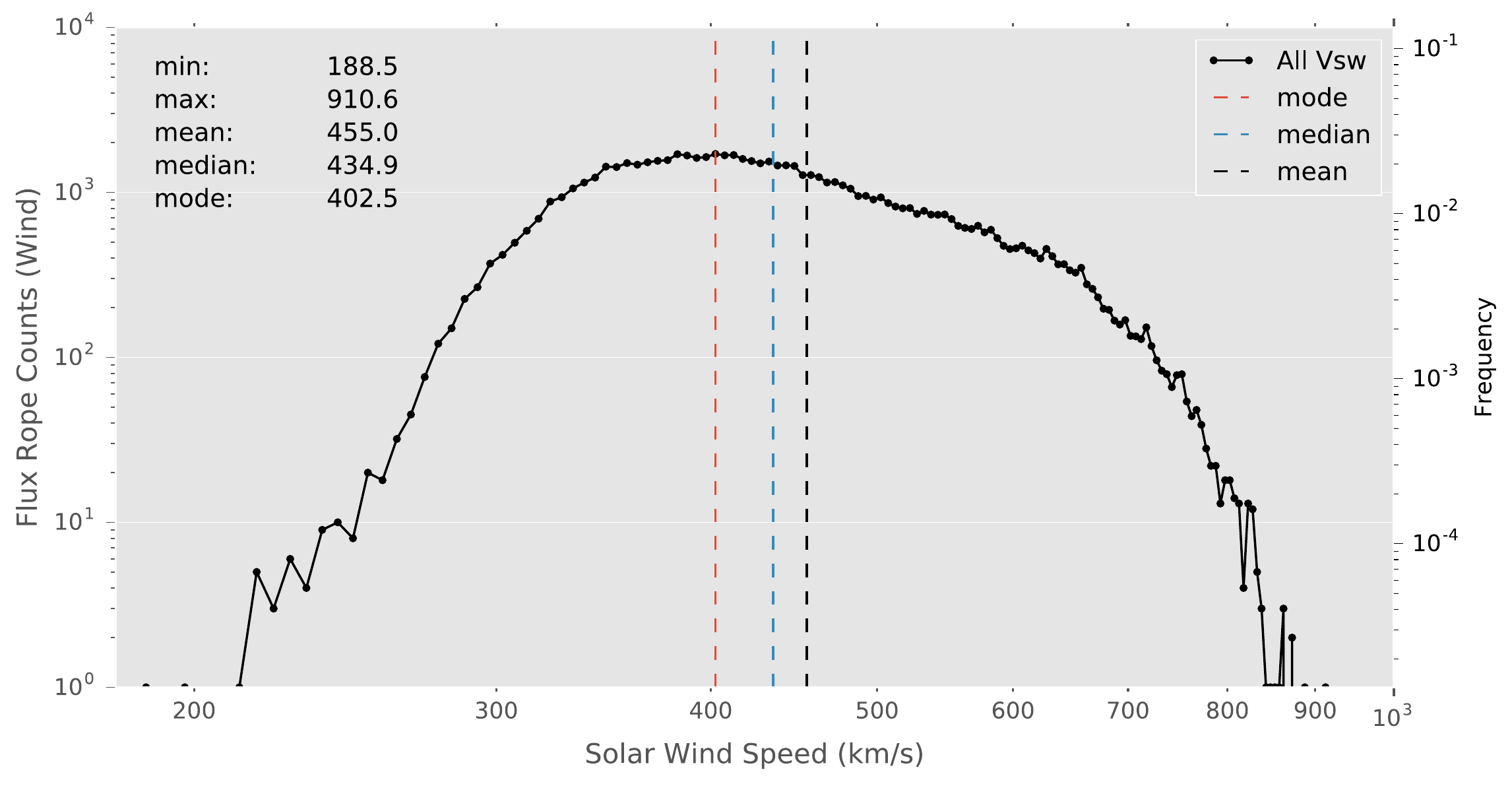}
  \end{center}
  \caption{The histogram of average solar wind speed within each small-scale flux rope interval.}\label{STA_hist_Vsw}
\end{figure}
 Figure~\ref{STA_hist_Vsw} is the histogram of
average solar wind speed $\overline{V}_{sw}$ within each flux rope
interval. One can see a peak near $\overline{V}_{sw}=$400 km/s
(the mode of the distribution is 402 km/s) and three approximately
linear sections in the curve with different slopes on this log-log
plot. The first section of the curve is from
$\overline{V}_{sw}=$200 km/s to $\overline{V}_{sw}=$400 km/s. The
slope of the first section is positive. The second section of the
curve is from $\overline{V}_{sw}=$400 km/s to
$\overline{V}_{sw}=$650 km/s, with a negative slope. The third
section of the curve is from $\overline{V}_{sw}=$650 km/s to
$\overline{V}_{sw}=$900 km/s, with a steep negative slope. Three
distinct slopes in log-log plot indicate three distinct power law
distributions with different power indices. Since the solar wind
speed is a key factor in space plasma dynamic processes, different
power law distributions may imply different physical processes
involved in different plasma flow streams. Therefore, in the
following analysis, we split the entire event database into two
subsets according to the corresponding average solar wind speed
either greater or less than 400 km/s. Note that this value is
nearly identical  to the mode of the distribution, 402 km/s,
corresponding to the peak in the distribution of average solar
wind speed (Figure~\ref{STA_hist_Vsw}).

\begin{figure}[ht] 
  \begin{center}
    \includegraphics[width=1.0\textwidth]{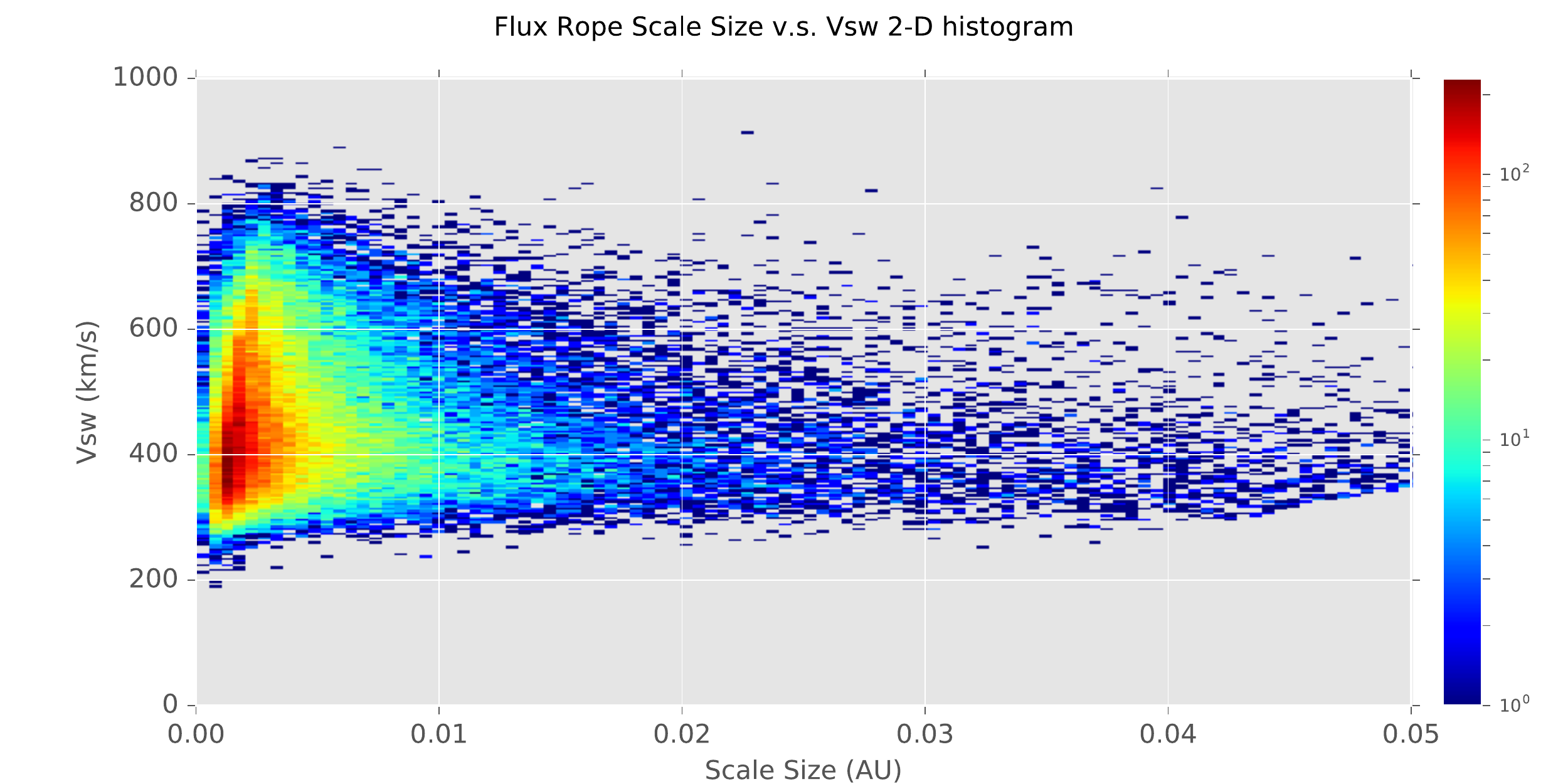}
  \end{center}
  \caption{The 2-D histogram of flux rope average solar wind speed ($\overline{V}_{sw}$) $versus$ scale size.}\label{STA_hist_Vsw_szie}
\end{figure}

Figure~\ref{STA_hist_Vsw_szie} is the 2-D histogram of flux rope
$\overline{V}_{sw}$  $versus$ scale size. It shows that the flux
ropes with larger scale sizes tend to appear in slow solar wind
($\overline{V}_{sw}<$400 km/s), and for the smaller scale size
flux ropes, the solar wind speed associated with them spreads
widely (from $\sim$ $\overline{V}_{sw}=$200 km/s to  $\sim$
$\overline{V}_{sw}=$800 km/s). The line of $\overline{V}_{sw}=$400
km/s divides the plotted shape into two triangles. Above the line
of $\overline{V}_{sw}=$400 km/s, the solar wind speed goes down as
the scale size becomes large, while below the line of
$\overline{V}_{sw}=$400 km/s, the solar wind speed goes up as the
scale size increases. This plot is another basis based on which we
split the entire event set at $\overline{V}_{sw}=$400 km/s.

\subsection{Statistical Results of Flux Rope Properties}
As discussed in Section~\ref{introduction}, there is a
long-standing debate on the origination of small-scale magnetic
flux ropes. In this section, we are going to present some
statistical analysis of the flux ropes properties based on our
database in the hope of shedding some light on the origin of these
structures in the solar wind.

\begin{figure}[ht] 
  \begin{center}
    \begin{minipage}{17pc}
      \includegraphics[width=17pc]{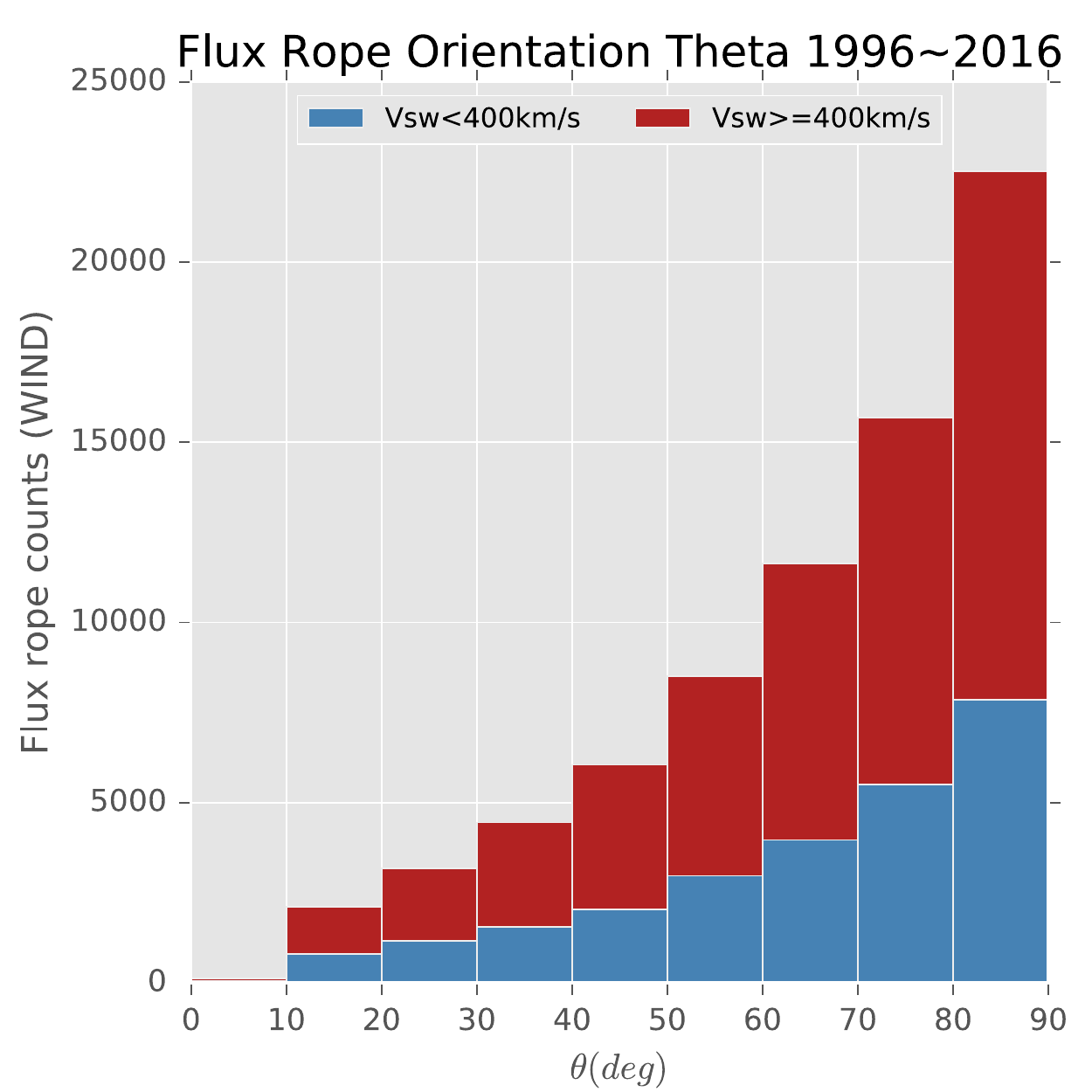}
      \begin{center}
        (a)
      \end{center}
    \end{minipage}\hspace{0pc}
    \begin{minipage}{17pc}
      \includegraphics[width=17pc]{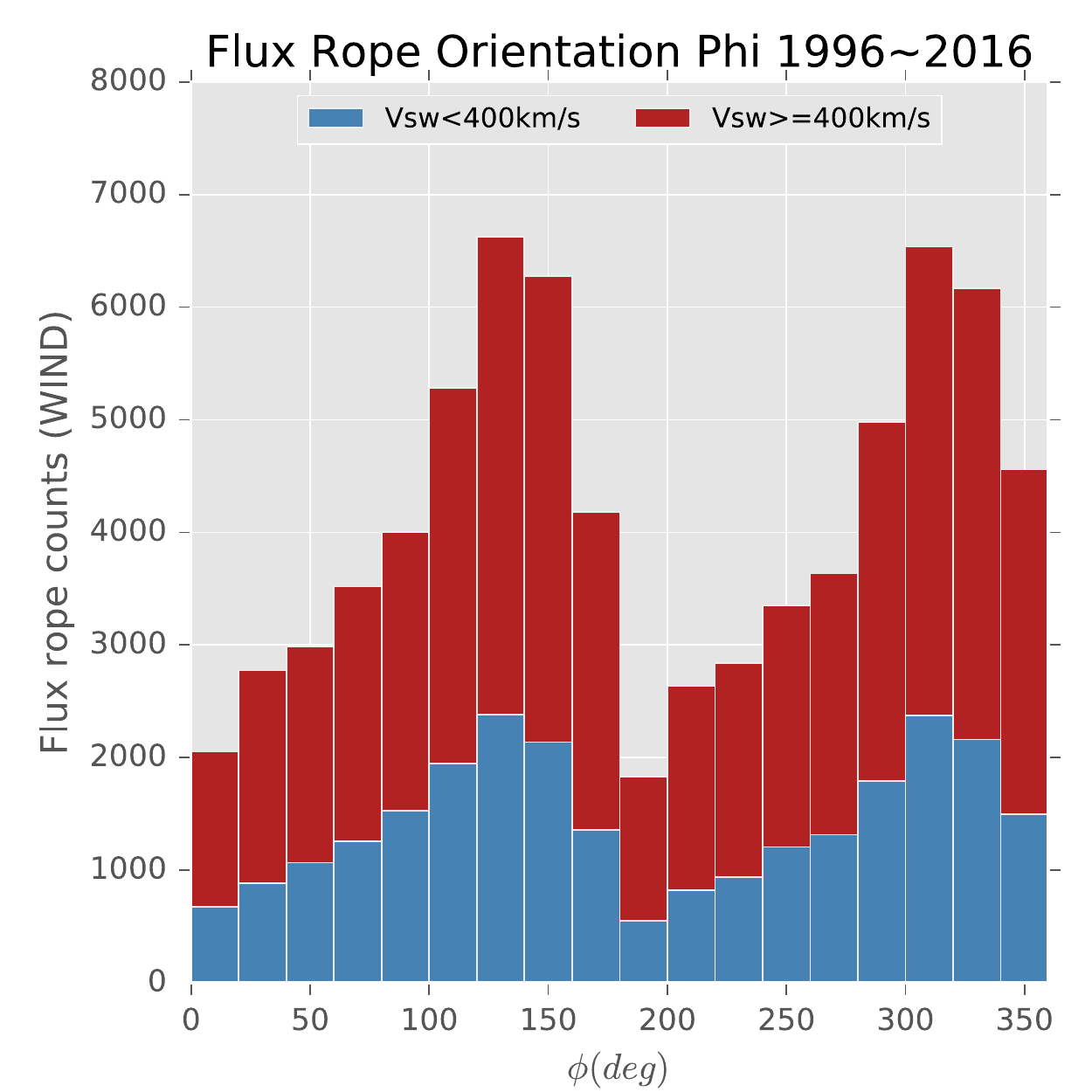}
      \begin{center}
        (b)
      \end{center}
    \end{minipage}
    \caption{(a) Flux rope axial orientation: polar angle $\theta$ histogram.
    The actual polar angle range is from $0^{\circ}$ to $180^{\circ}$. We restrict end points of the direction vectors in the upper
     hemisphere, and do not distinguish the vectors with opposite directions from each other.  (b) Flux rope axial orientation: azimuthal angle $\phi$ histogram. This angle is measured from the positive GSE-$X$ axis toward the projection of the flux rope axis onto the ecliptic plane.}\label{theta_phi}
  \end{center}
\end{figure}

  Figure~\ref{theta_phi} (a) and (b) show the histograms of the small-scale
flux rope axial orientations in the GSE spherical coordinates
based on the last column of the event list. Figure~\ref{theta_phi}
(a) is the polar angle histogram which is binned by $10^{\circ}$,
and Figure~\ref{theta_phi} (b) is the azimuthal angle histogram
which is binned by $20^{\circ}$. These two plots indicate that the
small-scale flux ropes located near the ecliptic plane  have
preferential axial orientations. From Figure~\ref{theta_phi} (a),
one can see that the distributions are skewed toward  large polar
angels, i.e., most of them tend to lie on the ecliptic plane.
Figure~\ref{theta_phi} (b) shows two peaks located at bin
$120^{\circ}\sim140^{\circ}$ and the bin
$300^{\circ}\sim320^{\circ}$. In fact, these two bins represent
two parallel but opposite directions in the GSE $XY$ plane, so
they differ by about $180^{\circ}$. Either one of these directions
happens to be the tangential direction of the Parker spiral at 1
AU (corresponding to $\phi \approx 135^\circ$ or $315^\circ$).
This indicates that the projection of the flux rope axis tends to
align with the Parker spiral direction on the ecliptic plane. The
red and blue bars represent the events occurring under different
solar wind speed conditions (blue: $\overline{V}_{sw}<$ 400 km/s;
red: $\overline{V}_{sw}\ge$ 400 km/s). Figure~\ref{theta_phi} (a)
and (b) indicate that the small-scale flux ropes have similar
orientation preferences in both fast and slow solar wind in the
ecliptic.

\begin{figure} 
    \begin{center} (a)\\
            \includegraphics[width=1\textwidth]{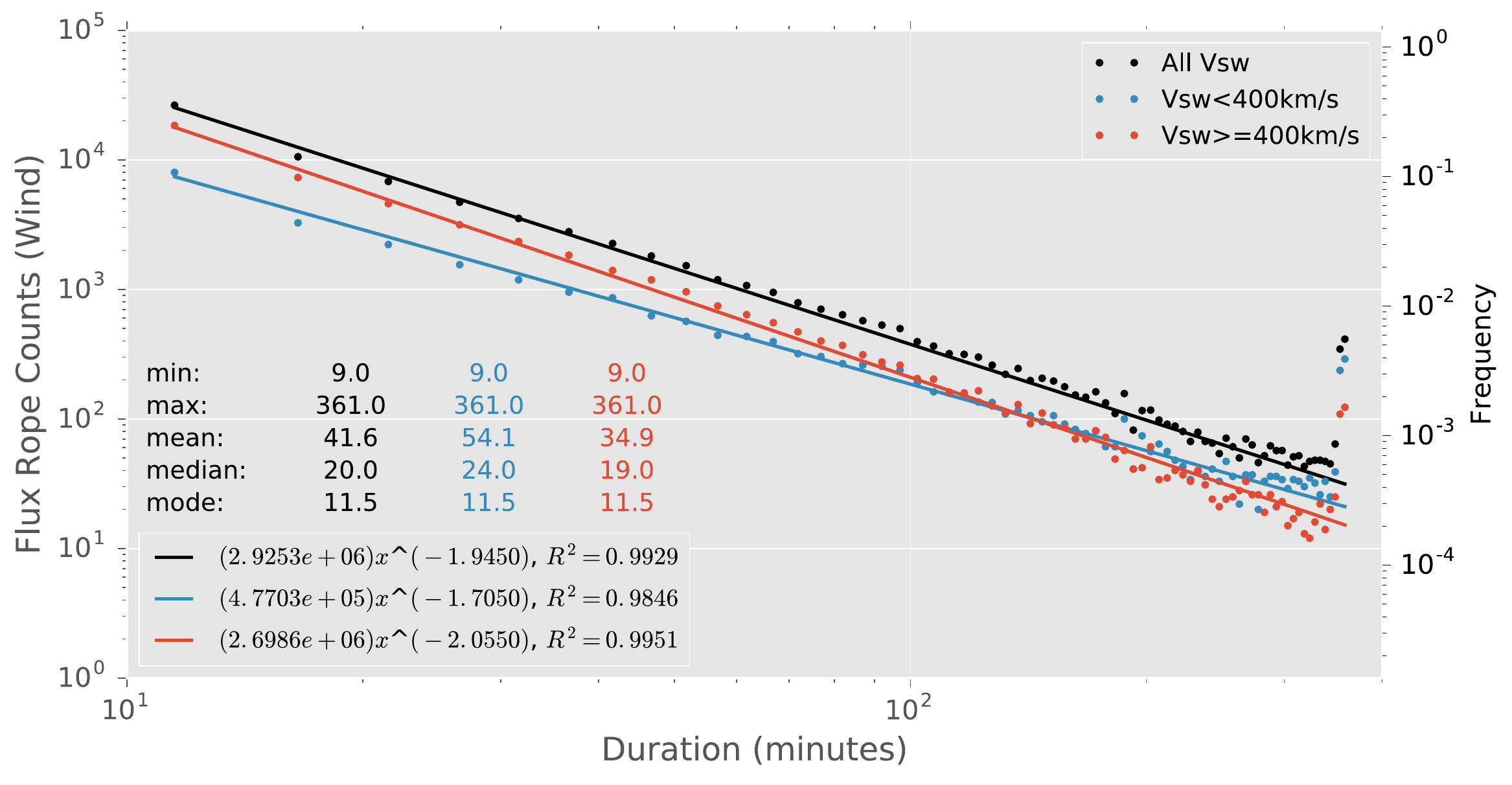}
       (b)\\
                \includegraphics[width=1\textwidth]{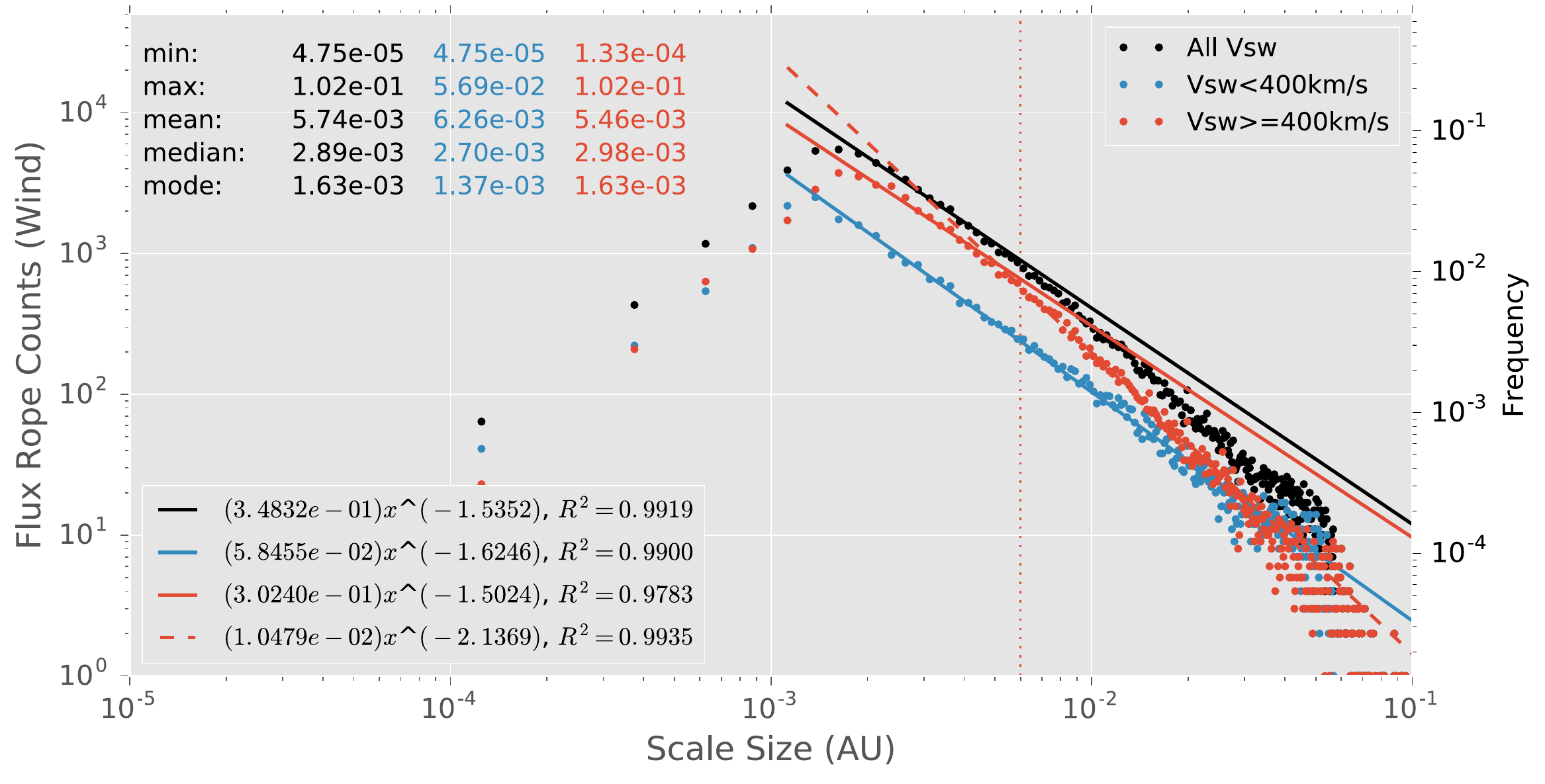}
                    \end{center}
    \caption{(a) The histogram of small-scale flux rope duration plotted in logarithmic scales.
    The time range is from 9 minutes to 361 minutes, with 5 minutes bin size.
    Note that the high tails to the right ends of the curves are due to duration boundary cutoff effect
    (see explanations in text).  (b) The histogram of small-scale flux rope cross section scale sizes plotted in logarithmic scales,
    with 0.00025 AU bin size. The flux rope scale size is calculated from the flux rope duration taking into account the axial
    orientation. Again, the low points to the left and right ends of the curves are due to boundary cutoff effect
    (see explanations in text). For both (a) and (b), the basic statistical quantities and linear regression parameters for each curve are listed, respectively. }\label{duration_size}
\end{figure}

Figure~\ref{duration_size} (a) and (b) show the duration and scale
size distributions of small-scale flux ropes in our database. The
data points in black in each plot represent the histograms of the
entire event set, and the points in blue or red represent the
histogram of the the subsets for slow ($\overline{V}_{sw}<$400
km/s) and fast ($\overline{V}_{sw}\ge$400 km/s) solar wind speed,
respectively. In Figure~\ref{duration_size} (a), the shapes of
these three curves are close to straight lines on the log-log
scale except for the high tails to the right. After examining the
generation process of our flux rope database, we find that the
high tail is due to cutoff effect of a finite range of window
widths. As described in Section~\ref{sec:algorithm}, the last step
of the flux rope detecting process is to combine all flux rope
candidate lists with different duration ranges. In this process,
some short duration flux ropes will be absorbed into the longer
duration flux ropes which enclose them. In the present database,
the longest duration range is 354$\sim$361 minutes, which means
that the flux ropes within this duration range will not be merged
into longer flux ropes, since there is no longer duration allowed
beyond this range, thus resulting in the cutoff. This cutoff leads
to the enhanced fluctuation and a high tail near the end to the
right, corresponding to the last duration range in the current
search. To confirm this effect, we manually shifted the cutoff
boundary to shorter duration, the high tail also shifted
accordingly. In Figure~\ref{duration_size} (b), there are low
tails at both ends in each curve, which are also caused by the
same cutoff effect of finite duration ranges. Due to the boundary
cutoff in duration ranges, when converting the duration to scale
size, there is a lack of events near the lower and upper scale
size boundaries as shown. We exclude these abnormal sections of
the data points in the subsequent analyses.

Except for the high tails due to the cutoff effect,
Figure~\ref{duration_size} (a) shows a linear relation between the
flux rope occurrence counts and the duration under logarithmic
scales, which indicates a power law distribution of flux rope
duration. The color coded straight lines are the corresponding
fitted power law functions as denoted. The flux ropes under the
slow solar wind condition obey the power law with a power index
$\sim-1.70$, while the ones in fast solar wind obey the power law
with power index $\sim-2.06$. One can see that the red curve
($\overline{V}_{sw}\ge$400 km/s) has larger absolute slope value
than the blue curve ($\overline{V}_{sw}<$400 km/s). The
intersection of the blue and red curves is located at about 100
minutes. To the left side of the intersection point, there are
more flux rope events in fast solar wind, while to the right,
there are more flux rope events with low solar wind speed. This
indicates that the longer duration flux ropes (duration$>$100
minutes) tend to occur under the slow solar wind speed condition.

Figure~\ref{duration_size} (b) also exhibits approximately linear
relations (in logarithmic scales) between flux rope scale size and
occurrence counts, excluding the low ends due to the cutoff
effect. This indicates a power law distribution of flux rope scale
sizes, properly calculated taking into account the axial
orientations. At a glance, the absolute slope of the red curve
($\overline{V}_{sw}\ge$400 km/s) seems to be greater than that of
the blue curve ($\overline{V}_{sw}<$400 km/s), especially for the
section of larger scale sizes. Comparing with
Figure~\ref{duration_size} (a), one can see that the flux ropes
ranging from 9$\sim$361 minutes have the approximately
corresponding scale-size range of 0.002$\sim$0.05 AU (excluding
the lower end). When we use power law functions to fit the data,
we find that the data points in blue color are fitted very well by
a single power function with power index $\sim-1.62$, but for the
red and black data points, the tails show noticeable deviations
from  single fitted lines. Apparently, the deviations in the black
data points are due to the deviations in the red ones. We use two
power law functions with different power indices to fit the red
data points, and find that in the scale size range $0.001\sim0.01$
AU, the red data points are well fitted by a power law function
(solid line) with a power index $\sim-1.50$, while in the range
$0.01\sim0.05$ AU, the red data points are well fitted by a power
law function (dashed line) with a power index $\sim-2.14$. The
breakpoint of the two power laws is at $\sim0.006$ AU.


It is interesting to note that both Figure~\ref{duration_size} (a)
and (b) obey the power law distribution.
The power law distributions are fairly common in nature. These
analysis results are relatively new, concerning these small-scale
flux ropes from a large sample. We will discuss these  results in
the following from the perspective of self-organized criticality
theory.

The self-organized criticality (SOC) theory was first proposed by
\citet{Bak1987}, and then applied by \citet{Lu1991,Lu1993} to
solar physics to explain the power law distributions of flare
occurrence rate over flare energy, peak flux, and duration. This
model is usually referred to as the avalanche model and has been
widely used to explain the statistical characteristics of hard
x-ray (HXR) flares
\citep{Lu1993,Vlahos1995,Mackinnon1997,Georgoulis1998}. The
avalanche model predicts a power law distribution for the total
energy, the peak luminosity, and the duration of individual
events. \citet{Li2016} studied the solar flares  and CMEs during
the solar cycle 23. They found that the solar flare duration
distribution obeys a power law with a power index $\sim-2.55$. In
Figure~\ref{duration_size} (a), we also produce a power law with
an index close to -2 for all events (black line), implying that
the occurrence of small-scale flux ropes may be explained by SOC
theory. Note that the absolute values of the power indices from
our fitted functions are generally smaller than those in
\citet{Li2016}, indicating solar flares or CMEs and small-scale
flux ropes are probably corresponding to two different kinds of
processes. Although they share the similar statistical
characteristics in terms of the power law distributions in certain
properties, complying with the SOC theory, the underlying physical
mechanism responsible for generating such behavior still cannot be
revealed.


\begin{figure} 
    \begin{center} (a)\\
            \includegraphics[width=1\textwidth]{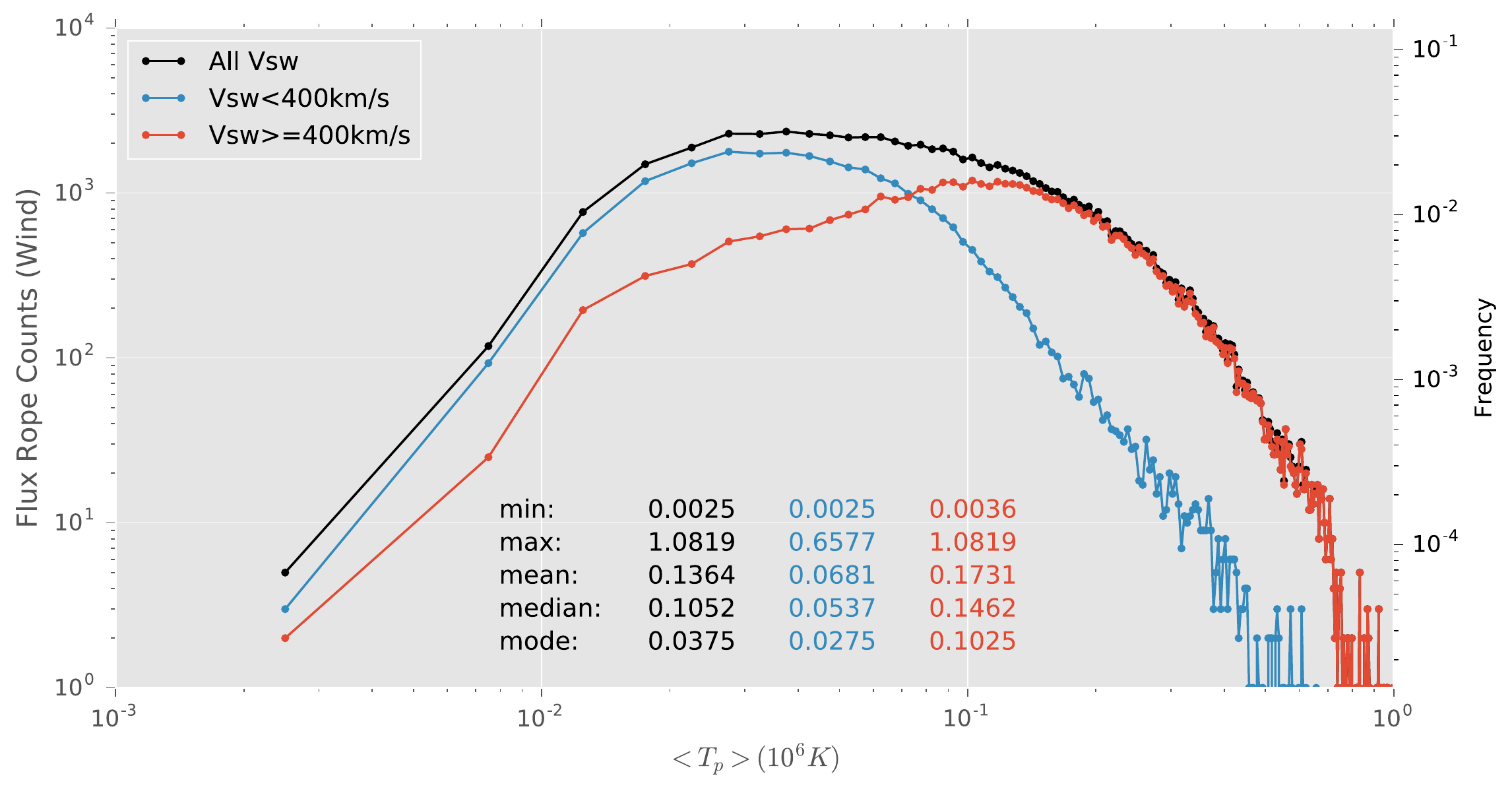}
        (b)\\
     \includegraphics[width=1\textwidth]{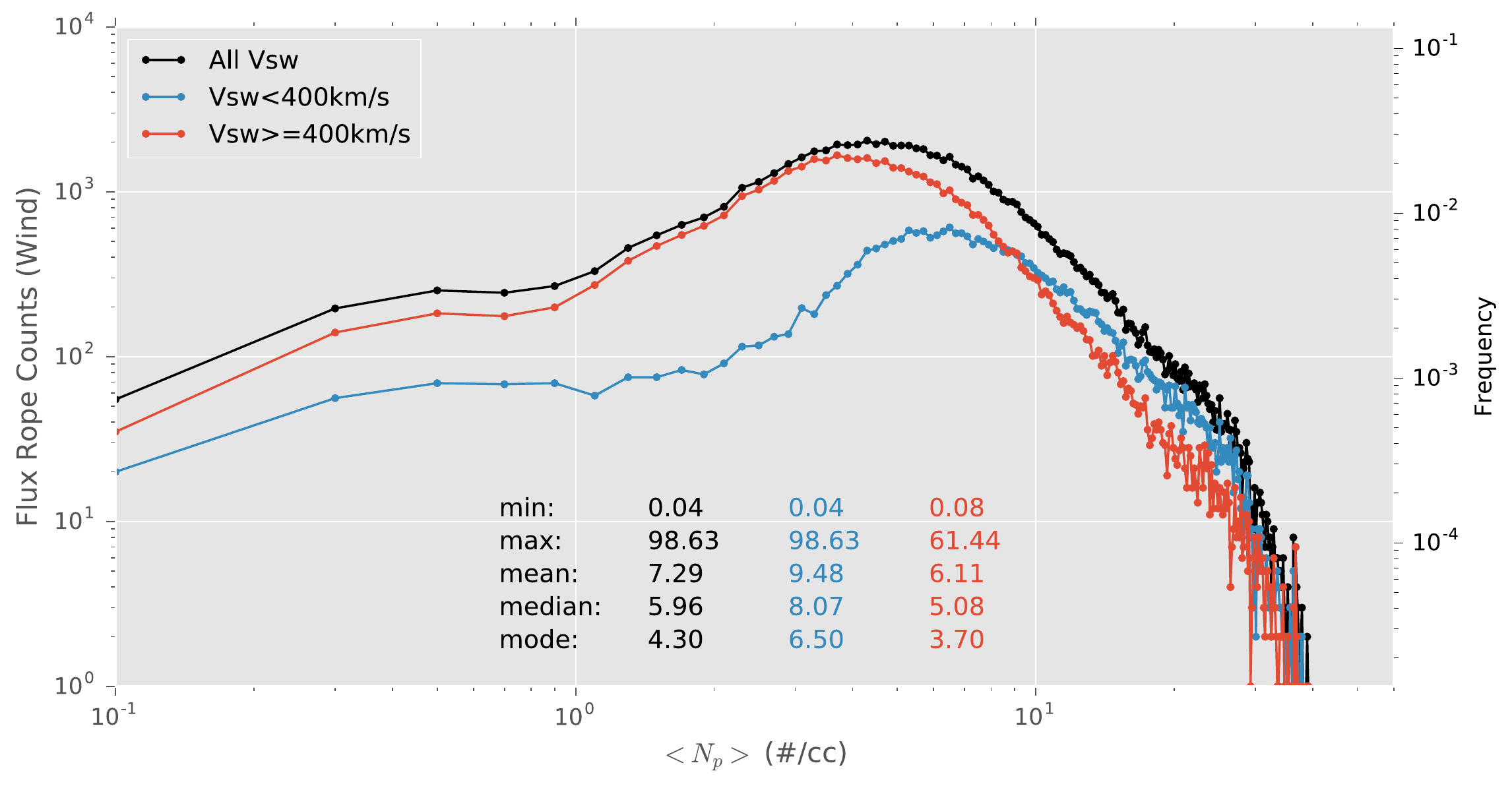} 
    \end{center}
    \caption{(a) The histogram of average proton temperature within flux ropes in logarithmic scales,
    with 0.005 ($10^6$ K) bin size. (b) The histogram of average proton number density within flux ropes in
    logarithmic scales, with 0.2 $\#/cc$ bin size. For both (a) and (b), the blue curve represents the flux rope events
    with solar wind speed $\overline{V}_{sw}<$400 km/s, and the red curve with solar wind speed $\overline{V}_{sw}\ge$400 km/s.
    The black curve represents the entire event set.}\label{Tp_Np}
\end{figure}

Figure~\ref{Tp_Np} (a) is the histogram of average proton
temperature within flux ropes plotted in logarithmic scales. We
can see that the peaks of blue and red curves are separated,
corresponding to different modes. The peak of the red curve
($\overline{V}_{sw}\ge$400 km/s) is near $\overline{T}_p$
$\approx$ 0.1 ($10^6$ K), while the peak of the blue curve
($\overline{V}_{sw}<$400 km/s) is near $\overline{T}_p$ $\approx$
0.03 ($10^6$ K). Therefore, the small-scale flux ropes in low
speed solar wind ($\overline{V}_{sw}<$400 km/s) tend to have low
proton temperature, while the ones under medium and high speed
solar wind ($\overline{V}_{sw}\ge$400 km/s) tend to have high
proton temperature. Note that the black curve and the red curve
are overlapping beyond $\overline{T}_p$ = 0.2 ($10^6$K), which
means that the small-scale flux ropes with proton temperature
greater than 0.2 ($10^6$ K) occur dominantly under medium and high
speed solar wind conditions. Figure~\ref{Tp_Np} (b) is the
histogram of average proton number density within flux ropes
plotted in logarithmic scales. Note that the appearance in
Figure~\ref{Tp_Np} (b) is opposite to that in Figure~\ref{Tp_Np}
(a). In Figure~\ref{Tp_Np} (b), the flux ropes with medium and
high solar wind speed tend to have lower proton number density
while  the flux ropes in slow solar wind tend to have higher
density. To summarize Figure~\ref{Tp_Np} (a) and (b), the flux
ropes in slow speed solar wind tend to have low proton temperature
and high proton number density, while the flux ropes in medium and
high speed solar wind tend to have high proton temperature and low
proton number density.

\begin{figure} 
    \begin{center} (a)\\
            \includegraphics[width=1\textwidth]{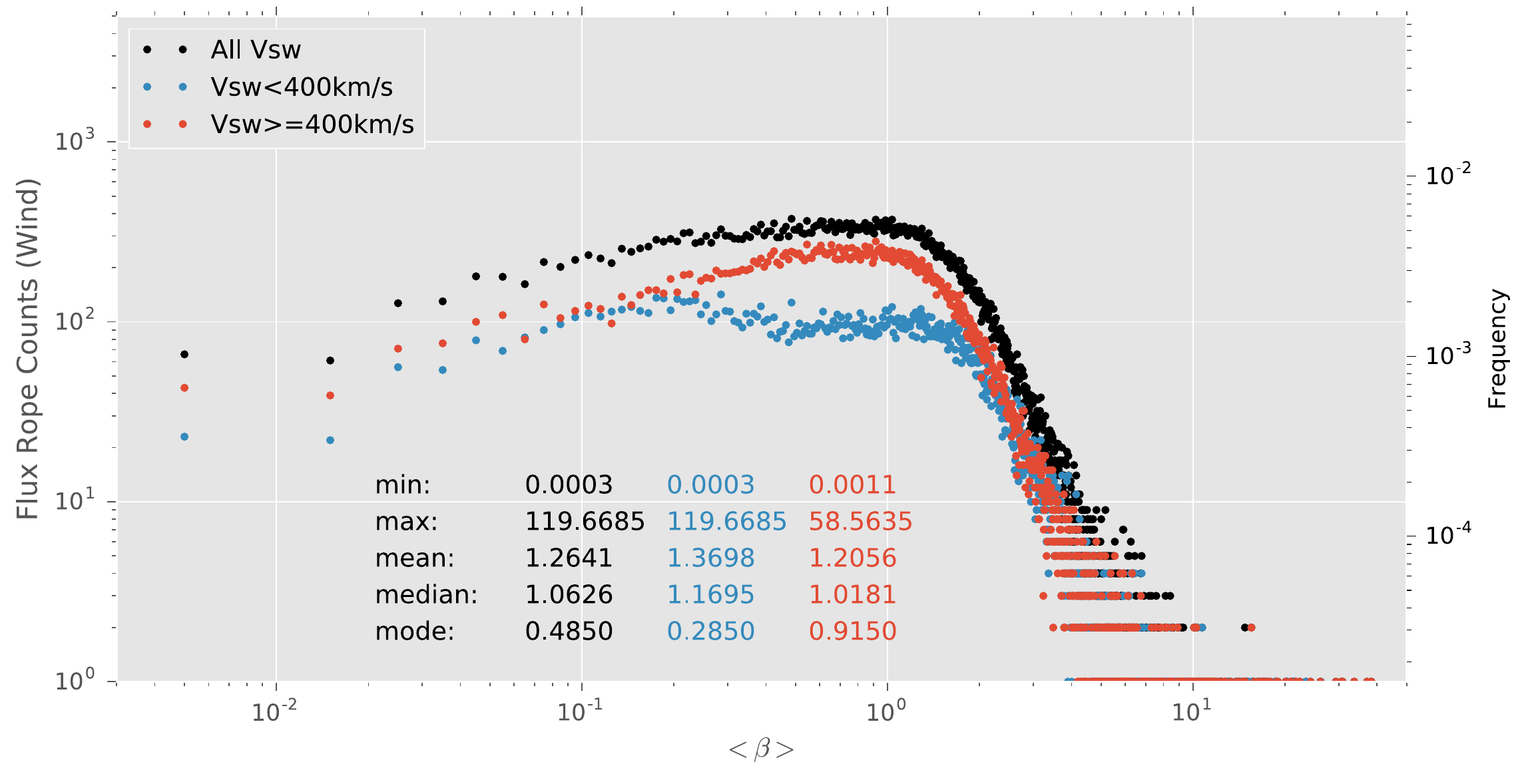}
       (b)\\
                \includegraphics[width=1\textwidth]{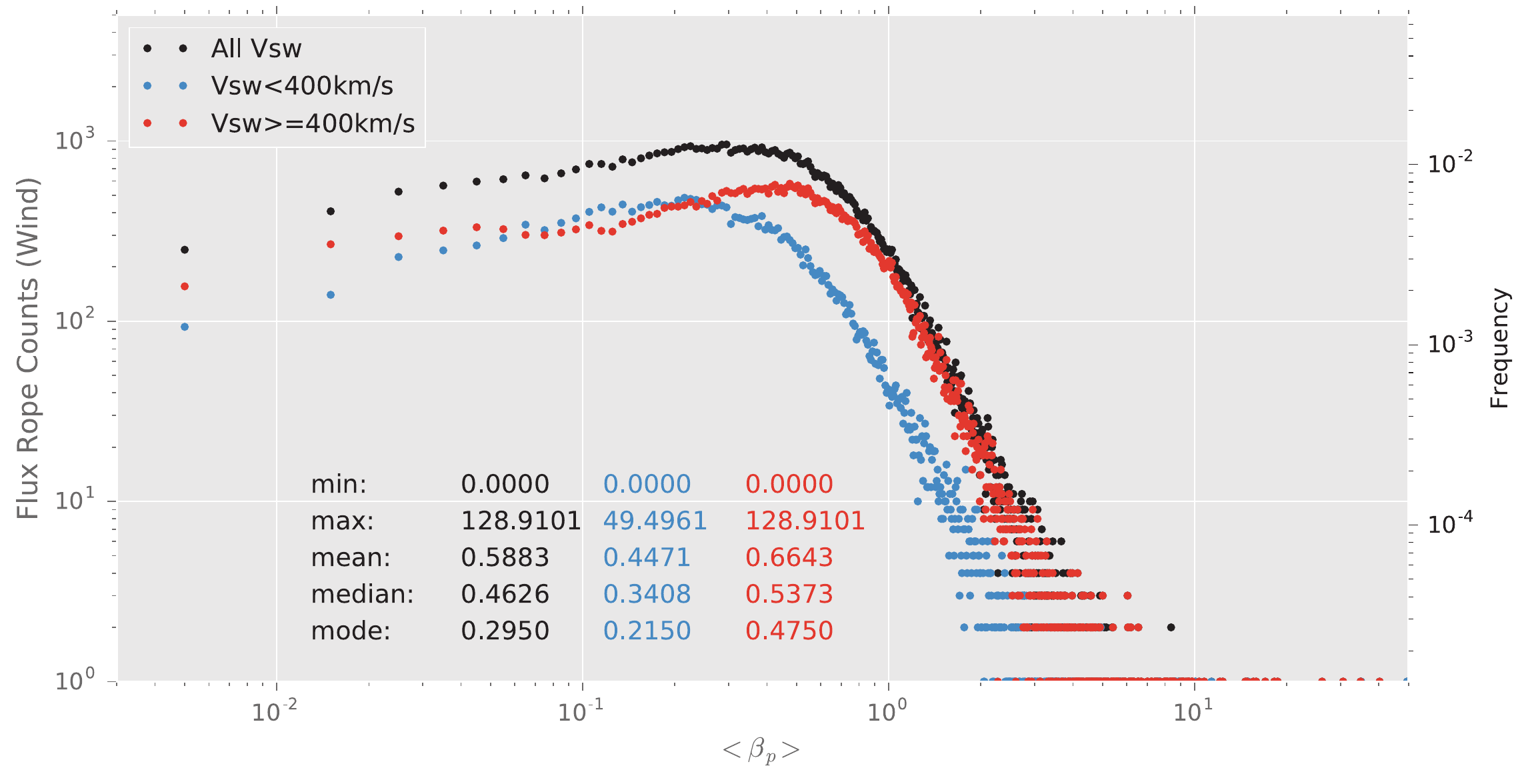}
                     \end{center}
    \caption{(a) The histogram of average plasma $\beta$ within flux ropes, with bin width 0.01. (b) The histogram of average
    proton  $\beta_p$ within flux ropes, with bin width 0.01. For both (a) and (b), the blue curve represents the flux rope
    events with solar wind speed $\overline{V}_{sw}<$400 km/s, and the red curve with solar wind speed $\overline{V}_{sw}\ge$400 km/s.
     The black curve represents the entire event set.}\label{Beta}
\end{figure}

Figure~\ref{Beta} (a) is the histogram of average plasma $\beta$
(electron temperature $T_e$ included) within flux ropes plotted in
log-log scales. The black curve shows that the occurrence peak of
all small-scale flux ropes is close to $\overline{\beta}=1$. The
red curve has the same trend as the black curve, while the blue
curve has a flat top. From the shapes of these curves, and
especially the medians, we find that the numbers of small-scale
flux rope with $\overline{\beta}<1$ and the ones with
$\overline{\beta}>1$ are about the same. However, in magnetic
clouds, the magnetic pressure always dominates over thermal
pressure, which causes ultra low plasma $\beta$. Figure~\ref{Beta}
(b) is the histogram of average proton plasma $\beta_p$ (excluding
$T_e$) within flux ropes plotted in log-log scales. Because the
electron temperature data quality is generally poor, they are not
always available. To overcome the large data gaps in $T_e$, we
also plot the histogram of the average proton plasma $\beta_p$, in
which only the contribution of protons temperature $T_p$ is
included when calculating the thermal pressure. Figure~\ref{Beta}
(b) shows the similar distributions as Figure~\ref{Beta} (a). All
curves in Figure~\ref{Beta} (b) are shifted to the left compared
with Figure~\ref{Beta} (a), indicating that the plasma $\beta$ is
significantly enhanced by the electron temperature contribution,
as also indicated by the various statistical quantities denoted on
each plot.

 \begin{figure}         
  \begin{center}
    \includegraphics[width=1.0\textwidth]{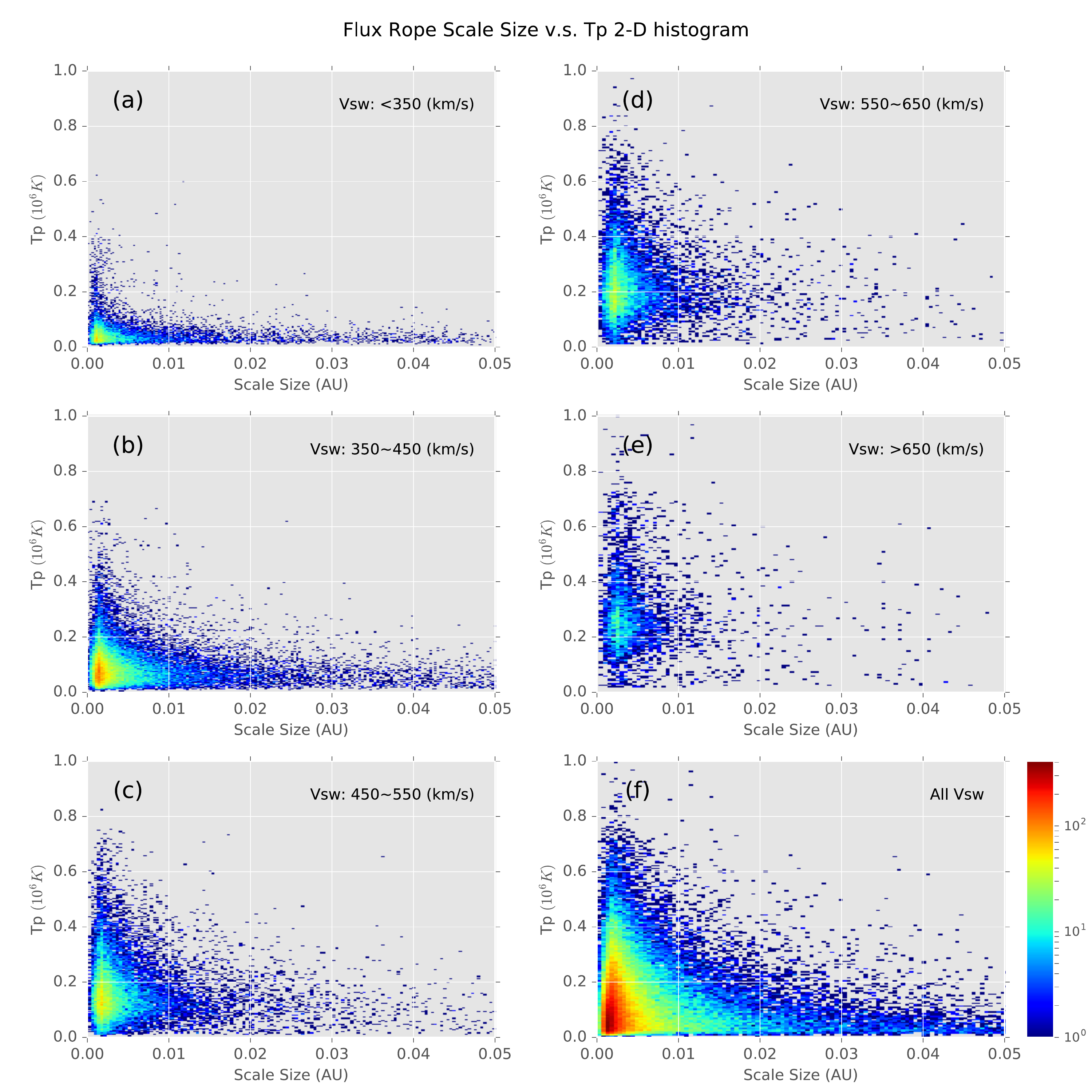}
  \end{center}
  \caption{The 2-D histogram of flux rope proton temperature $versus$ scale size. The bin grids are 200$\times$200.
  Subplots (a) to (e) are histograms under different solar wind speeds as denoted, and subplot (f) is the histogram for the entire event set.
  The color bar represents the small-scale flux rope counts.}\label{Size_vs_Tp}
\end{figure}

The low proton temperature ($T_p$) is a key characteristic of
magnetic clouds. However, for small-scale magnetic flux ropes, the
$T_p$ varies case by case. Figure~\ref{Size_vs_Tp} is the 2-D
histogram of flux rope $T_p$ $versus$ scale size. The triangle
shape distribution stretching down to the right in
Figure~\ref{Size_vs_Tp} (f) indicates that the flux ropes with
larger scale size ($\geq0.02$ AU) usually have lower $T_p$
($\leq0.1$ $\times10^6$ K). Given that the large-scale magnetic
flux ropes (MCs) have low $T_p$, there seems to be a smooth
transition from the small-scale magnetic flux ropes to their
larger counterparts. For the smaller size small-scale flux ropes,
the range of $T_p$ spreads widely. Figure~\ref{Size_vs_Tp} (f)
shows that the range of $T_p$ of the flux ropes with scale size
less than 0.05 AU spreads from near 0 to 0.8 ($10^6$ K). When
looking into the distribution under different solar wind speed
conditions, we find that most of the relatively larger size
small-scale flux ropes ($\geq0.02$ AU) appear in the slow solar
wind with low $T_p$ (see Figure~\ref{Size_vs_Tp} (a), (b), and
(c)). As for the relatively smaller size small-scale flux ropes
($\leq0.01$ AU), they appear in both fast and slow solar wind. The
different locations of the high frequency regions (in red and
yellow colors) in each subplot ( Figure~\ref{Size_vs_Tp} (a)
$\sim$ (e)) show that the flux ropes in high (low) speed solar
wind tend to have higher (lower) proton temperature, which is
consistent with the result shown in Figure~\ref{Tp_Np} (a).



\subsection{Cycle-to-cycle Variations}\label{sec:cycle}
\begin{figure}[ht] 
  \begin{center}
    \includegraphics[width=0.49\textwidth]{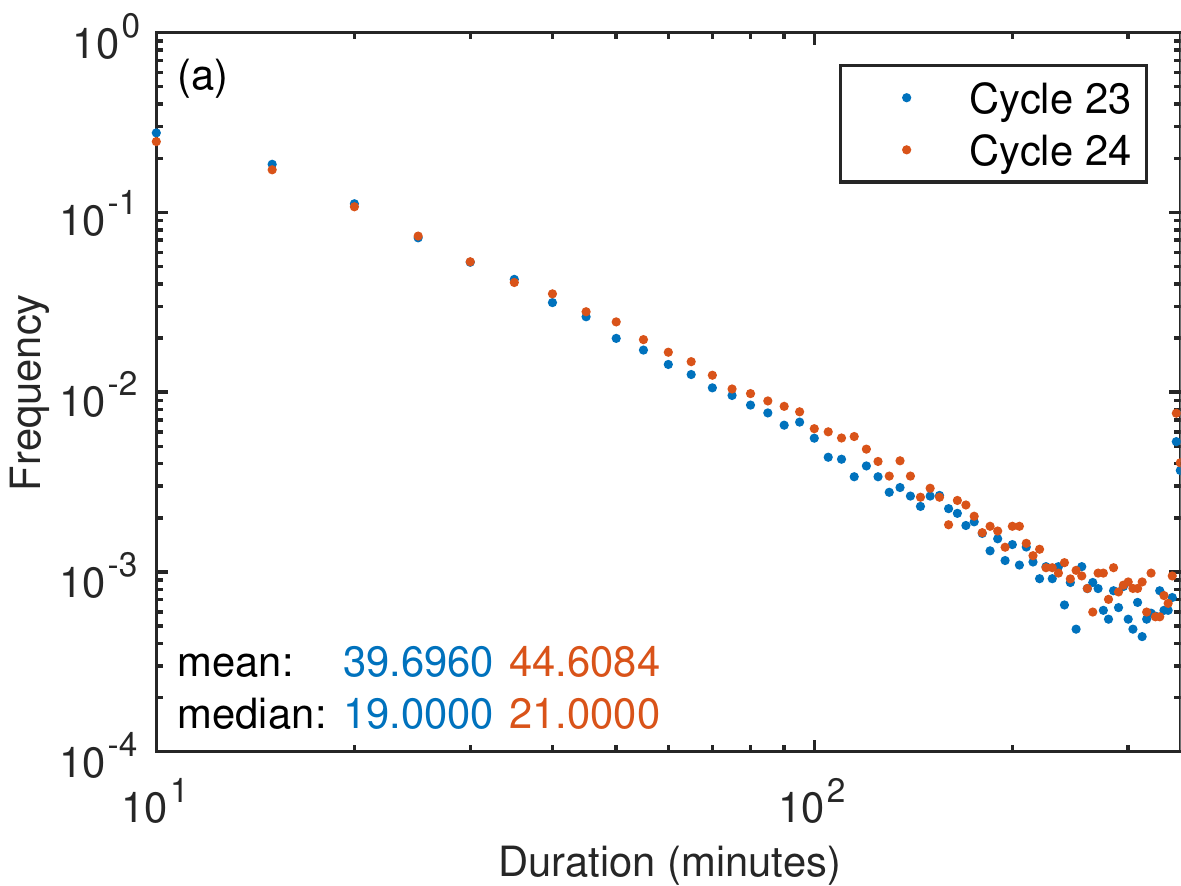}
    \includegraphics[width=0.49\textwidth]{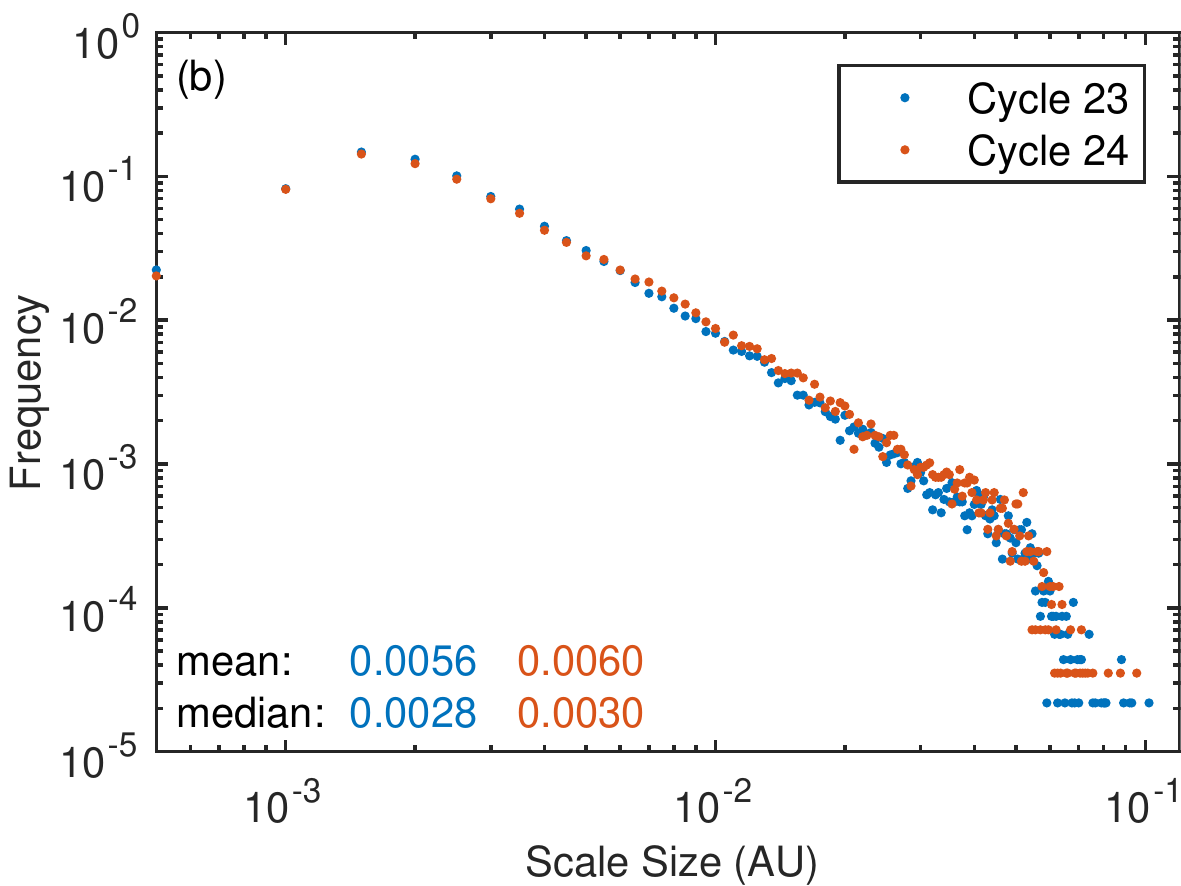}
      \includegraphics[width=0.49\textwidth]{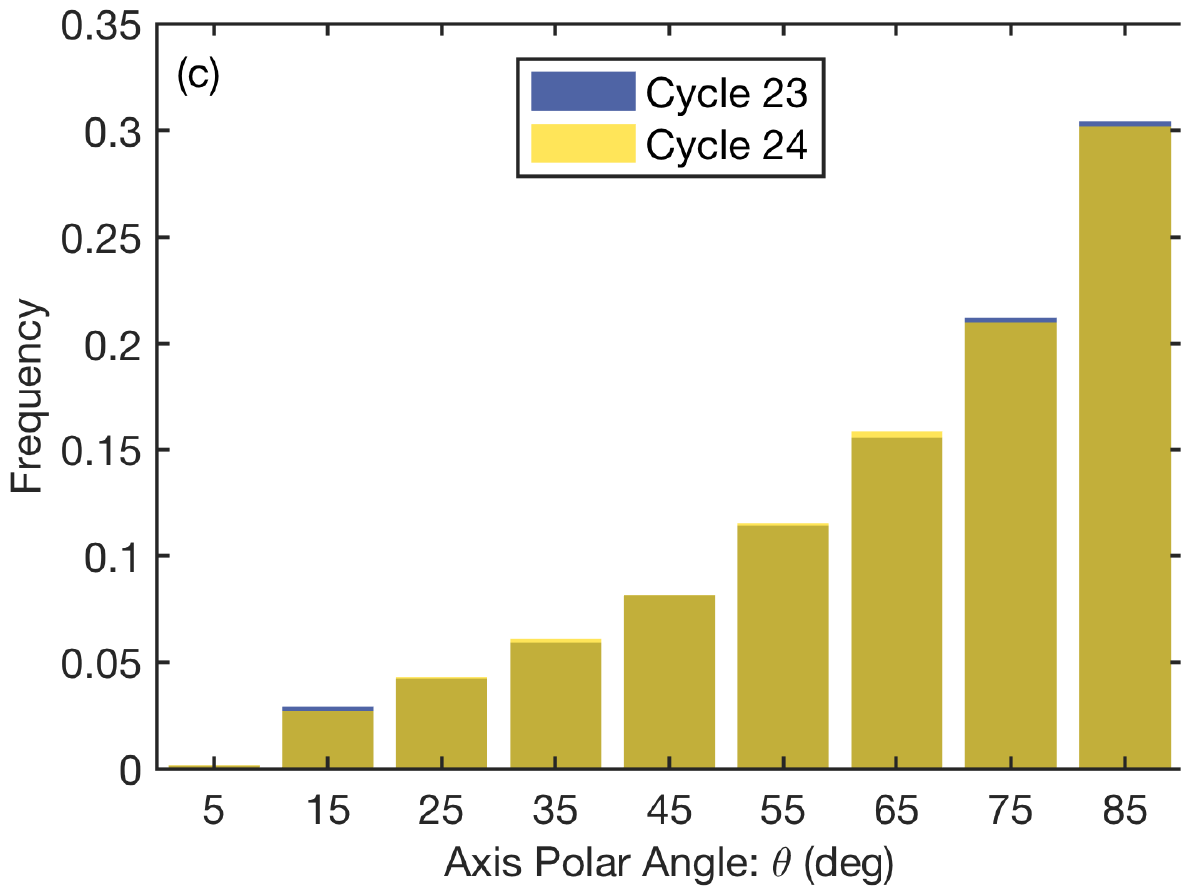}
    \includegraphics[width=0.49\textwidth]{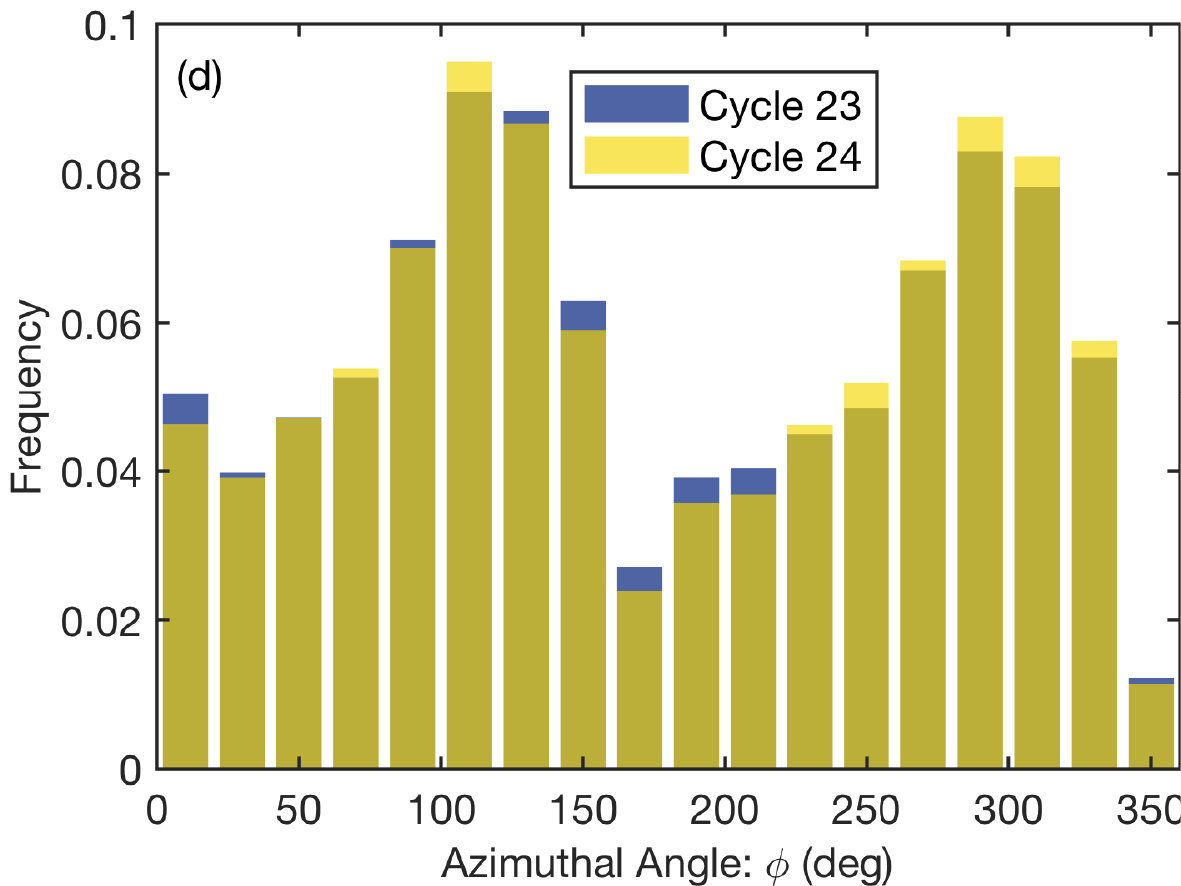}
  \end{center}
  \caption{The histograms of (a) duration, (b) scale size, (c) axial orientation angle $\theta$,
  and (d) $\phi$ of small-scale flux rope events for the solar cycles 23 and 24.}\label{cycle2c2}
\end{figure}

\citet{Gopalswamy2015} investigated the correlation between the
annual number of magnetic clouds events and the frontside halo
CMEs, and compared both with the sunspot numbers. They found
significant difference in the properties of MCs between solar
cycles 23 and 24. It is natural to perform a similar analysis on
the small-scale flux ropes by simply repeating the above
statistical analyses on the two cycles separately. Unlike the
findings of \citet{Gopalswamy2015}, who compared the MC properties
and their geo-effectiveness for the past two solar cycles and
concluded that significant differences were found, we conducted a
similar analysis by comparing the statistics of the same set of
parameters between the two solar cycles. We do not find any
noticeable differences among the majority of parameters, except
for the following two quantities: the average field magnitude and
the average plasma $\beta$, within each flux rope interval. {{For
example, Figure~\ref{cycle2c2} shows the distributions of selected
parameters for each cycle separately. Little difference is seen
between the two cycles.}}

\begin{figure}[ht] 
  \begin{center}
    \includegraphics[width=0.49\textwidth]{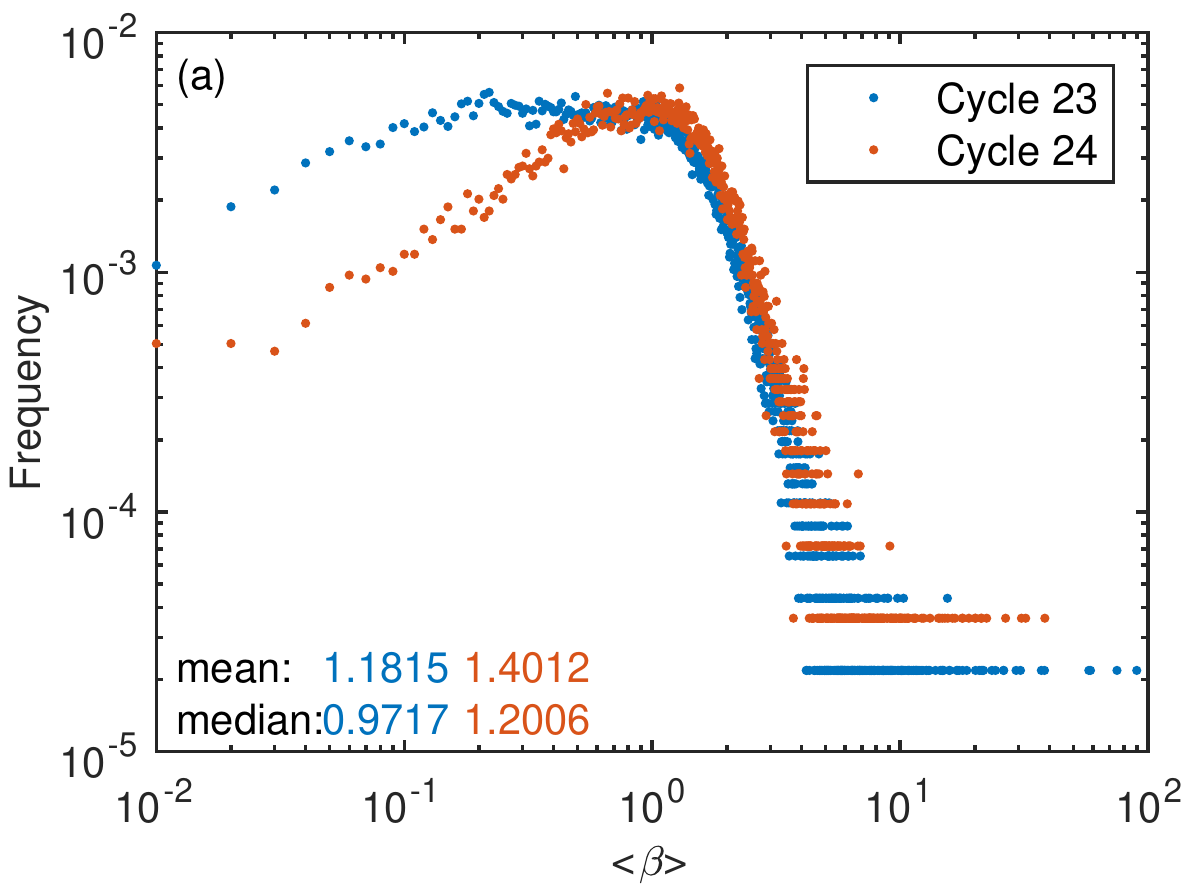}
    \includegraphics[width=0.49\textwidth]{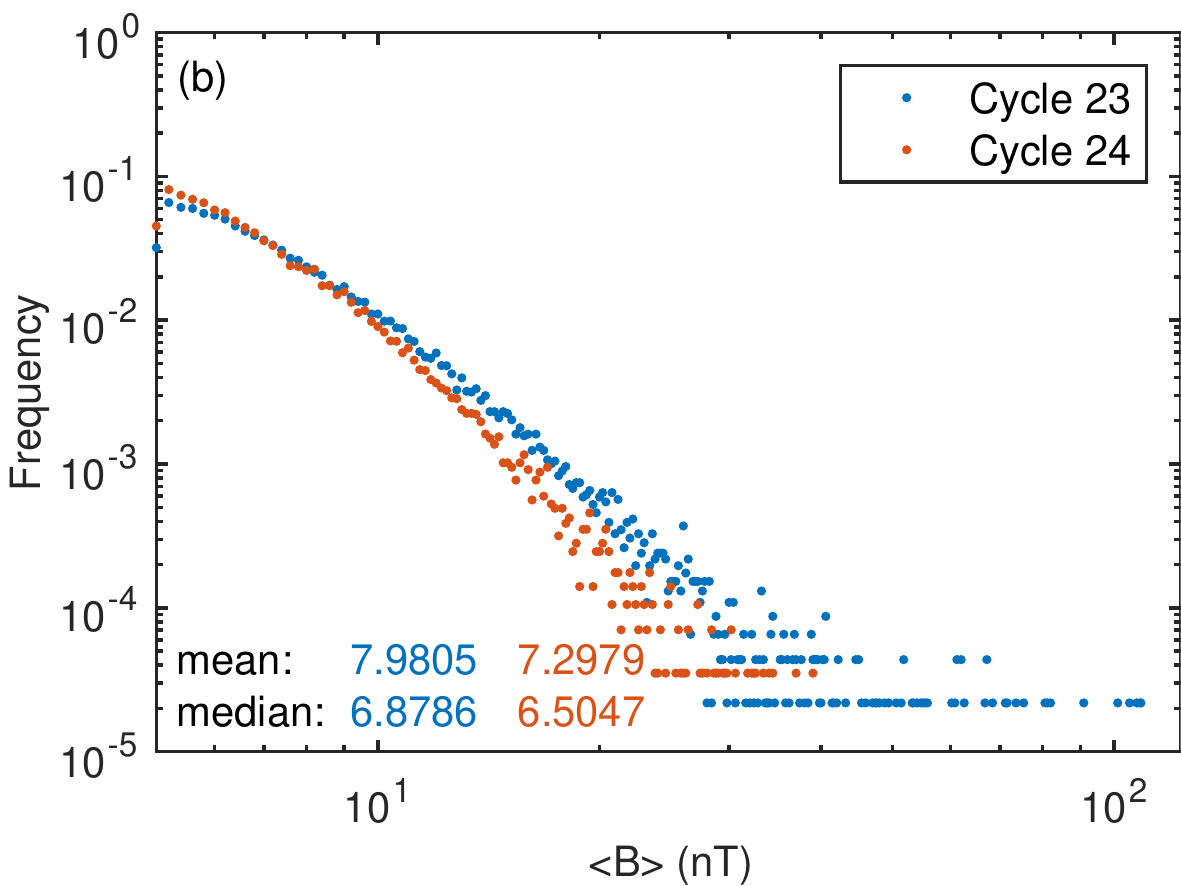}
  \end{center}
  \caption{The histograms of (a) average $\beta$  and (b) average magnetic field magnitude within small-scale flux rope intervals for the solar cycles 23 and 24.}\label{cycle2c}
\end{figure}
Figure~\ref{cycle2c} shows the histograms of the average plasma
$\beta$ and the average magnetic field magnitude for solar cycles
23 and 24, respectively. The overall distributions are similar for
the two cycle. For solar cycle 23, the plasma $\beta$ has slightly
higher mean and median values than cycle 24. Correspondingly the
average field magnitude has a smaller mean in cycle 24, and does
appear to be weaker as shown in Figure~\ref{cycle2c} (b). These
variations can be explained by the known cycle-to-cycle variation
in the interplanetary magnetic field magnitude because the cycle
24 has a weaker magnetic field magnitude than cycle 23
\citep[e.g.,][]{2018Z}.

\section{Waiting Time Distributions}\label{sec:WTD}
 The waiting time
distribution (WTD) is defined by the distribution of time
intervals or separations between discrete events. The WTD of the
successive discrete events reveals whether they occur
independently. Many models on solar eruptive processes predict
definitive WTDs, so the WTD analysis based on observational data
is a powerful tool to validate these models. The WTD analysis is
widely used in analyzing space plasma processes such as CMEs
\citep{Li2016}, solar flares \citep{Wheatland1998,Li2016}, current
sheets \citep{Miao2011}, gamma ray burst \citep{Wang2013}, and
solar energetic particles \citep{Li2014}, and also in other
 discrete time random processes such as earthquakes
\citep{Sotolongo-Costa2000}. In this section, we apply the WTD
analysis on the occurrence of  small-scale magnetic flux ropes, in
order to investigate the underlying mechanism governing the flux
rope origination process. We discuss the possible implications on
the flux rope origination by comparing the WTDs of flux ropes with
the WTDs of some relevant processes, especially those from the
analysis of current sheets \citep{Bruno2001, Greco2008,
Greco2009a,Miao2011}. Part of the analysis results from the flux
rope wall-to-wall time distribution had been reported in
\citet{zhengandhu2018}, which will not be repeated here.

As predicted by the avalanche model \citep{Bak1987, Lu1991,
Lu1993}, the occurrence of solar flares is a Poisson process,
i.e., the flare WTD is a simple exponential function in waiting
time, $\Delta t$. In fact, the occurrence rate of many processes
is non-constant, and as an observational result, the flare WTD
usually shows time-dependent Poisson distributions
\citep{Wheatland2000, Aschwanden2010, Guidorzi2015, Li2016}.
However, some observational results showed the deviations of flare
WTD from a simple Poisson process. \citet{Pearce1993} studied 8319
HXR solar flares during solar minimum (1980$\sim$1985), and
pointed out that the WTD ($\Delta t$ ranging from 0 to 60 minutes)
of flares has large deviation from a stationary Poisson process,
but is well fitted by a power law function with a power index
-0.75 (See Figure 4 in \citet{Pearce1993}), indicating that flares
did not occur purely randomly. \citet{Li2016} investigated the
statistical properties of CMEs and solar flares during solar cycle
23. They adopted the non-stationary Poisson distribution functions
from \citet{Li2014} and \citet{Guidorzi2015} to fit the WTDs of
solar flares and CMEs, and obtained good fitting results. In
\citet{Li2014,Guidorzi2015} the non-stationary Poisson
distribution functions and the asymptotic behavior of the longer
waiting time near the tail lead to power law functions.

In this section, we try various fitting functions mentioned above
to fit the flux ropes WTDs. Primarily we use exponential and/or a
power law function fittings as we present below, which yields the
optimal outcome. The functions used by \citet{Li2014,Guidorzi2015}
do not work, partially due to the limited range of waiting time,
$\Delta t$, in out database. Here we use two different kinds of
definition for waiting time \citep[e.g.,][]{Greco2009b}. The first
kind of waiting time is defined by the elapsed time from the end
of one flux rope to the start of the next one, i.e., the time
interval between adjacent flux ropes. And the second kind of
waiting time is defined by the elapsed time from the starting time
of one flux rope to the starting time of the next, i.e., the time
interval between the starting times of two successive flux ropes.

\begin{figure}[ht] 
  \begin{center}
    \includegraphics[width=0.9\textwidth]{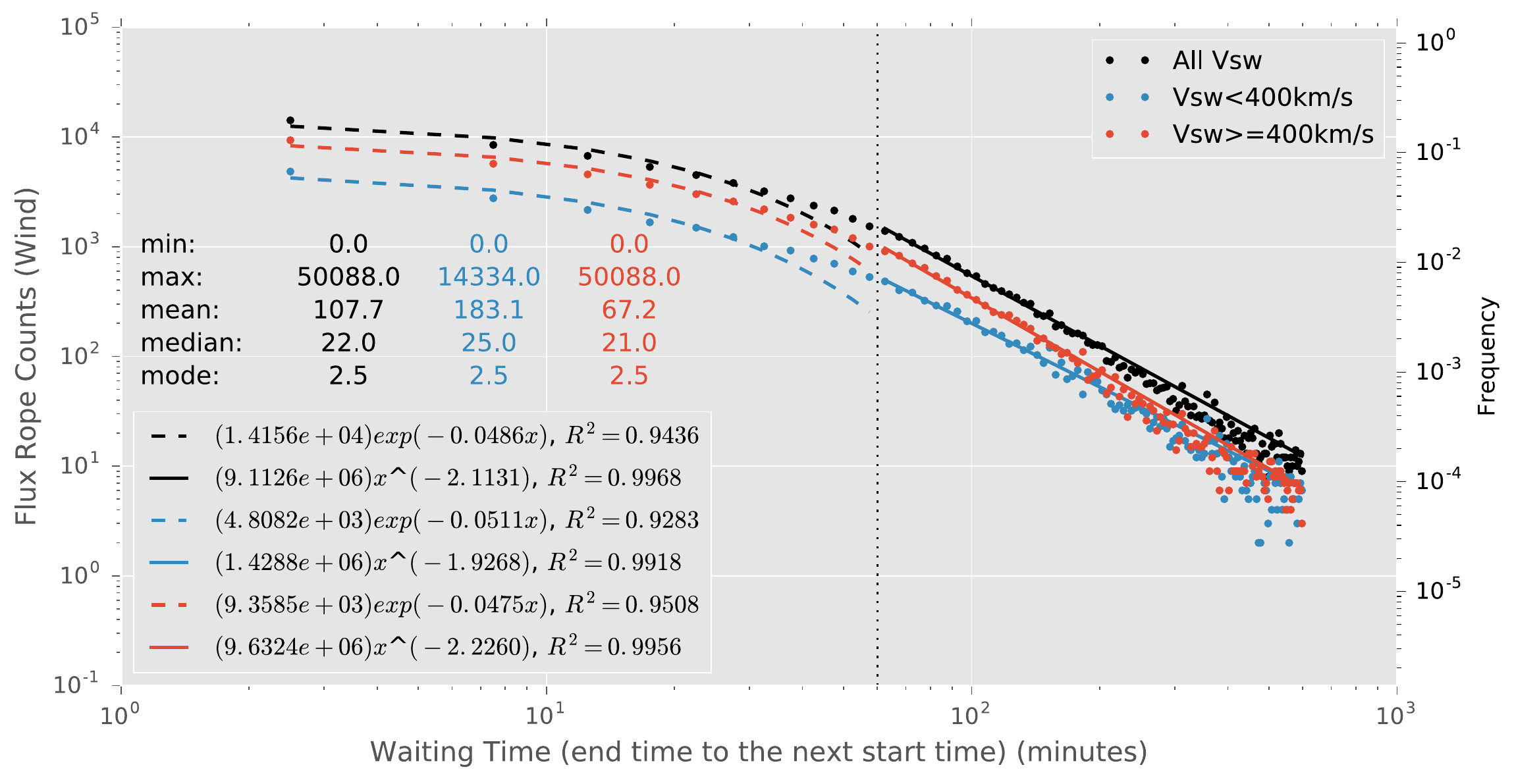}
  \end{center}
  \caption{The waiting time distribution (1st kind) of small-scale magnetic flux ropes. In this plot,
  the waiting time is defined by the interval between the end time of one flux rope to the start time of the next one.
  The dots in black represent the entire event set, and the dots in blue and red represent the subsets with low solar wind speed
  ($\overline{V}_{sw}<$400 km/s) and medium as well as high solar wind speed ($\overline{V}_{sw}\ge$400 km/s), respectively.
  The bin size is 5 minutes, and the data point for each bin is located in the bin center.
  The dashed lines are fitted curves by exponential functions, and the solid lines are fitted curves by power law functions.
  The dotted vertical line denotes the break point, which is located at 60 minutes. The statistical quantities and fitting
  parameters are denoted, where the function forms and the fitting quality metrics $R^2$ are listed.}\label{STA_hist_waitTime}
\end{figure}

Figure~\ref{STA_hist_waitTime} is the flux rope WTD of the first
kind. For each color-coded data set, we use an exponential
function to fit the data points with waiting time less than 60
minutes. The exponential function fits the data well where waiting
time is less than 40 minutes. Then the deviation grows as the
waiting time becomes larger. The section beyond 60 minutes is
fitted by a power law function. The power law function fits the
tail very well, with the coefficient of determination $R^2>0.99$
for each data set. An exponential WTD indicates a random Poisson
process of the event occurrence, while a power law WTD suggests
the clustering behavior of the events. The fitting results show
that the overall WTD shown in Figure~\ref{STA_hist_waitTime} is
neither a simple Poisson distribution nor a power law
distribution.

\begin{figure}[ht] 
  \begin{center}
    \includegraphics[width=0.9\textwidth]{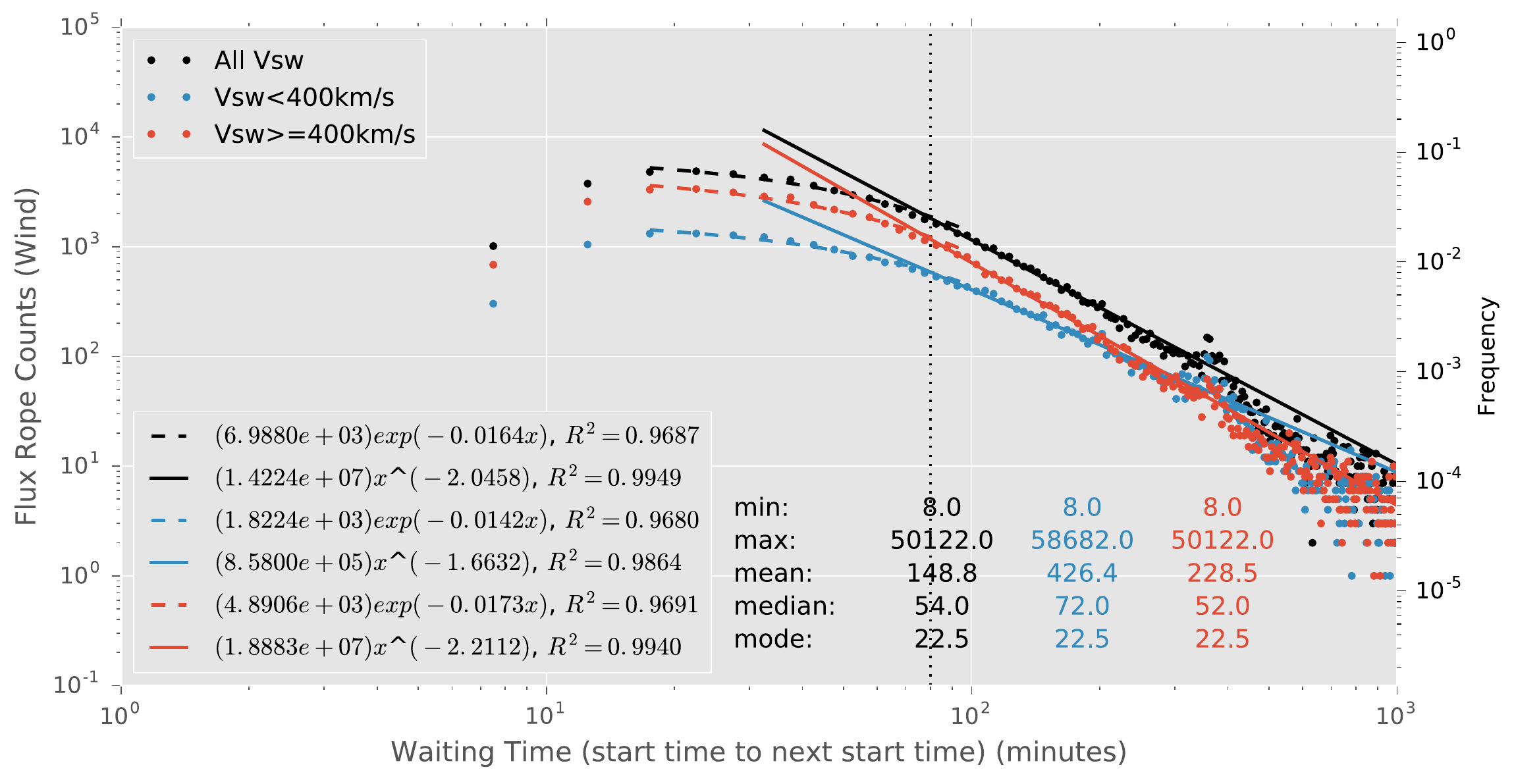}
  \end{center}
  \caption{The waiting time distribution (2nd kind) of small-scale magnetic flux ropes. In this plot,
   the waiting time is defined by the interval between the start time of one flux rope to the start time of the next one.
   The format is the same as in Figure~\ref{STA_hist_waitTime}. The vertical dotted line is at 80 minutes.}\label{STA_hist_waitTime2}
\end{figure}

Considering that the starting time of one flux rope may correspond
to the onset of solar eruptive process back on the Sun, according
to the definition by \citet{Wheatland2000}, we perform the WTD
analysis based on an alternative definition of waiting time. In
this definition, the interval between two consecutive starting
times is considered as the waiting time. We call the WTD under
this definition the WTD of the second kind.
Figure~\ref{STA_hist_waitTime2} shows the flux rope WTD of the
second kind. One can notice the outliers near the two ends. The
outliers on the tail locate between 300 minutes to 400 minutes,
which is the same location as the outliers appearing in
Figure~\ref{duration_size} (a) and (b). Apparently, these outliers
are due to boundary cutoff effect since the flux rope duration
contributes to the waiting time of the second kind. As for the the
outliers with the shortest waiting time, we also attribute them to
boundary cutoff effect. Because in our database, the minimum flux
rope duration is set at 9 minutes, there are no flux rope events
with duration less than 9 minutes. However, the separation between
two flux ropes can be of any time length. When we use the
definition of the first kind, the shortest waiting time can be
zero, but for the definition of the second kind, the shortest
waiting time has to be 9 minutes. As a result, there is a lack of
events with short waiting time in the WTD of the second kind.
Excluding these abnormalities, we apply the same fitting approach
as in Figure~\ref{STA_hist_waitTime}. In
Figure~\ref{STA_hist_waitTime2}, for each curve, we can see that
the section less than 80 minutes is well fitted by an exponential
function, and the  section greater than 80 minutes is well fitted
by a power law function. The exponential fittings in
Figure~\ref{STA_hist_waitTime2} are better than those in
Figure~\ref{STA_hist_waitTime}, indicating that the events
associated with short waiting time ($<80$ minutes) are subject to
the simple Poisson process, while the ones associated with longer
waiting time ($>80$ minutes) seem to have clustering behavior. The
combination of an exponential and a power law distribution implies
there may be two distinct mechanisms responsible for the event
occurrence.

It is worth noting that the fitting result given in
Figure~\ref{STA_hist_waitTime2} is similar to the result on
current sheets WTD obtained by \citet{Bruno2001,Miao2011}.
\citet{Bruno2001} showed an exponential fitting to the waiting
time of intermittent events (current sheets) identified from the
Helios 2 spacecraft measurements at 0.9 AU in a high speed stream.
The exponential fit persisted for a range of waiting time between
12 minutes and $\sim$ 2 hours, compatible with the
exponential-fitting range of $\Delta t$ in
Figure~\ref{STA_hist_waitTime2}. \citet{Miao2011} studied the
statistical characteristics of current sheets observed by Ulysses
in 1997, 2004, and 2005 (also including a few days in 1996 and
2006). They found that the current sheets WTD can be well fitted
by an exponential function and a power law function as well.
However, the break point in Miao et al.'s fitting is at $\sim
e^{10}$ seconds (367 minutes) (see Figure 9 in \citet{Miao2011}),
and the power law index is -1.85, which is different from the
power law index in our results. A word of caution is that Miao et
al.'s results were obtained from Ulysses observations near the
ecliptic plane, but at radial distances $\sim$ 4-5 AU from the
Sun. Therefore significant evolution, at least passively, has
occurred between the two sets of results. Although not totally
comparable, the similar piecewise behaviors of the flux rope WTD
and the current sheet WTD imply that they may share similar
generation mechanisms or have some kind of association.

\citet{Greco2009a, Greco2009b} showed the consistency between the
WTDs of the solar wind discontinuities observed by the ACE
spacecraft at 1 AU and the corresponding WTDs of intermittent
structures (current sheets) from the MHD turbulence simulations,
suggesting that the solar wind magnetic structures may be created
locally by MHD turbulence. Here the structures referred to by
Greco et al. consist of ``small random currents", ``current cores"
(i.e. flux ropes), and ``intermittent current sheets"
\citep{Greco2008, Greco2009a}. In their study, the WTD of MHD
simulation result agrees well with the WTD of ACE observational
data in the short waiting time range ($<50$ minutes, the
correlation length scale of solar wind turbulence). For the
departure from the power law beyond 50 minutes, they attributed
that to the limited length scales, thus no large-scale features
allowed in the MHD simulations. In general space plasma scenario,
the current sheets can be considered as boundaries of flux ropes
with negligible thickness. To make a direct comparison with the
WTD of current sheets in Greco et al.'s study \citep{Greco2009a,
Greco2009b}, where current sheets were identified with zero
thickness, we provide a proxy to those current sheets from our
small-scale magnetic flux rope database. Considering that the flux
ropes are bounded by current sheets, we can assume that there are
current sheets existing at the starting time and the end time of
each flux rope interval. We call these current sheets flux rope
``walls". Then the time interval between adjacent walls can be
considered as a proxy to the waiting time of current sheets. We
showed in \citet{zhengandhu2018} the consistent results agreeing
with \citet{Greco2009b} in terms of the wall-to-wall time
(equivalent to the waiting time of current sheets) and axial
current density distributions from our GS-based analysis of
small-scale magnetic flux ropes. Such analysis provided the direct
evidence, through our unique approach, in supporting the view of
locally generated coherent structures intrinsic to the dynamic
processes in the solar wind as manifested either by magnetic
reconnection or turbulence cascade.

\section{Locations of SSMFR with Respect to the HCS}\label{sec:HCS}
 The argument
that the small-scale magnetic flux ropes are created locally is
partly based on the fact that some events were found near the
heliospheric current sheet (HCS). Upon the first discovery of
small-scale flux ropes, \citet{Moldwin1995} interpreted their
findings in terms of multiple magnetic reconnection of previously
open field lines at the HCS. Later, \citet{Moldwin2000} reported
several additional small-scale flux ropes observed by both IMP 8
and Wind spacecraft. They suggested that these small-scale flux
ropes were created in the HCS instead of the solar corona because
of  the following reasons: (1) bimodal size distribution, (2) lack
of expansion, (3) different plasma characteristics, and (4)
similar radial scale size with estimated HCS thickness.
\citet{Cartwright2010} studied the distribution of small-scale
flux ropes location with respect to the HCS, and found that most
events were observed near HCS, although, as we pointed out
earlier, their event sample size is extremely small. In this
section, we redo the same analysis based on our database, which
contains much more number of events than Cartwright and Moldwin's
database.

 \begin{figure}[ht]   
  \begin{center}
    \includegraphics[width=0.9\textwidth]{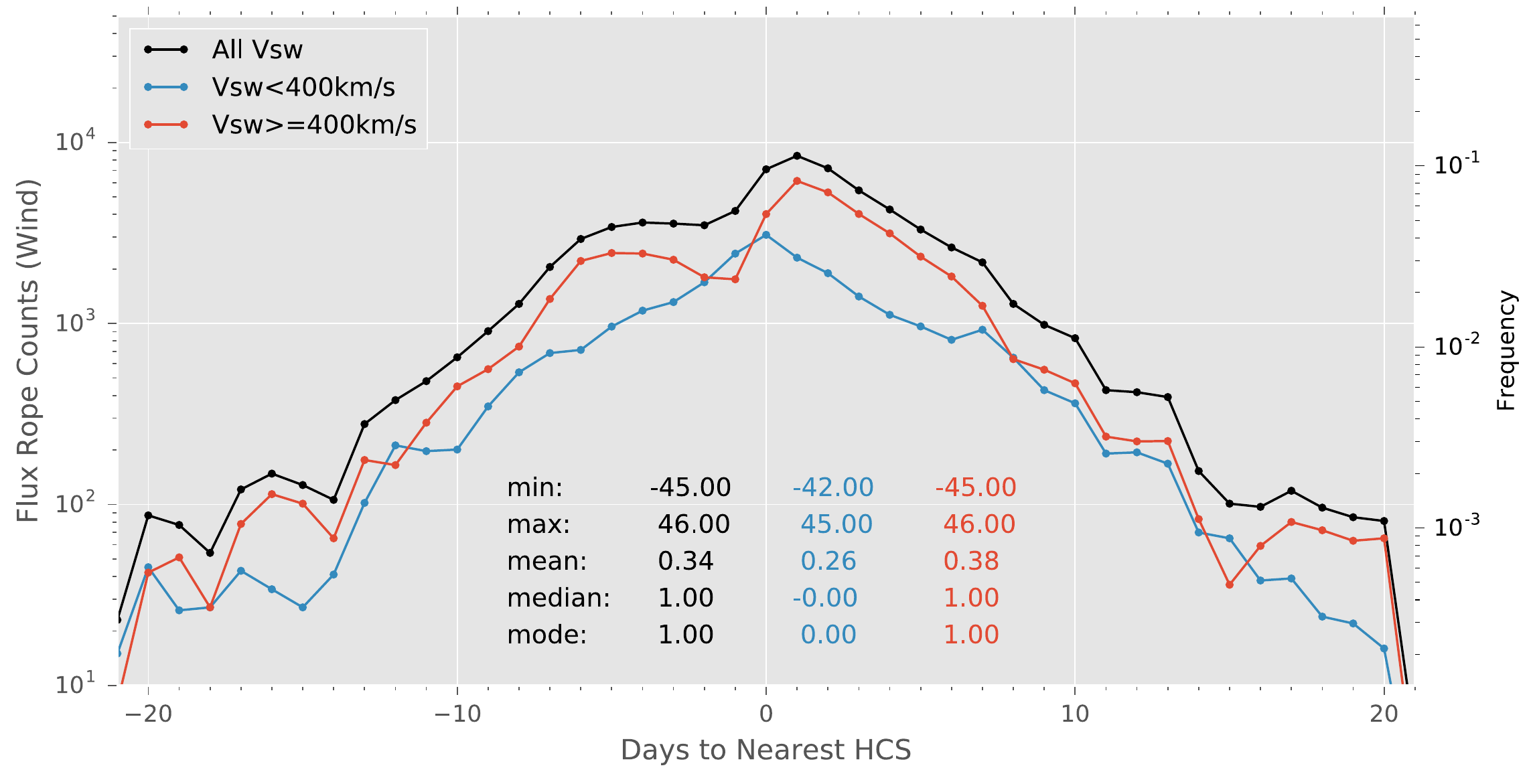}
  \end{center}
  \caption{Small-scale flux rope distribution with respect to the time to nearest heliospheric current sheet.
   The events are binned by 1 day. Negative value means the time ahead of the HCS crossing.
   The red and blue curves represent the events occurred under different solar wind speed conditions (blue curve:
   $\overline{V}_{sw}<$ 400 km/s; red curve: $\overline{V}_{sw}\ge$ 400 km/s). Black curve is for all events.}\label{STA_hist_daysToHCS}
\end{figure}

Figure~\ref{STA_hist_daysToHCS} is the histogram of the time to
the nearest HCS for the small-scale flux ropes in our database.
The HCS crossing times are taken from L. Svalgaard's list of
sector boundaries in the solar wind
(\textit{http://www.leif.org/research/sblist.txt}). This plot
indicates that the small-scale flux ropes tend to appear near the
sector boundary crossings. This result is consistent with
Cartwright and Moldwin's result (see Figure 9 in
\citet{Cartwright2010}). However, from
Figure~\ref{STA_hist_daysToHCS} one can see that the peak of the
black curve is located at 1 day after the HCS crossings, instead
of 0 day given in \citet{Cartwright2010}. After we split the
entire event set into two subsets based on solar wind speed, we
find that the events with solar wind speed less than 400 km/s tend
to occur within $1$ day with respect to HCS with a central peak
right at $0$ day. The events with solar wind speed greater than or
equal to 400 km/s tend to occur at 1 day after HCS crossing, with
two broad and asymmetric peaks at $\sim -5$ days and $\sim 1$ day.
Correspondingly, a broad secondary peak is also located at $-5$
days for the black curve. The similar trends of black and red
curves  imply that the secondary peak in black curve is
contributed by the events with solar wind speed
$\overline{V}_{sw}\ge$ 400 km/s.  The peak at $\sim 1$ day is more
pronounced, indicating a close association to HCS crossings for
relatively fast solar wind. We caution not to over-interpret the
peak preceding the HCS crossings, which is weak and the separation
is large so that the association is much less certain. We offer an
alternative explanation which is that the flux ropes near $-5$ day
to HCS possibly occur right after other HCS crossings. However,
these crossings are not observed due to unknown reasons. As a
result, these flux ropes appear to occur at a relatively larger
separation time preceding the nearest HCS as identified. This can
happen when the sector boundaries are not in the ecliptic plane,
so that they are not observed by spacecraft in the ecliptic plane.
The blue curve ($\overline{V}_{sw}<400$ km/s) is symmetric, which
is different from the red curve. The peak of the blue curve is at
0 day, indicating that the small-scale flux ropes tend to occur in
the same day with HCS crossings under slow solar wind speed
condition.

 \begin{figure}         
  \begin{center}
    \includegraphics[width=1.0\textwidth]{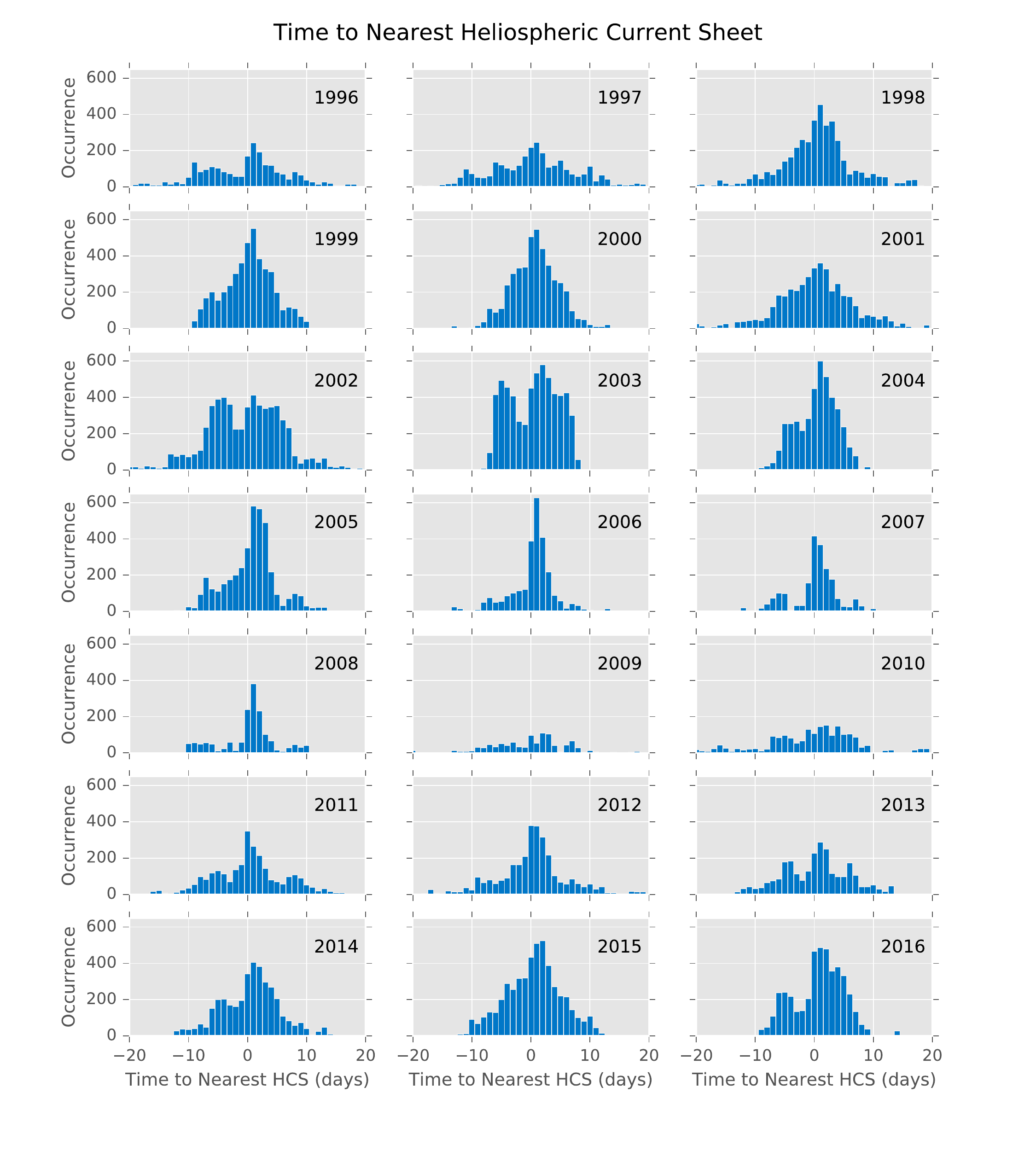}
  \end{center}
  \caption{Annual small-scale flux rope counts distribution with respect to the time to the nearest heliospheric current sheet.
  The events are binned by 1 day. Negative days mean the occurrence time ahead of HCS crossing.}\label{TimeToHCS_yearly}
\end{figure}


In order to look into more details on the distribution of flux
rope occurrence time/location with respect to HCS, we plot the
histograms of days to nearest HCS for each year in
Figure~\ref{TimeToHCS_yearly}. From Figure~\ref{TimeToHCS_yearly},
one can see that the small-scale flux ropes do appear near HCS in
each year, although the spread in time can be wide, reaching
$\pm10$ days. In addition, the distributions have year by year
variations. During solar cycle 23 (1996$\sim$2008), the histograms
show a triangle distribution in the years 1998, 1999, 2000, and
2001, all of which are during ascending phase of solar activity.
The histograms show an additional peak near $-5$ days in the years
2002, 2003, and 2004, all of which are during descending phase.
However, in solar cycle 24, there is no such a clear
classification. The double peaks show up in years 2011, 2013,
2014, and 2016, in which 2011 and 2013 are during the ascending
phase, while 2014 and 2016 are during the descending phase. From
the analysis on Figure~\ref{STA_hist_daysToHCS}, we concluded that
the secondary peaks near $-5$ day are more directly associated
with medium and high speed solar wind. In
Figure~\ref{TimeToHCS_yearly}, most of those years that have
secondary peaks are during the descending phase of each solar
cycle, which is usually dominated by high speed solar wind
streams.

 \begin{figure}  
  \begin{center}
    \includegraphics[width=1.0\textwidth]{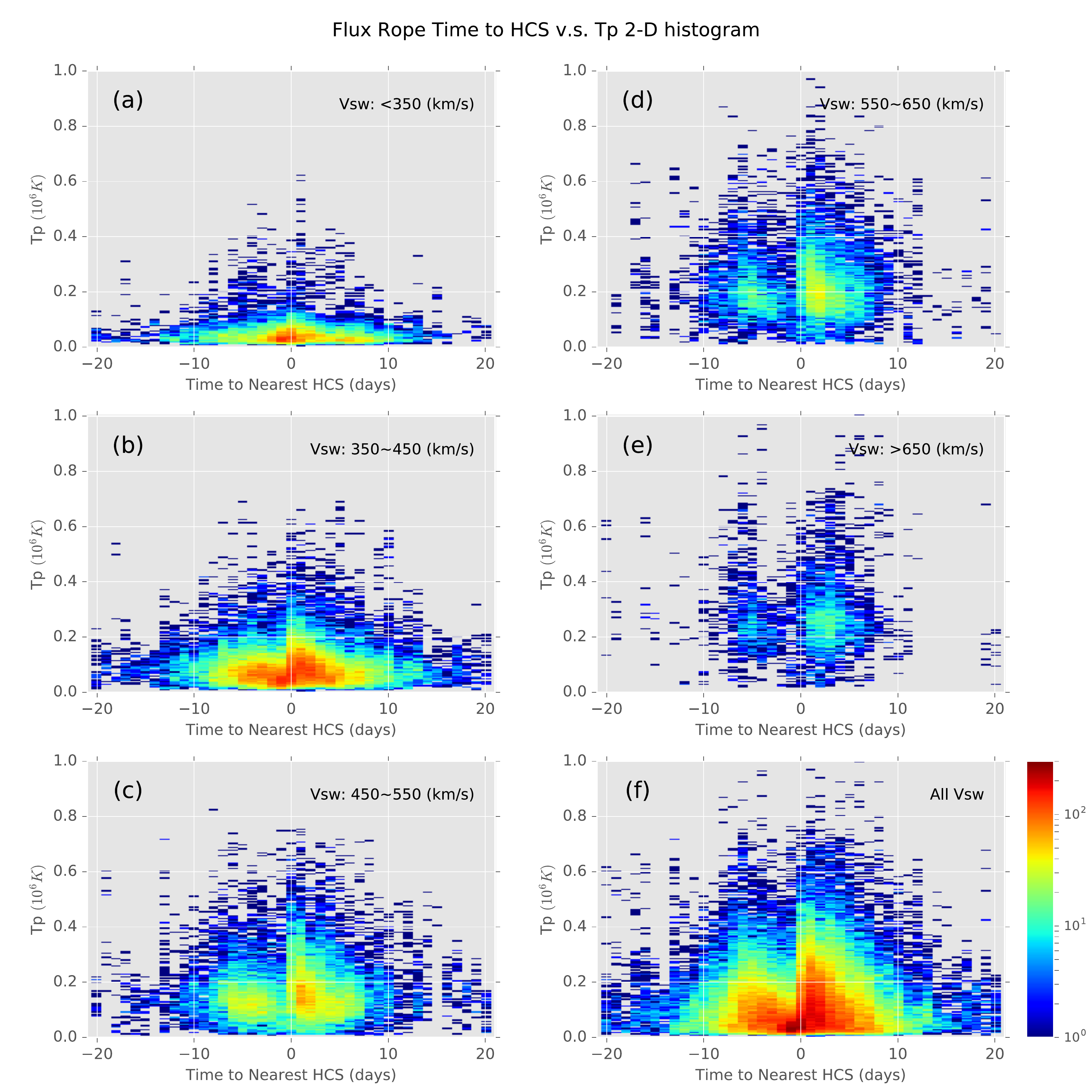}
  \end{center}
  \caption{The 2-D histograms of flux rope average proton temperature $\textit{versus}$ time to the nearest HCS. The horizontal axis is
   binned by 1 day. Negative values mean the time ahead of HCS. Subplots (a) to (e) are histograms under different solar wind speed
   conditions, and subplot (f) is the histogram for the entire event set. The color bar represents the small-scale flux rope counts.}\label{DaysToHCS_vs_Tp}
\end{figure}

 \begin{figure}     
  \begin{center}
    \includegraphics[width=1.0\textwidth]{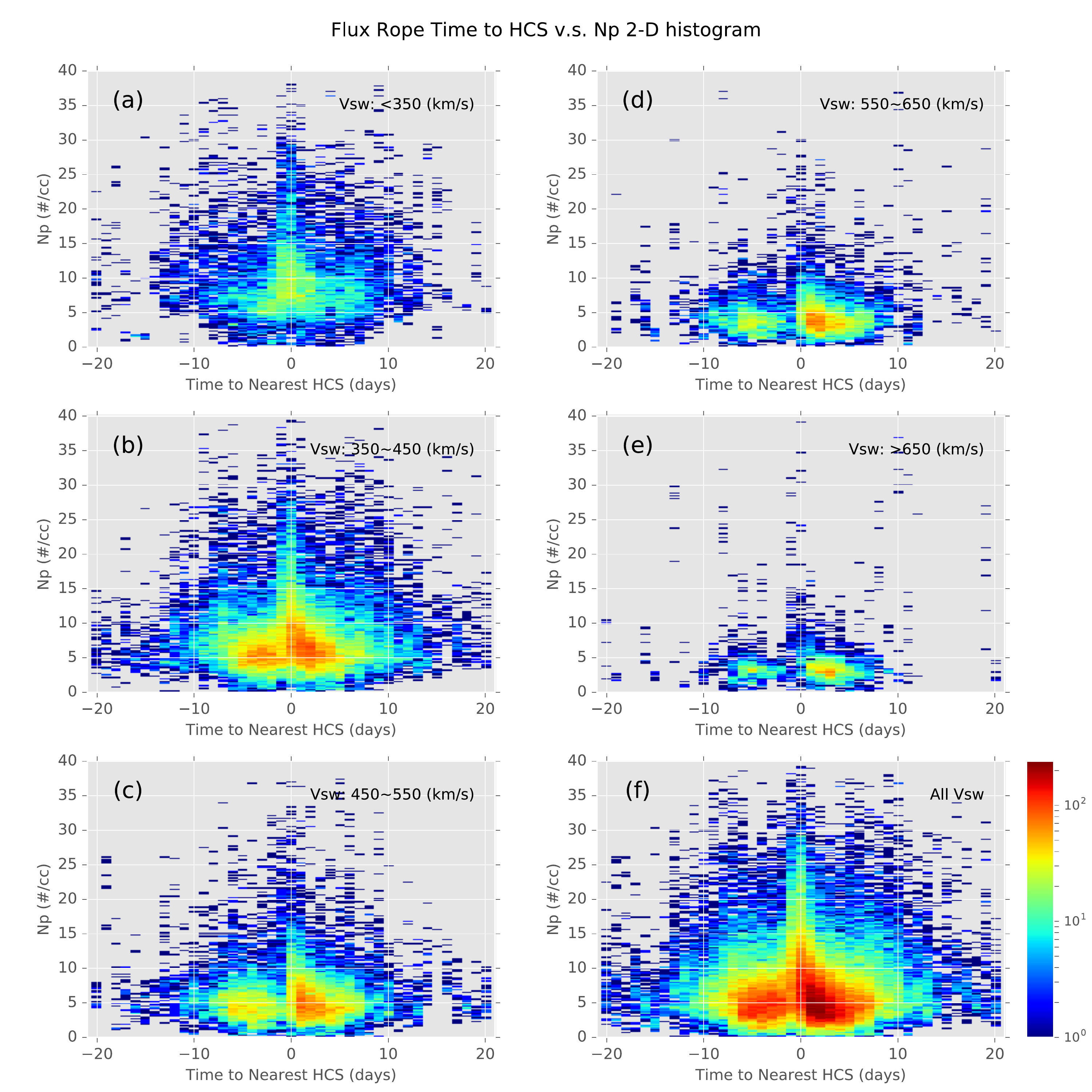}
  \end{center}
  \caption{The 2-D histograms of flux rope average proton number density $\textit{versus}$ time to nearest HCS. The format is the same as in Figure~\ref{DaysToHCS_vs_Tp}.}\label{DaysToHCS_vs_Np}
\end{figure}

Figures~\ref{DaysToHCS_vs_Tp} and \ref{DaysToHCS_vs_Np} are the
2-D histograms of time to nearest HCS \textit{versus} average
proton temperature and  average proton number density,
respectively. Figure~\ref{DaysToHCS_vs_Tp} (a) and (b) show that
the small-scale flux ropes with lower proton temperature spread
widely around the HCS. Figure~\ref{DaysToHCS_vs_Tp} (c), (d), and
(e) show that in medium and high speed solar wind, there are fewer
number of small-scale flux ropes appearing far from HCS, and the
proton temperature is elevated. The triangle outline in
Figure~\ref{DaysToHCS_vs_Tp} (f) shows a pattern consistent with
Figure~\ref{STA_hist_daysToHCS}. Each panel of
Figure~\ref{DaysToHCS_vs_Np} shows that the flux ropes with higher
proton number density are more likely to appear near the HCS. We
also note the location of the colors denoting higher occurrence
rate (red  and yellow color). In Figure~\ref{DaysToHCS_vs_Np} (a),
the yellow color region is within the range $5\leq N_p \leq 15$.
In Figure~\ref{DaysToHCS_vs_Np} (b), the upper boundary of the red
and yellow color region is at about $N_p=15$, but the lower
boundary exceeds beyond $N_p=5$. In Figure~\ref{DaysToHCS_vs_Np}
(c) and (d) the red and yellow color region moves down to $0\leq
N_p \leq 10$, and in Figure~\ref{DaysToHCS_vs_Np} (e), the red and
yellow color region is below $N_p = 5$. The different locations of
the red and yellow color regions with different solar wind speed
indicate that for higher solar wind speed, the flux ropes tend to
have lower proton number density.

Combining Figures~\ref{DaysToHCS_vs_Tp} and \ref{DaysToHCS_vs_Np},
they seem to indicate that small-scale flux ropes occurring at the
HCS crossings (0 day) tend to have lower average proton
temperature and higher average density, consistent with the
general plasma property of the HCS, embedded in relatively low
speed streams. The opposite seems to be the case for flux ropes
occurring in the vicinity of HCS, within $\pm 10$ days.


\section{Summary and Discussion}

In summary, we have developed a new small-scale magnetic flux rope
detection algorithm based on the GS equation, and applied this
approach to Wind spacecraft in-situ measurements to detect
small-scale magnetic flux ropes. We successfully detected 74,241
small-scale flux rope events from 1996 to 2016, covering two solar
cycles. We have described, in length, the detailed automated
detection algorithm and the statistical properties resulted from
the detected events. This large number of small-scale flux ropes
has not been discovered by any other detection approach or in any
other previous studies, owing to the implementation of a highly
computerized and automated algorithm rooted on the most general
theoretical considerations. We build and maintain a flux rope
database online to provide scientific community open access to
this database. In practice, we have relied on high-performance
computing facilities to fulfill this non-trivial task.  By
developing this database, we contribute to the study of relevant
physical processes throughout the heliosphere. It is expected that
the database will be expanded by including additional results from
other spacecraft missions, covering a range of helio-latitudes and
heliocentric distances. We hope that this database will benefit
the broader scientific community of space plasma physics and
astrophysical  research.

Using this database, we have performed the statistical analysis as
well as individual case studies on small-scale magnetic flux ropes
in the solar wind, and obtained a number of significant results as
summarized below:

\begin{enumerate}

\item The occurrence of small-scale magnetic flux ropes has strong solar cycle
dependency. The occurrence count is about $3,500$ events per year
on average over the past two solar cycles.

\item The small-scale magnetic  flux ropes  tend to align along the Parker spiral direction, in the ecliptic plane.
This is consistent with the orientation pattern of ``flux tubes".
This indicates that they belong to the general population of
``flux tubes" identified in the same space plasma regime.

\item The small-scale magnetic flux ropes show
 different statistical properties under different solar wind speed conditions.
 In low speed ($<400$ km/s) solar wind, the flux ropes tend to have lower proton temperature, and higher proton number density,
 and the event counts peak at the HCS. In high speed
 ($\ge400$ km/s) solar wind, they tend to have higher proton
 temperature and lower proton number density. The event counts
 peak near the vicinity of the HCS, $\sim$ 1 day later.

\item 
Both the duration and scale size distributions obey the power law.
The power index close to -2 suggests that the larger scale size
flux ropes (0.01 $\sim$ 0.05 AU) in our database under the
condition of relatively high speed solar wind
($\overline{V}_{sw}\ge$400 km/s) may have closer relation to a
solar source than the small-scale flux ropes under other
conditions, considering the similar distributions of events with
clear solar origin, i.e., flares and CMEs.

\item From the waiting time analysis, we found that the distribution for the shorter waiting time can be fitted by an
exponential function, indicating that these flux ropes may undergo
the pure Poisson process. The distribution associated with longer
waiting time can be fitted by a power law function, indicating the
clustering behavior of these flux ropes. The break point is at 60
$\sim$ 80 minutes.

\item The wall to wall time distribution obeys double power law with the break point at about 60 minutes as reported separately
 by \citet{zhengandhu2018} which corresponds to the
 typical correlation length scale in solar wind turbulence
\citep{Matthaeus2005}. This result is consistent with the WTDs of
intermittent structures (current sheets) from MHD turbulence
simulations for the inertia range that is covered by the scale
size range in our database.

\item The study of the locations of small-scale magnetic flux ropes with respect to the HCS shows that the small-scale magnetic
 flux ropes  tend to accumulate near the HCS, especially for the flux ropes under slow solar wind condition ($\overline{V}_{sw}<$400 km/s).

\item Some additional statistics show that the small-scale magnetic flux ropes with larger scale sizes tend to have
low proton temperature, and tend to appear in slow speed solar
wind. These behaviors, especially the former, are similar to MCs.

\item There is little  cycle-to-cycle variation in the statistics of the majority
of selected bulk parameters for solar cycles 23 and 24, except for
the average magnetic field magnitude and plasma $\beta$. The
average magnetic field magnitude within flux rope intervals is
weaker in cycle 24.
\end{enumerate}

We had reported in \citet{zhengandhu2018} that the wall-to-wall
time distribution  obeys double power laws with the break point at
60 minutes (corresponding to the correlation length in solar wind
turbulence), which is consistent with the waiting time
distributions of intermittent structures from magnetohydrodynamic
(MHD) turbulence simulations and the related observations. Through
our unique approach, we provided the direct evidence in supporting
the view of locally generated coherent structures intrinsic to the
dynamic processes in the solar wind as manifested by magnetic
reconnection and inverse turbulence cascade. We also performed
case studies on the small-scale magnetic flux ropes downstream of
interplanetary shocks, and found that the peaks of enhanced ions
flux correspond to the merging X-line between two adjacent flux
ropes, indicating that the merging flux ropes are able to energize
particles within certain energy bands \citep{Zheng2017}.

These results support the notion of a ``sea of flux ropes", i.e.,
the ubiquitous  existence of SSMFRs in the solar wind medium. They
are expected to play an important role in fundamental space plasma
processes. For example, a recent study by \citet{zhao2018uly}
revealed  detailed observational signatures of particle
energization by small-scale magnetic flux ropes, which fit the
theory of
\citet{Zank2014,Zank2015b,le_Roux2015,le_Roux2016,le_Roux2018}.
Over a hundred SSMFR events were identified based on the GS
reconstruction approach described here from the Ulysses spacecraft
in-situ measurements over a time period of two weeks. They appear
to co-locate with the energetic proton flux enhancement and the
theoretic predictions fit the observed time-intensity profile and
particle spectra well. It is therefore promising to further pursue
more advanced study of flux rope dynamics involving contraction
and merging that can be better addressed by other GS-type and full
MHD-based approaches \citep{2012MEEP....1...71H,Hu2017GSreview}.

As a point for discussion, Table~\ref{last} lists, as a first
attempt,  the observational evidence obtained from our analysis,
in support of the two competing views on the origin of small-scale
magnetic flux ropes. Each column contains the results that have
been considered well established and exclusive for the
corresponding origination mechanism. This is nowhere near to be
final and we expect the table to be updated when further analyses
are to be carried out, especially when the new discoveries are
expected to be returned by the  Parker Solar Probe and Solar
Orbiter missions. Additionally, based on the present analysis
results, the event occurrence rate, the axial orientations, and
the power-law distributions in duration and scale sizes seem to be
arguably supporting both hypotheses, at least not in direct
contradiction to either view. For example, \citet{Borovsky2008}
analyzed the orientations of 65,860 flux tubes from 1998 to 2004
identified by searching for their boundaries, usually
corresponding to discontinuities from the ACE spacecraft in-situ
measurements. He found that the axial orientations are mostly
aligned with the nominal Parker spiral direction at 1 AU. He
therefore advocates the view that these flux tubes are distinct
from one another and still rooted on the Sun. They form the
``spaghetti-like" structure of the solar wind, connecting back to
the source. On the other hand, such axial orientations are also
consistent with the view that these structures are generated from
solar wind turbulence
\citep[e.g.,][]{2007ApJ...667..956M,2008PhRvL.100i5005S,Zank2017}
which has a prominent 2D component and the associated
(perpendicular) guide field along the Parker spiral direction.
Another controversial issue concerns the phenomenon of the
accumulation of these small-scale magnetic flux ropes near HCS.
This had long been considered as an evidence supporting the view
that they originate locally across the HCS, presumably through
magnetic reconnection. On the other hand, these structures may
also originate from streamer belts in low corona, as shown by the
latest numerical simulations of generation of ``plasma blobs"
\citep{2018ApJ...859....6H}. Therefore the HCS could serve as a
conduit for them to propagate out and reach 1 AU,  although
detailed studies on their configurations suitable for the GS
reconstruction approach may be able to discern these two possible
scenarios.
\begin{table}[h]
\caption{Observational evidence in support of two
competing views on the origin of small-scale magnetic flux
ropes.}\label{last}
\begin{tabular}{ll}
\hline
Solar Origin & Local Origin\\
\hline
 Flux rope configuration similar to MCs& --- \\
--- & Properties similar to 2D MHD turbulence\tablenotemark{a}  \\
--- & Plasma properties different from MCs\\
--- & No significant cycle-to-cycle variations\\
 \hline
\end{tabular}
 \tablenotetext{a}{See, e.g., \citet{zhengandhu2018}}
\end{table}

It is worth noting that in our database there may still exist
events with relatively small duration overlapping with large-scale
ICMEs identified by other sources, either published or archived
online. We did not take the effort in eliminating those from our
current database, since the impact is minimal, especially for the
small-duration events we are most interested in with duration less
than 1 hour \citep{zhengandhu2018}. They occur at a rate of a few
hundreds a month (see
Figure~\ref{STA_fluxrope_monthly_counts_duration}), compared with
the occurrence of ICMEs at a rate of no more than 2-3 a month, on
average \citep{2016JGRA..121.5005Y}. Therefore the possible error
introduced in the event count is $\lesssim$1\%. Of course
individual users can take the further steps of eliminating these
events by cross-checking with lists of well-identified ICMEs,
depending on their own needs. A remedy to this is to extend our
search window sizes to include longer duration up to tens of
hours. Such an effort is currently undertaken for the ACE
spacecraft measurements, so that large-scale flux rope ICMEs can
also be identified through our automated approach.

\acknowledgments We are grateful to our colleagues at SPA/CSPAR,
UAH, Drs.~Laxman Adhikari,  Gang Li, Gary Webb, Gary Zank, and
Lingling Zhao for on-going collaborations. We also thank Dr.~Olga
Khabarova for stimulating discussions. The Wind spacecraft data
are provided by the NASA CDAWeb. We acknowledge NASA grants
NNX12AH50G, NNX14AF41G, NNX15AI65G, NNX17AB85G, subawards NRL
N00173-14-1-G006 and
 SAO SV4-84017, NSF-DOE grant PHY-1707247, and NSF grant AGS-1650854 for support. Special
 thanks also go to the SCOSTEP/VarSITI program for support of the
 development and maintenance of the on-line small-scale magnetic flux rope
 database website. Besides the in-house computer cluster Bladerunner, part of the work is also performed on the BlueShark Cluster of Florida Institute of Technology,
 supported by the National Science Foundation under Grant No. CNS 09-23050.

%








\begin{thebibliography}{}
\expandafter\ifx\csname natexlab\endcsname\relax\def\natexlab#1{#1}\fi
\providecommand{\url}[1]{\href{#1}{#1}}

\bibitem[{Aschwanden \& McTiernan(2010)}]{Aschwanden2010}
Aschwanden, M.~J., \& McTiernan, J.~M. 2010, The Astrophysical Journal, 717,
  683.
\newblock \url{http://stacks.iop.org/0004-637X/717/i=2/a=683}

\bibitem[{Bak {et~al.}(1987)Bak, Tang, \& Wiesenfeld}]{Bak1987}
Bak, P., Tang, C., \& Wiesenfeld, K. 1987, Phys. Rev. Lett., 59, 381.
\newblock \url{https://link.aps.org/doi/10.1103/PhysRevLett.59.381}

\bibitem[{{Borovsky}(2008)}]{Borovsky2008}
{Borovsky}, J.~E. 2008, Journal of Geophysical Research (Space Physics), 113,
  A08110

\bibitem[{{Bruno} {et~al.}(2001){Bruno}, {Carbone}, {Veltri}, {Pietropaolo}, \&
  {Bavassano}}]{Bruno2001}
{Bruno}, R., {Carbone}, V., {Veltri}, P., {Pietropaolo}, E., \& {Bavassano}, B.
  2001, Planetary Space Science, 49, 1201

\bibitem[{{Cartwright} \& {Moldwin}(2010)}]{Cartwright2010}
{Cartwright}, M.~L., \& {Moldwin}, M.~B. 2010, Journal of Geophysical Research
  (Space Physics), 115, A08102, a08102.
\newblock \url{http://dx.doi.org/10.1029/2009JA014271}

\bibitem[{Feng {et~al.}(2007)Feng, Wu, \& Chao}]{Feng2007}
Feng, H.~Q., Wu, D.~J., \& Chao, J.~K. 2007, Journal of Geophysical Research:
  Space Physics, 112, a02102.
\newblock \url{http://dx.doi.org/10.1029/2006JA011962}

\bibitem[{Feng {et~al.}(2008)Feng, Wu, Lin, Chao, Lee, \& Lyu}]{Feng2008}
Feng, H.~Q., Wu, D.~J., Lin, C.~C., {et~al.} 2008, Journal of Geophysical
  Research: Space Physics, 113, A12105, a12105.
\newblock \url{http://dx.doi.org/10.1029/2008JA013103}

\bibitem[{{Georgoulis} \& {Vlahos}(1998)}]{Georgoulis1998}
{Georgoulis}, M.~K., \& {Vlahos}, L. 1998, Astronomy and Astrophysics, 336, 721

\bibitem[{Gopalswamy {et~al.}(2015)Gopalswamy, Yashiro, Xie, Akiyama, \&
  Mäkelä}]{Gopalswamy2015}
Gopalswamy, N., Yashiro, S., Xie, H., Akiyama, S., \& Mäkelä, P. 2015,
  Journal of Geophysical Research: Space Physics, 120, 9221, 2015JA021446.
\newblock \url{http://dx.doi.org/10.1002/2015JA021446}

\bibitem[{Greco {et~al.}(2008)Greco, Chuychai, Matthaeus, Servidio, \&
  Dmitruk}]{Greco2008}
Greco, A., Chuychai, P., Matthaeus, W.~H., Servidio, S., \& Dmitruk, P. 2008,
  Geophysical Research Letters, 35, doi:10.1029/2008GL035454, l19111.
\newblock \url{http://dx.doi.org/10.1029/2008GL035454}

\bibitem[{Greco {et~al.}(2009{\natexlab{a}})Greco, Matthaeus, Servidio,
  Chuychai, \& Dmitruk}]{Greco2009a}
Greco, A., Matthaeus, W.~H., Servidio, S., Chuychai, P., \& Dmitruk, P.
  2009{\natexlab{a}}, The Astrophysical Journal Letters, 691, L111.
\newblock \url{http://stacks.iop.org/1538-4357/691/i=2/a=L111}

\bibitem[{Greco {et~al.}(2009{\natexlab{b}})Greco, Matthaeus, Servidio, \&
  Dmitruk}]{Greco2009b}
Greco, A., Matthaeus, W.~H., Servidio, S., \& Dmitruk, P. 2009{\natexlab{b}},
  Phys. Rev. E, 80, 046401.
\newblock \url{https://link.aps.org/doi/10.1103/PhysRevE.80.046401}

\bibitem[{Guidorzi {et~al.}(2015)Guidorzi, Dichiara, Frontera, Margutti,
  Baldeschi, \& Amati}]{Guidorzi2015}
Guidorzi, C., Dichiara, S., Frontera, F., {et~al.} 2015, The Astrophysical
  Journal, 801, 57.
\newblock \url{http://stacks.iop.org/0004-637X/801/i=1/a=57}

\bibitem[{{Hasegawa}(2012)}]{2012MEEP....1...71H}
{Hasegawa}, H. 2012, Monographs on Environment, Earth and Planets, 1, 71

\bibitem[{Hau \& Sonnerup(1999)}]{Hau1999}
Hau, L.-N., \& Sonnerup, B. U.~O. 1999, Journal of Geophysical Research: Space
  Physics, 104, 6899.
\newblock \url{http://dx.doi.org/10.1029/1999JA900002}

\bibitem[{{Higginson} \& {Lynch}(2018)}]{2018ApJ...859....6H}
{Higginson}, A.~K., \& {Lynch}, B.~J. 2018, \apj, 859, 6

\bibitem[{{Hu}(2017)}]{Hu2017GSreview}
{Hu}, Q. 2017, Sci.~China Earth Sciences, 60, 1466

\bibitem[{Hu {et~al.}(2004)Hu, Smith, Ness, \& Skoug}]{Hu2004}
Hu, Q., Smith, C.~W., Ness, N.~F., \& Skoug, R.~M. 2004, Journal of Geophysical
  Research: Space Physics, 109, doi:10.1029/2003JA010101, a03102.
\newblock \url{http://dx.doi.org/10.1029/2003JA010101}

\bibitem[{Hu \& Sonnerup(2000)}]{Hu2000}
Hu, Q., \& Sonnerup, B. U.~O. 2000, Geophysical Research Letters, 27, 1443.
\newblock \url{http://dx.doi.org/10.1029/1999GL010751}

\bibitem[{{Hu} \& {Sonnerup}(2001)}]{Hu2001}
{Hu}, Q., \& {Sonnerup}, B.~U.~{\"O}. 2001, Geophysical Research Letters, 28,
  467

\bibitem[{Hu \& Sonnerup(2002)}]{Hu2002}
Hu, Q., \& Sonnerup, B. U.~O. 2002, Journal of Geophysical Research: Space
  Physics, 107, SSH 10.
\newblock \url{http://dx.doi.org/10.1029/2001JA000293}

\bibitem[{{Khrabrov} \& {Sonnerup}(1998)}]{Khrabrov1998}
{Khrabrov}, A.~V., \& {Sonnerup}, B.~U.~{\"O}. 1998, ISSI Scientific Reports
  Series, 1, 221

\bibitem[{le~Roux {et~al.}(2018)le~Roux, Zank, \& Khabarova}]{le_Roux2018}
le~Roux, J.~A., Zank, G.~P., \& Khabarova, O. 2018, The Astrophysical Journal,
  in press

\bibitem[{le~Roux {et~al.}(2015)le~Roux, Zank, Webb, \&
  Khabarova}]{le_Roux2015}
le~Roux, J.~A., Zank, G.~P., Webb, G.~M., \& Khabarova, O. 2015, The
  Astrophysical Journal, 801, 112.
\newblock \url{http://stacks.iop.org/0004-637X/801/i=2/a=112}

\bibitem[{le~Roux {et~al.}(2016)le~Roux, Zank, Webb, \&
  Khabarova}]{le_Roux2016}
le~Roux, J.~A., Zank, G.~P., Webb, G.~M., \& Khabarova, O.~V. 2016, The
  Astrophysical Journal, 827, 47.
\newblock \url{http://stacks.iop.org/0004-637X/827/i=1/a=47}

\bibitem[{{Lepping} {et~al.}(1995){Lepping}, {Ac{\~u}na}, {Burlaga}, {Farrell},
  {Slavin}, {Schatten}, {Mariani}, {Ness}, {Neubauer}, {Whang}, {Byrnes},
  {Kennon}, {Panetta}, {Scheifele}, \& {Worley}}]{1995SSRv...71..207L}
{Lepping}, R.~P., {Ac{\~u}na}, M.~H., {Burlaga}, L.~F., {et~al.} 1995, \ssr,
  71, 207

\bibitem[{{Li} {et~al.}(2016){Li}, {Yang}, {Li}, {Tian}, {Yang}, \&
  {Huang}}]{Li2016}
{Li}, C., {Yang}, J., {Li}, C., {et~al.} 2016, Scientia Sinica Physica,
  Mechanica and Astronomica, 46, 29501.
\newblock
  \url{http://phys.scichina.com:8083/sciG/EN/abstract/article\_519739.shtml}

\bibitem[{Li {et~al.}(2014)Li, Zhong, Wang, Su, \& Fang}]{Li2014}
Li, C., Zhong, S.~J., Wang, L., Su, W., \& Fang, C. 2014, The Astrophysical
  Journal Letters, 792, L26.
\newblock \url{http://stacks.iop.org/2041-8205/792/i=2/a=L26}

\bibitem[{Lu {et~al.}(1993)Lu, Hamilton, Mctiernan, \& Bromund}]{Lu1993}
Lu, E., Hamilton, R., Mctiernan, J., \& Bromund, K. 1993, Astrophysical
  Journal, 412, 841

\bibitem[{{Lu} \& {Hamilton}(1991)}]{Lu1991}
{Lu}, E.~T., \& {Hamilton}, R.~J. 1991, The Astrophysical Journal Letters, 380,
  L89

\bibitem[{{MacKinnon} \& {MacPherson}(1997)}]{Mackinnon1997}
{MacKinnon}, A.~L., \& {MacPherson}, K.~P. 1997, Astronomy and Astrophysics,
  326, 1228

\bibitem[{{Marubashi} {et~al.}(2010){Marubashi}, {Cho}, \&
  {Park}}]{2010AIPC.1216..240M}
{Marubashi}, K., {Cho}, K.-S., \& {Park}, Y.-D. 2010, Twelfth International
  Solar Wind Conference, 1216, 240

\bibitem[{{Matthaeus} {et~al.}(2007){Matthaeus}, {Bieber}, {Ruffolo},
  {Chuychai}, \& {Minnie}}]{2007ApJ...667..956M}
{Matthaeus}, W.~H., {Bieber}, J.~W., {Ruffolo}, D., {Chuychai}, P., \&
  {Minnie}, J. 2007, \apj, 667, 956

\bibitem[{Matthaeus {et~al.}(2005)Matthaeus, Dasso, Weygand, Milano, Smith, \&
  Kivelson}]{Matthaeus2005}
Matthaeus, W.~H., Dasso, S., Weygand, J.~M., {et~al.} 2005, Phys. Rev. Lett.,
  95, 231101.
\newblock \url{https://link.aps.org/doi/10.1103/PhysRevLett.95.231101}

\bibitem[{Miao {et~al.}(2011)Miao, Peng, \& Li}]{Miao2011}
Miao, B., Peng, B., \& Li, G. 2011, Annales Geophysicae, 29, 237.
\newblock \url{https://www.ann-geophys.net/29/237/2011/}

\bibitem[{{Moldwin} {et~al.}(2000){Moldwin}, {Ford}, {Lepping}, {Slavin}, \&
  {Szabo}}]{Moldwin2000}
{Moldwin}, M.~B., {Ford}, S., {Lepping}, R., {Slavin}, J., \& {Szabo}, A. 2000,
  Journal of Geophysical Research: Space Physics, 27, 57

\bibitem[{Moldwin {et~al.}(1995)Moldwin, Phillips, Gosling, Scime, McComas,
  Bame, Balogh, \& Forsyth}]{Moldwin1995}
Moldwin, M.~B., Phillips, J.~L., Gosling, J.~T., {et~al.} 1995, Journal of
  Geophysical Research: Space Physics, 100, 19903.
\newblock \url{http://dx.doi.org/10.1029/95JA01123}

\bibitem[{{Ogilvie} {et~al.}(1995){Ogilvie}, {Chornay}, {Fritzenreiter},
  {Hunsaker}, {Keller}, {Lobell}, {Miller}, {Scudder}, {Sittler}, {Torbert},
  {Bodet}, {Needell}, {Lazarus}, {Steinberg}, {Tappan}, {Mavretic}, \&
  {Gergin}}]{1995SSRv...71...55O}
{Ogilvie}, K.~W., {Chornay}, D.~J., {Fritzenreiter}, R.~J., {et~al.} 1995,
  \ssr, 71, 55

\bibitem[{{Paschmann} \& {Sonnerup}(2008)}]{Paschmann2008}
{Paschmann}, G., \& {Sonnerup}, B.~U.~O. 2008, ISSI Scientific Reports Series,
  8, 65

\bibitem[{Pearce {et~al.}(1993)Pearce, Rowe, \& Yeung}]{Pearce1993}
Pearce, G., Rowe, A.~K., \& Yeung, J. 1993, Astrophysics and Space Science,
  208, 99.
\newblock \url{https://doi.org/10.1007/BF00658137}

\bibitem[{{Press} {et~al.}(2007){Press}, {Teukolsky}, {Vetterling}, \&
  {Flannery}}]{2007nrca.book.....P}
{Press}, W.~H., {Teukolsky}, S.~A., {Vetterling}, W.~T., \& {Flannery}, B.~P.
  2007, {Numerical recipes : the art of scientific computing} (Cambridge
  University Press)

\bibitem[{{Servidio} {et~al.}(2008){Servidio}, {Matthaeus}, \&
  {Dmitruk}}]{2008PhRvL.100i5005S}
{Servidio}, S., {Matthaeus}, W.~H., \& {Dmitruk}, P. 2008, Physical Review
  Letters, 100, 095005

\bibitem[{Sonnerup \& Guo(1996)}]{Sonnerup1996}
Sonnerup, B. U.~O., \& Guo, M. 1996, Geophysical Research Letters, 23, 3679.
\newblock \url{http://dx.doi.org/10.1029/96GL03573}

\bibitem[{{Sonnerup} \& {Scheible}(1998)}]{Sonnerup1998}
{Sonnerup}, B.~U.~{\"O}., \& {Scheible}, M. 1998, ISSI Scientific Reports
  Series, 1, 185

\bibitem[{Sotolongo-Costa {et~al.}(2000)Sotolongo-Costa, Antoranz, Posadas,
  Vidal, \& Vázquez}]{Sotolongo-Costa2000}
Sotolongo-Costa, O., Antoranz, J.~C., Posadas, A., Vidal, F., \& Vázquez, A.
  2000, Geophysical Research Letters, 27, 1965.
\newblock \url{http://dx.doi.org/10.1029/2000GL011394}

\bibitem[{{Vlahos} {et~al.}(1995){Vlahos}, {Georgoulis}, {Kluiving}, \&
  {Paschos}}]{Vlahos1995}
{Vlahos}, L., {Georgoulis}, M., {Kluiving}, R., \& {Paschos}, P. 1995,
  Astronomy and Astrophysics, 299, 897

\bibitem[{{Wang} \& {Dai}(2013)}]{Wang2013}
{Wang}, F.~Y., \& {Dai}, Z.~G. 2013, Nature Physics, 9, 465

\bibitem[{Wheatland(2000)}]{Wheatland2000}
Wheatland, M.~S. 2000, The Astrophysical Journal Letters, 536, L109.
\newblock \url{http://stacks.iop.org/1538-4357/536/i=2/a=L109}

\bibitem[{Wheatland {et~al.}(1998)Wheatland, Sturrock, \&
  McTiernan}]{Wheatland1998}
Wheatland, M.~S., Sturrock, P.~A., \& McTiernan, J.~M. 1998, The Astrophysical
  Journal, 509, 448.
\newblock \url{http://stacks.iop.org/0004-637X/509/i=1/a=448}

\bibitem[{{Yu} {et~al.}(2016){Yu}, {Farrugia}, {Galvin}, {Lugaz}, {Luhmann},
  {Simunac}, \& {Kilpua}}]{2016JGRA..121.5005Y}
{Yu}, W., {Farrugia}, C.~J., {Galvin}, A.~B., {et~al.} 2016, Journal of
  Geophysical Research (Space Physics), 121, 5005

\bibitem[{{Yu} {et~al.}(2014){Yu}, {Farrugia}, {Lugaz}, {Galvin}, {Kilpua},
  {Kucharek}, {M{\"o}stl}, {Leitner}, {Torbert}, {Simunac}, {Luhmann}, {Szabo},
  {Wilson}, {Ogilvie}, \& {Sauvaud}}]{Yu2014}
{Yu}, W., {Farrugia}, C.~J., {Lugaz}, N., {et~al.} 2014, \jgra, 119, 689

\bibitem[{Zank {et~al.}(2017)Zank, Adhikari, Hunana, Shiota, Bruno, \&
  Telloni}]{Zank2017}
Zank, G.~P., Adhikari, L., Hunana, P., {et~al.} 2017, The Astrophysical
  Journal, 835, 147.
\newblock \url{http://stacks.iop.org/0004-637X/835/i=2/a=147}

\bibitem[{Zank {et~al.}(2014)Zank, le~Roux, Webb, Dosch, \&
  Khabarova}]{Zank2014}
Zank, G.~P., le~Roux, J.~A., Webb, G.~M., Dosch, A., \& Khabarova, O. 2014, The
  Astrophysical Journal, 797, 28.
\newblock \url{http://stacks.iop.org/0004-637X/797/i=1/a=28}

\bibitem[{{Zank} {et~al.}(2015){Zank}, {Hunana}, {Mostafavi}, {Le Roux}, {Li},
  {Webb}, {Khabarova}, {Cummings}, {Stone}, \& {Decker}}]{Zank2015b}
{Zank}, G.~P., {Hunana}, P., {Mostafavi}, P., {et~al.} 2015, The Astrophysical
  Journal, 814, 137

\bibitem[{{Zhao} {et~al.}(2018{\natexlab{a}}){Zhao}, {Adhikari}, {Zank}, {Hu},
  \& {Feng}}]{2018Z}
{Zhao}, L.-L., {Adhikari}, L., {Zank}, G.~P., {Hu}, Q., \& {Feng}, X.~S.
  2018{\natexlab{a}}, \apj, 856, 94

\bibitem[{{Zhao} {et~al.}(2018{\natexlab{b}}){Zhao}, {Zank}, {Khabarova}, {Du},
  {Chen}, {Adhikari}, \& {Hu}}]{zhao2018uly}
{Zhao}, L.-L., {Zank}, G.~P., {Khabarova}, O., {et~al.} 2018{\natexlab{b}},
  \apjl, 864, L34

\bibitem[{{Zheng} \& {Hu}(2018)}]{zhengandhu2018}
{Zheng}, J., \& {Hu}, Q. 2018, \apjl, 852, L23

\bibitem[{{Zheng} {et~al.}(2017){Zheng}, {Hu}, {Chen}, \& {le
  Roux}}]{Zheng2017}
{Zheng}, J., {Hu}, Q., {Chen}, Y., \& {le Roux}, J. 2017, in Journal of Physics
  Conference Series, Vol. 900, Journal of Physics Conference Series, 012024

\end{thebibliography}
\end{document}